\documentclass[11pt,a4paper]{article}
\usepackage[pagella]{preprint-style}
\usepackage{amsthm,changepage,needspace}
\usepackage{tcolorbox}
\tcbuselibrary{breakable,skins}
\newtcolorbox{manuscriptalgorithm}[2][10]{enhanced,breakable,
  colback=white,colframe=black,frame hidden,boxrule=0pt,arc=0pt,left=5pt,right=5pt,
  borderline north={.6pt}{0pt}{black},borderline south={.6pt}{0pt}{black},
  fonttitle={\bfseries\raggedright\hyphenpenalty=10000\relax},colbacktitle=white,coltitle=black,
  titlerule=.4pt,titlerule style={black},title={#2},title after break={#2},
  before skip=\medskipamount,after skip=\medskipamount,
  before={\par\Needspace{#1\baselineskip}\addvspace{\medskipamount}\noindent}}
\newtheoremstyle{authorstatement}{\medskipamount}{\medskipamount}{\itshape}{}{\bfseries}{.}{.5em}{\thmname{#1}\thmnumber{ #2}\thmnote{ {\normalfont(#3)}}}
\theoremstyle{authorstatement}
\newtheorem{assumption}{Assumption}[section]
\newtheorem{theorem}{Theorem}[section]

\newtheorem{proposition}{Proposition}[section]

\theoremstyle{definition}
\newtheorem{remark}{Remark}[section]
\numberwithin{figure}{section}
\numberwithin{table}{section}
\usepackage{newunicodechar}
\newcommand{\manuscriptauthor}[1]{#1}
\newcommand{\manuscriptaffiliation}[1]{#1}
\usepackage[a4paper,margin=25mm]{geometry}
\newunicodechar{✓}{\ensuremath{\checkmark}}
\newunicodechar{β}{\ensuremath{\beta}}
\newunicodechar{（}{(}
\newunicodechar{）}{)}
\usepackage{graphicx,booktabs,longtable,array,calc,float,placeins}
\usepackage{xurl,hyperref}

\hypersetup{colorlinks=true,linkcolor=ppslink,citecolor=ppslink,urlcolor=ppslink,pdftitle={Balancing Weights for Causal Mediation Analysis},pdfauthor={Kentaro Kawato}}
\journal{Journal of Econometrics}

\newsavebox{\bodymathbox}
\newlength{\bodylistshift}
\makeatletter
\newcommand{\BodyDisplay}[2]{%
  \begingroup
  \sbox{\bodymathbox}{$\displaystyle #2$}%
  \typeout{BODY-DISPLAY-#1-WIDTH: \the\wd\bodymathbox\space/ \the\linewidth}%
  \setlength{\bodylistshift}{\z@}%
  \ifdim\wd\bodymathbox>\linewidth
    \ifdim\@totalleftmargin>\z@
      \setlength{\bodylistshift}{\dimexpr\wd\bodymathbox-\linewidth\relax}%
      \ifdim\bodylistshift>\@totalleftmargin \setlength{\bodylistshift}{\@totalleftmargin}\fi
      \typeout{BODY-DISPLAY-#1-LIST-SHIFT: \the\bodylistshift\space/ \the\@totalleftmargin}%
    \fi
  \fi
  \ifdim\dimexpr\wd\bodymathbox-\bodylistshift\relax>\linewidth
    \typeout{BODY-DISPLAY-#1-EXCESS: \the\dimexpr\wd\bodymathbox-\bodylistshift-\linewidth\relax}%
  \fi
  \[\hspace*{-\bodylistshift}\usebox{\bodymathbox}\]%
  \endgroup\ignorespaces}
\makeatother
\newcommand{\BodyInlineGroup}[2]{%
  \sbox{\bodymathbox}{$\textstyle #2$}%
  \typeout{BODY-INLINE-GROUP-#1-WIDTH: \the\wd\bodymathbox\space/ \the\linewidth}%
  \ifdim\wd\bodymathbox>\linewidth
    \typeout{BODY-INLINE-GROUP-#1-EXCESS: \the\dimexpr\wd\bodymathbox-\linewidth\relax}%
  \fi
  #2}
\newcommand{\BodyEquation}[4]{%
  \begingroup
  \setcounter{equation}{\numexpr#3-1\relax}%
  \sbox{\bodymathbox}{$\displaystyle #4$}%
  \typeout{BODY-EQUATION-#1-WIDTH: \the\wd\bodymathbox\space/ \the\linewidth}%
  \ifdim\wd\bodymathbox>\linewidth
    \typeout{BODY-EQUATION-#1-EXCESS: \the\dimexpr\wd\bodymathbox-\linewidth\relax}%
  \fi
  \begin{equation}\tag{#3}\label{#2}\usebox{\bodymathbox}\end{equation}%
  \endgroup\ignorespaces}

\providecommand{\doi}[1]{\url{https://doi.org/#1}}

\makeatletter
\newcommand{\AppendixDisplay}[2]{%
  \begingroup
  \sbox{\bodymathbox}{$\displaystyle #2$}%
  \typeout{APPENDIX-DISPLAY-#1-WIDTH: \the\wd\bodymathbox\space/ \the\linewidth}%
  \setlength{\bodylistshift}{\z@}%
  \ifdim\wd\bodymathbox>\linewidth
    \typeout{APPENDIX-DISPLAY-#1-EXCESS: \the\dimexpr\wd\bodymathbox-\linewidth\relax}%
    \ifdim\@totalleftmargin>\z@
      \setlength{\bodylistshift}{\dimexpr\wd\bodymathbox-\linewidth\relax}%
      \ifdim\bodylistshift>\@totalleftmargin \setlength{\bodylistshift}{\@totalleftmargin}\fi
    \else
      \setlength{\bodylistshift}{\dimexpr(\wd\bodymathbox-\linewidth)/2\relax}%
    \fi
  \fi
  \[\hspace*{-\bodylistshift}\usebox{\bodymathbox}\]%
  \endgroup\ignorespaces}
\makeatother
\newcommand{\AppendixRaggedParagraph}{\rightskip=0pt plus 1fil\relax}
\newcommand{\AppendixQED}{\leavevmode\unskip\penalty9999 \hbox{}\nobreak\hfill\quad\hbox{\(\square\)}}
\begin{document}
\begin{frontmatter}
\title{Balancing Weights for Causal Mediation Analysis\tnoteref{authornote}}
\tnotetext[authornote]{Date: \today. I am grateful to Ryo Okui for his continued guidance and support. All errors are my own.}
\author{\manuscriptauthor{Kentaro Kawato}\fnref{contact}}
\address{\manuscriptaffiliation{Graduate School of Economics, The University of Tokyo}}
\fntext[contact]{\raggedright Graduate School of Economics, The University of Tokyo, 7-3-1 Hongo, Bunkyo-ku, Tokyo 113-0033, Japan.
\\ Email address: \textnormal{\href{mailto:kawato-kentaro380@g.ecc.u-tokyo.ac.jp}{kawato-kentaro380@g.ecc.u-tokyo.ac.jp}}}
\begin{abstract}
Causal mediation analysis decomposes the average treatment effect into components that operate through intermediate variables and a component that does not. This paper studies weighting estimation of generalized mediation functionals, whose contrasts include natural direct and indirect effects, for any fixed finite number of causally ordered mediators. Inverse propensity score weights are vulnerable to (a) weight instability, because they are products of estimated probability ratios, and (b) finite-sample covariate and mediator imbalance, which can bias the estimator. We instead propose sequential balancing weights, which address (a) by minimizing weight dispersion and (b) by imposing exact balance. Stage one balances covariates between a treatment group and the full sample. Each later stage adds the next mediator and balances a treatment group against the preceding weighted group. A finite-sample bias decomposition of the augmented weighting estimator determines these variables and targets. When exact balance cannot be attained in a small sample, the procedure relaxes the balancing conditions by tolerances selected without using the outcome. Under the stated assumptions, the estimated weights converge at every stage to their population inverse-probability counterparts, with errors from preceding stages entering additively. We also establish asymptotic normality and semiparametric efficiency of the augmented weighting estimator, together with conditions for consistent variance estimation. In the main simulation design with two ordered mediators, the proposed weights outperform the compared methods in bias, root mean squared error, and coverage. In a small media-framing experiment with two ordered mediators, the proposed weights reduce the imbalance left by propensity-score weighting.
\end{abstract}
\end{frontmatter}
\FloatBarrier

\section{Introduction}\label{sec-1}

Researchers who estimate the effect of a treatment often want to know how the effect arises. Causal mediation analysis addresses this question by decomposing the average treatment effect (ATE) into components that operate through intermediate variables, called mediators, and a component that does not \citep{Imai-GeneralApproachAnalysis-2010o,VanderWeele-ExplanationCausalInteraction-2015c,Huber-MediationAnalysis-2020y,Nguyen-CausalMediationEffects-2023j}. Such decompositions are useful for policy because improving an intervention often requires knowing why it works, not only whether it works. In many applications, the treatment can affect the outcome through several mediators, and an earlier mediator can affect a later one. For example, a job training program may raise participants' skills, which may in turn help them find employment, and both may affect their later earnings. Mediation analyses of the US Job Corps program have studied employment as a mediator of its effects on health \citep{Huber-IdentifyingCausalMechanisms-2014x} and hours worked as a mediator of its effects on earnings \citep{FrolichHuber-DirectIndirectIV-2017}. When the mediators follow such a known causal order, contrasts of nested potential outcomes identify, under suitable assumptions, a class of path-specific effects that separate the contributions of individual mediators \citep{Zhou-SemiparametricEstimationMediators-2022}. This paper studies the resulting estimation problem through weighting. Inverse probability weighting estimators are already used in causal mediation analysis \citep{Huber-IdentifyingCausalMechanisms-2014x}, and inverse-probability weights have also been used to augment outcome regression \citep{TchetgenTchetgen-SemiparametricTheoryAnalysis-2012y}. This paper instead develops an estimator that constructs the weights directly, rather than by inverting estimated propensity scores, in order to improve finite-sample performance in causal mediation analysis.

Before turning to mediation, we recall inverse probability weighting (IPW) for standard treatment effects. IPW estimates target parameters, such as the ATE or the average treatment effect on the treated (ATT), by combining observed outcomes with weights constructed from treatment probabilities. A common implementation first estimates propensity scores and then forms inverse-probability weights. This approach does not require specifying an outcome model, allows the comparison to be designed and diagnosed before examining outcomes, and is easy to interpret. For example, when estimating the ATT, weighting the control group by the treatment odds has the intuitive interpretation of aligning its covariate distribution with that of the treated group.

Despite these advantages, limitations of inverse probability weighting have been extensively documented \citep{Kang-DemystifyingDoubleRobustness-2007,Hirshberg-ApproachesWeightingInference-2017l,Chattopadhyay-BalancingVsPractice-2020m}. We focus on two related concerns, (a) \emph{weight instability} and (b) \emph{finite-sample covariate imbalance}. The first concern arises because standard propensity score estimation targets treatment probabilities without directly controlling the dispersion of the inverse weights ultimately used for estimation. The second concern is the connection between sample covariate imbalance and finite-sample estimation bias, as emphasized by \citet{Ben-Michael-BalancingActInference-2021x} and \citet{Cohn-BalancingWeightsInference-2023j}. These concerns motivate viewing inverse-probability weights through their balancing properties and estimating weights that target those properties more directly. Related approaches to weighting for covariate balance have been developed for a range of econometric methods, including instrumental-variable estimation \citep{Heiler-EfficientCovariateBalancing-2022}, difference-in-differences \citep{Cefalu-ReducingBiasDiDEntropyBalancing-2020}, regression discontinuity designs \citep{ma2026generic}, and synthetic control methods \citep{BenMichael-AugmentedSyntheticControl-2021}.

In causal mediation analysis, however, it is not immediately clear which variables should be balanced across which groups, or in what order. For the ATT, aligning the covariate distribution of the control group with that of the treated group provides a natural target. Even with a single mediator, mediation analysis requires determining whether the relevant target is the treated group, the control group, or the overall population, and how the marginal and joint distributions of covariates and the mediator should be aligned. These choices become more complex with multiple mediators. This paper derives the relevant balancing targets and their order from a bias decomposition and proposes a unified sequential balancing procedure in which each stage solves a convex optimization problem that directly controls weight dispersion. Our main analysis focuses on multiple mediators with a prespecified causal ordering.

To clarify the motivation and the proposed method, we first introduce notation for the single-mediator case and define the estimands. Let \(\mathcal X\), \(\mathcal M\), and \(\mathcal Y\) denote the supports of the covariates, mediator, and outcome, respectively. The observed data are \((Y,M,D,X)\in\mathcal Y\times\mathcal M\times\{0,1\}\times\mathcal X\), and \((Y_i,M_i,D_i,X_i)_{i=1}^{N}\) are independent and identically distributed observations from a common population. For \(d\in\{0,1\}\), let \(M(d)\) denote the mediator that would be observed if treatment were set to \(d\), and let \(Y(d,m)\) denote the outcome that would be observed if treatment and the mediator were set to \(d\) and \(m\), respectively. We write \(Y(d):=Y(d,M(d))\). Central to causal mediation analysis is the nested potential outcome \(Y(d,M(d'))\), which sets treatment to \(d\) and the mediator to the value it would take under treatment \(d'\).

We consider two key estimands, the natural direct effect (NDE) and the natural indirect effect (NIE). Define \(\theta_{a,b}:=\mathbb E[Y(a,M(b))],a,b\in\{0,1\}\). The NDE is \(\mathrm{NDE}(d) :=\mathbb E[Y(1,M(d))-Y(0,M(d))] =\theta_{1,d}-\theta_{0,d}\). It measures the effect of changing treatment from \(0\) to \(1\) while holding the mediator at its natural potential value under treatment \(d\). It therefore captures the component of the treatment effect that does not operate through the mediator. The NIE is \(\mathrm{NIE}(d) :=\mathbb E[Y(d,M(1))-Y(d,M(0))] =\theta_{d,1}-\theta_{d,0}\). It measures the effect of changing the mediator from its potential value under treatment \(0\) to its potential value under treatment \(1\), while holding treatment fixed at \(d\). Figure 1.1 illustrates these direct and indirect pathways. Because each effect is a difference of these nested means, the ATE decomposes as \(\mathrm{ATE} :=\mathbb E[Y(1)-Y(0)] =\theta_{1,1}-\theta_{0,0} =\mathrm{NDE}(1)+\mathrm{NIE}(0) =\mathrm{NDE}(0)+\mathrm{NIE}(1).\) These definitions follow the potential-outcomes framework for mediation analysis \citep{Imai-IdentificationInferenceEffects-2010a}. Their identification requires the assumptions stated in Section 2.

\begin{figure}[htbp]
\centering
\includegraphics[scale=1.2]{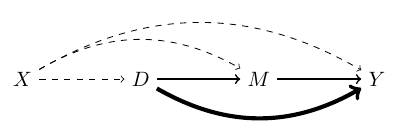}
\setcounter{figure}{0}
\caption{Standard mediation diagram. \(D\) denotes treatment, \(M\) the mediator, \(Y\) the outcome, and \(X\) pretreatment covariates. The thick arrow represents the NDE, and the thin solid arrows represent the NIE pathway. Dashed arrows indicate possible confounding by the covariates.}
\label{block-fig-mediation-setting}
\end{figure}

Following \citet{TchetgenTchetgen-SemiparametricTheoryAnalysis-2012y}, we use an augmented weighting estimator that is based on the efficient influence function (EIF) and combines weighting with outcome regression. Our proposal keeps this estimator and changes only how its weights are constructed. To illustrate its structure, consider \(\theta_{1,0}\). Estimating a single treatment-specific mean such as \(\theta_{1,1}\) or \(\theta_{0,0}\) uses one inverse-probability weight function, whereas estimating this nested mean uses two linked weight functions because it adjusts for both the baseline covariates and the mediator. The first weight function reweights the control observations to represent the full sample, and the second reweights the treated observations to represent the reweighted controls. Write \(\pi_0(d;x):=P(D=d\mid X=x)\) and \(\pi_1(d;m,x):=P(D=d\mid M=m,X=x)\). In the propensity-score representation, the estimated weights are
\BodyDisplay{1}{
\widehat w^{\mathrm{PS}}_1(X_i)
=\frac{1}{\widehat\pi_0(0;X_i)},\qquad
\widehat w^{\mathrm{PS}}_2(M_i,X_i)
=\frac{\widehat\pi_1(0;M_i,X_i)}
{\widehat\pi_0(0;X_i)\widehat\pi_1(1;M_i,X_i)},
}
where hats denote estimators of the corresponding conditional treatment probabilities. Estimators that use these weights, including the estimator based on the efficient influence function, can be sensitive to their quality. The two concerns mentioned above take the following forms in the mediation setting.

\begin{itemize}
\item
  (a) \emph{Weight instability.} The weights can become highly variable when estimated probabilities in their denominators approach zero. In the mediation setting, the second weight \(\widehat w^{\mathrm{PS}}_2(M_i,X_i)\) contains a product of two probabilities in its denominator. For example, if \(\widehat\pi_0(0;X_i)=\widehat\pi_1(1;M_i,X_i)=0.01\), then \(\widehat\pi_1(0;M_i,X_i)=0.99\) and the resulting weight is \(9{,}900\). This amplification motivates directly controlling weight dispersion, rather than relying solely on the fit of the propensity score models.
\item
  (b) \emph{Finite-sample covariate and mediator imbalance.} The population counterparts of the two propensity-score-based weights are \BodyDisplay{2}{
  w_1^*(X)=\frac{1}{\pi_0(0;X)},\qquad
  w_2^*(M,X)=\frac{\pi_1(0;M,X)}
  {\pi_0(0;X)\pi_1(1;M,X)}.
  }

  They satisfy the balancing identities \BodyEquation{3}{eq-intro-population-balance}{1}{
  \begin{aligned}
  \mathbb E[(1-D)w_1^*(X)h(X)]
  &=\mathbb E[h(X)],\\
  \mathbb E[Dw_2^*(M,X)h(M,X)]
  &=\mathbb E[(1-D)w_1^*(X)h(M,X)] \end{aligned}
  } for every bounded measurable function of \(X\) in the first identity and every bounded measurable function of \((M,X)\) in the second. Conversely, if unknown weights \(w_1(X)\) and \(w_2(M,X)\) satisfy these identities for all such functions, they equal \(w_1^*(X)\) and \(w_2^*(M,X)\) almost surely, under the positivity conditions stated in Section 2 (Proposition 3.2). Thus, the identities in \eqref{eq-intro-population-balance} characterize what inverse-probability weighting balances. The control group's covariates are aligned with the overall population, and the treated group's joint mediator--covariate distribution is aligned with that of the reweighted control group.
\end{itemize}

Replacing the population probabilities with consistent estimators can recover these identities asymptotically under appropriate regularity conditions, but it does not generally impose their empirical counterparts. In particular, for selected functions \(h(X_i)\) and \(h(M_i,X_i)\), we may have
\BodyEquation{4}{eq-intro-stage1-imbalance}{2}{
\frac1N\sum_{i=1}^{N}(1-D_i)\widehat w_1^{\mathrm{PS}}(X_i)h(X_i)
\neq\frac1N\sum_{i=1}^{N}h(X_i) }

and
\nopagebreak[4]
\BodyEquation{5}{eq-intro-stage2-imbalance}{3}{
\frac1N\sum_{i=1}^{N}D_i\widehat w_2^{\mathrm{PS}}(M_i,X_i)h(M_i,X_i)
\neq\frac1N\sum_{i=1}^{N}(1-D_i)\widehat w_1^{\mathrm{PS}}(X_i)h(M_i,X_i). }

As shown in Section 3 (Proposition 3.1), the finite-sample bias of the augmented estimator depends on the empirical discrepancies illustrated in \eqref{eq-intro-stage1-imbalance} and \eqref{eq-intro-stage2-imbalance}, computed for the errors in the outcome and nested regression functions. This motivates enforcing covariate and mediator balance directly in the observed sample.

These two concerns, (a) \emph{weight instability} and (b) \emph{finite-sample covariate and mediator imbalance}, can also arise in the multiple-mediator setting that is the main focus of this paper. The corresponding inverse-probability weights involve products of conditional-probability ratios. Additional stages can lengthen these products, and small denominator probabilities can produce highly concentrated or extreme weights. The balancing problem also requires determining which covariates and mediator histories should be aligned with which target distributions at each stage. The relevant targets and the opportunities for finite-sample imbalance therefore depend on both the mediator history and the treatment inputs defining the estimand.

We call our approach to these problems \emph{sequential balancing weights}. To describe the general construction, let \(M_1,\ldots,M_J\) denote a fixed finite sequence of causally ordered mediators. The index follows the causal order, so \(M_1\) is the earliest mediator, and each mediator may depend on the preceding mediators but not on later ones. In the job training example, \(M_1\) is the participants' skills and \(M_2\) is their employment. Write \(\overline M_j=(M_1,\ldots,M_j)\) for \(j=1,\ldots,J\), with \(\overline M_0=\varnothing\). For a given target, let \(d_j\in\{0,1\}\) denote the treatment input assigned to mediator \(M_j\) for \(j=1,\ldots,J\), and let \(d_{J+1}\in\{0,1\}\) denote the treatment input assigned to the outcome. The single-mediator example above is recovered by setting \(J=1\), \(d_1=0\), and \(d_2=1\). The first stage assigns nonnegative weights to observations with \(D=d_1\). Let \(K_1\) be the number of prespecified basis functions and let \(B_1(X)=(b_{1,1}(X),\ldots,b_{1,K_1}(X))^\top\) collect those functions. The procedure constrains their weighted sample moments to match their full-sample counterparts exactly. These functions may be selected using substantive knowledge about features relevant to outcome prediction. Among weights satisfying the constraints, the procedure minimizes a convex dispersion criterion, such as \(\ell(w)=w\log w-w\) or \(\ell(w)=w^2/2\). At each subsequent stage \(s=2,\ldots,J+1\), let \(K_s\) denote the number of basis functions of the available history and let \(B_s(\overline M_{s-1},X)=(b_{s,1}(\overline M_{s-1},X),\ldots,b_{s,K_s}(\overline M_{s-1},X))^\top\) collect them. The preceding weights are held fixed, and the weights for observations with \(D=d_s\) are chosen so that the selected moments of \((\overline M_{s-1},X)\) match those in the weighted \(D=d_{s-1}\) group. Thus, \(J\) mediators give rise to \(J+1\) sequential convex balancing problems. For example, with \(J=2\), \(d_1=1\), \(d_2=0\), and \(d_3=1\), a case studied in Section 6, the first stage weights the treated observations toward the full sample on functions of \(X\), the second weights the controls toward the weighted treated observations on functions of \((M_1,X)\), and the third weights the treated observations toward the weighted controls on functions of \((M_1,M_2,X)\). Whenever the constraints are feasible, the procedure therefore addresses (a) \emph{weight instability} by discouraging unnecessarily dispersed weights and (b) \emph{finite-sample covariate and mediator imbalance} by directly controlling the selected empirical imbalances.

This construction extends convex balancing methods developed for average treatment effects, including entropy balancing \citep{Hainmueller-EntropyBalancingStudies-2012c}, stable balancing weights \citep{Zubizarreta-StableWeightsData-2015v}, empirical balancing calibration weighting \citep{Chan-GloballyEfficientWeighting-2016v}, approximate residual balancing \citep{athey2018approximate}, and minimal-dispersion approximately balancing weights \citep{Wang-MinimalDispersionConsiderations-2019m}. Our extension accounts for the sequence of target distributions specific to mediation analysis, in which the weights estimated at each stage define the target distribution for the next stage, rather than imposing a single treated-versus-control balance condition. Balancing methods for mediation analysis have also been studied by \citet{Chan-EfficientNonparametricEffects-2016g} and \citet{liu2024two}. \citet{Chan-EfficientNonparametricEffects-2016g} develop weighting estimators based on empirical balancing conditions, establish asymptotic efficiency, and consider an extension with two causally ordered mediators. \citet{liu2024two} take a different approach. They develop bias-targeted, regularized nonparametric balancing estimators, including closed-form implementations in reproducing kernel Hilbert spaces, and also consider mediator-specific effects without requiring a prespecified causal ordering.

We develop a sequential exact-balancing procedure for any fixed finite number of causally ordered mediators. The finite-sample bias decomposition in Section 3 shows that the empirical imbalances in the selected basis functions enter the bias of the augmented weighting estimator, so our procedure removes these imbalances exactly at every stage and, among the weights that do so, minimizes weight dispersion. Exact balance can be infeasible in small samples, where the number of balancing conditions is large relative to the size of a target group. For such cases, our framework also accommodates approximate balance by introducing tolerances into the balancing conditions. The tolerances are selected stage by stage without using the outcome, following the trade-off between empirical balance and weight stability in approximate residual balancing \citep{athey2018approximate} and minimal-dispersion approximately balancing weights \citep{Wang-MinimalDispersionConsiderations-2019m}. Our approach is complementary to that of \citet{liu2024two}. While they develop a nonparametric estimation framework and examine its performance in relatively large samples, with sample sizes of \(N=1{,}000\), \(2{,}000\), and \(4{,}000\) in their simulations and \(N=2{,}685\) in their empirical application, our study places greater emphasis on exact finite-sample balance and its feasibility in substantially smaller samples. Specifically, we consider sample sizes of \(N=265\), \(500\), and \(1{,}000\) in our simulations and apply the proposed method to an empirical dataset with only \(N=265\) observations. Such a sample is small relative to the number of balancing conditions that one treatment group must satisfy. In the empirical application, for instance, one balancing problem weights 68 treated observations to balance 32 basis functions of the covariates and the two mediators, including the constant.

Our theoretical analysis first establishes convergence of the estimated balancing weights to their population inverse-probability counterparts. In the ordered multiple-mediator setting, for each stage \(s=1,\ldots,J+1\), the estimated weight \(\widehat w_s\) converges to the corresponding population weight \(w_s^*\) under regularity conditions (Theorem 5.2). These population weights are characterized by the sequence of population balancing identities induced by the estimand (Proposition 3.2). Consequently, when conventional propensity-score-based weights are themselves consistently estimated, the balancing weights and the propensity-score-based weights share the same probability limit, namely \(w_s^*\) at each stage. To prove this, we use sieve arguments related to \citet{Newey-ConvergenceRatesEstimators-1997k} and \citet{Chen-Chapter76Models-2007p}, together with asymptotic analyses from the balancing literature, including \citet{Wang-MinimalDispersionConsiderations-2019m} and \citet{Fan-OptimalCovariateEstimation-2023s}. A central difficulty in the multiple-mediator setting is that the target distribution at each stage is itself constructed using weights estimated at the preceding stage from the same sample. Thus, estimation error can propagate recursively through the sequence of balancing problems. By controlling this propagation, we obtain stagewise bounds in which errors from preceding stages enter additively. Under additional assumptions, we establish asymptotic linearity, normal limits, and semiparametric efficiency for the proposed estimator (Theorem 5.3), and show that its asymptotic variance can be estimated consistently (Theorem 5.4). The argument exploits the product structure of the influence-function representation, which is also central to the double machine learning of causal mediation in \citet{Farbmacher-CausalMediationLearning-2022c} and \citet{Zhou-SemiparametricEstimationMediators-2022}.

Our Monte Carlo simulations use a design with two binary mediators and weak overlap and sample sizes of \(N=265\), \(500\), and \(1{,}000\), where \(N=265\) matches the small empirical sample. The outcome contains a term that the regression and balancing bases omit, so the outcome regression is misspecified. We compare conventional propensity score (PS) weighting, covariate balancing propensity scores (CBPS; \citealp{Imai-CovariateBalancingScore-2014v}), and the proposed balancing weights (BW). The three weighting methods use the same OLS outcome and backward regressions, and we also report regression-only OLS using the same predictions. Every weighting method has a smaller bias than regression-only OLS. BW attains the smallest bias and RMSE among the four methods at every sample size, and the coverage rates of its confidence intervals are 9 to 11 percentage points closer to the nominal 95\% level than those of PS. This pattern is consistent with the finite-sample bias decomposition in Section 3, which motivates imposing the balancing constraints exactly. Appendix E repeats the comparison with elastic-net regressions and with stronger and weaker overlap.

We illustrate the sequential procedure using the media-framing experiment of \citet{brader2008triggers}, with 265 participants, of whom 68 are treated. Following earlier analyses of these data \citep{ImaiYamamoto-MultipleCausalMechanisms-2013,Tingley-MediationRPackage-2014}, we take perceived harm from immigration to precede anxiety and decompose the effect of a negatively framed news story about a Latino immigrant on attitudes toward immigration into a natural direct effect and the path-specific effects through anxiety alone and through perceived harm. The two ordered mediators give three balancing stages, and we compare PS, CBPS, and BW. Among the stage problems that the reported effects use, BW attains exact balance in all but one, where tolerances are introduced. PS leaves a maximum standardized imbalance of 0.308 at the second stage, where the 68 treated participants are weighted toward the weighted controls on functions of perceived harm and the covariates, CBPS reduces it to 0.137, and BW removes it. The three methods give similar estimates of the ATE and the direct effect but split the indirect effect differently, as PS and CBPS attribute it to perceived harm and BW attributes most of it to anxiety alone.

\subsection{Related literature}\label{sec-1-1}

This paper contributes to two strands of literature, causal mediation analysis and balancing weights. \citet{Imai-GeneralApproachAnalysis-2010o} and \citet{Imai-IdentificationInferenceEffects-2010a} establish general potential outcome frameworks for defining, identifying, and estimating mediation effects. Taking the identification assumptions as given, we focus on the estimation of these effects. \citet{TchetgenTchetgen-SemiparametricTheoryAnalysis-2012y} derive the semiparametric efficiency bound and develop multiply robust estimators for a single mediator. \citet{Huber-IdentifyingCausalMechanisms-2014x} develops inverse-probability-weighting representations, while \citet{Farbmacher-CausalMediationLearning-2022c} combine orthogonal scores with machine learning and cross-fitting. \citet{Zhou-SemiparametricEstimationMediators-2022} extends this literature to multiple causally ordered mediators, developing multiply robust and locally efficient estimators. \citet{Nguyen-CausalMediationEffects-2023j} provide a recent review of these estimation methods. This paper studies how to construct the weights used in mediation estimators, particularly when multiple mediators are causally ordered.

Within the balancing-weights literature, an important distinction is between estimating propensity scores using balance-oriented criteria and directly optimizing weights to achieve covariate balance. The first approach incorporates covariate balancing conditions into propensity score estimation. \citet{Imai-CovariateBalancingScore-2014v} propose the covariate balancing propensity score, which uses generalized method of moments estimation to incorporate balancing conditions, while \citet{Fan-OptimalCovariateEstimation-2023s} investigate how the choice of balancing conditions affects robustness and efficiency. Although these methods improve covariate balance, the resulting weights are still obtained by inverting estimated propensity scores, and their dispersion is not directly controlled. An alternative approach is to construct weights directly by optimizing an objective function subject to balancing constraints, without first estimating propensity scores. This approach includes \citet{Hainmueller-EntropyBalancingStudies-2012c}, \citet{Zubizarreta-StableWeightsData-2015v}, \citet{Chan-GloballyEfficientWeighting-2016v}, \citet{athey2018approximate}, and \citet{Wang-MinimalDispersionConsiderations-2019m}. Our proposed method follows this approach by directly constructing weights that attain exact balance with minimal dispersion, and it relaxes balance only when exact balance cannot be attained. For recent reviews of balancing methods, see \citet{Ben-Michael-BalancingActInference-2021x} and \citet{Cohn-BalancingWeightsInference-2023j}.

Related work includes dynamic covariate balancing, which studies estimation and inference for treatment histories in panel data when treatments change dynamically over time and constructs weights recursively across periods \citep{VivianoBradic-DynamicCovariateBalancing-2026}. In contrast, our method addresses the sequential structure arising from causally ordered mediators rather than longitudinal treatment assignment. Our approach is also related to the direct estimation of Riesz representers \citep{ChernozhukovNeweySingh-AutomaticDebiased-2022}. While sharing the objective of estimating the weights required for causal estimation without explicitly estimating propensity scores, our method emphasizes imposing balancing constraints exactly to reduce bias, together with the control of weight dispersion in finite samples.

\subsection{Outline}\label{sec-1-2}

The remainder of the paper is organized as follows. Section 2 presents the setup and identification assumptions for causally ordered mediators. Section 3 introduces the augmented weighting estimator and derives a finite-sample bias decomposition. Section 4 develops the sequential balancing weights algorithm, and Section 5 establishes the asymptotic theory for these weights and the augmented weighting estimator. Section 6 presents the simulation study, and Section 7 reports the empirical application. Section 8 concludes. The Appendix contains proofs and implementation details.

\FloatBarrier

\section{Causal mediation analysis with causally ordered mediators}\label{sec-2}

We now extend the single-mediator setting introduced in Section 1 to multiple causally ordered mediators, which constitute the main setting of this paper. This section defines the generalized mediation functional, states the assumptions under which it is identified, and presents its regression and efficient influence-function representations.

\subsection{Setup and notation}\label{sec-2-1}

Let \(O_i=(Y_i,D_i,\overline M_{J,i},X_i)\), \(i=1,\ldots,N\), be independent and identically distributed observations from a common population. Write \(O=(Y,D,\overline M_J,X)\) for a generic observation. Here, \(D_i\in\{0,1\}\) is the treatment indicator, \(X_i\in\mathcal X\) is a vector of pretreatment covariates, and \(Y_i\in\mathbb R\) is the outcome. We consider \(J\geq1\) causally ordered mediators \(M_1,\ldots,M_J\), where \(J\) is fixed and does not grow with the sample size. Each mediator may be multivariate and may depend on treatment, pretreatment covariates, and all preceding mediators, but not on subsequent mediators. Let \(\mathcal M_j\) denote the support of \(M_j\), and write \(\overline M_j=(M_1,\ldots,M_j)\), \(\overline m_j=(m_1,\ldots,m_j)\), and \(\overline M_0=\overline m_0=\varnothing\).

For \(j=1,\ldots,J\), let \(M_j(d,\overline m_{j-1})\) denote the potential value of mediator \(j\) when treatment is set to \(d\) and the preceding mediator history is set to \(\overline m_{j-1}\). Because \(\overline m_0=\varnothing\), we write \(M_1(d)\) for \(M_1(d,\overline m_0)\). Let \(Y(d,m_J,\ldots,m_1)\) denote the potential outcome when treatment is set to \(d\) and the complete mediator history is set to \((m_1,\ldots,m_J)\). We use \(Y(d,\overline m_J)\) as shorthand for \(Y(d,m_J,\ldots,m_1)\).

Following the framework of \citet{Zhou-SemiparametricEstimationMediators-2022}, we allow the treatment input to differ across the outcome and mediators. Fix \(\boldsymbol d=(d_{J+1},d_J,\ldots,d_1)\in\{0,1\}^{J+1}\), where \(d_{J+1}\) specifies the treatment input to the outcome and \(d_j\) specifies the treatment input to mediator \(M_j\). Define the nested potential mediator history recursively, starting with \(\overline M_1(d_1)=(M_1(d_1))\) and, for \(j=2,\ldots,J\), setting \(\overline M_j(d_1,\ldots,d_j)=\bigl(\overline M_{j-1}(d_1,\ldots,d_{j-1}),M_j(d_j,\overline M_{j-1}(d_1,\ldots,d_{j-1}))\bigr)\).

The target generalized mediation functional is \BodyDisplay{6}{
\theta_{\boldsymbol d}=\mathbb E\!\left[Y\!\left(d_{J+1},\overline M_J(d_1,\ldots,d_J)\right)\right].
} This construction passes the same realized potential value of each mediator to all subsequent mediators and the outcome. The treatment inputs in \(\boldsymbol d\) are written from the outcome backward through \(M_J,\ldots,M_1\); stagewise construction still follows mediator causal order. For example, when \(J=2\), the target becomes \(\theta_{(d_3,d_2,d_1)} =\mathbb E\!\left[ Y\!\left(d_3,M_2\{d_2,M_1(d_1)\},M_1(d_1)\right) \right].\) The single-mediator setting introduced in Section 1 is recovered by setting \(J=1\), \(M=M_1\), \(a=d_2\), and \(b=d_1\). In this case, \(\theta_{\boldsymbol d}=\theta_{a,b}=\mathbb E[Y(a,M(b))]\).

To express the generalized mediation functional in terms of the observed-data distribution, we introduce conditional treatment probabilities and a sequence of backward regression functions. For \(j=0,\ldots,J\) and \(d\in\{0,1\}\), define \(\pi_j(d;\overline m_j,x)=P(D=d\mid\overline M_j=\overline m_j,X=x)\). At \(j=0\), write \(\pi_0(d;x)=P(D=d\mid X=x)\). For \(j\geq1\), \(\pi_j\) is a conditional treatment probability given the observed mediator history \(\overline M_j\). These probabilities will be used to construct the sequential density ratios underlying the weighting representation. Unlike the conventional pretreatment propensity score \(\pi_0\), the probabilities \(\pi_j\) for \(j\geq1\) condition on post-treatment variables. When \(J=1\), they reduce to \(\pi_0(d;x)=P(D=d\mid X=x)\) and \(\pi_1(d;m,x)=P(D=d\mid M=m,X=x)\), as introduced in Section 1.

We next define the backward regression functions, which recursively integrate the outcome regression over the conditional distributions of the mediators. Starting with the outcome regression \(\mu_J^{(d_{J+1})}(\overline m_J,x)=\mathbb E[Y\mid D=d_{J+1},\overline M_J=\overline m_J,X=x]\), define, for \(j=J,J-1,\ldots,1\),
\BodyDisplay{7}{
\begin{aligned}
&\mu_{j-1}^{(d_j,\ldots,d_{J+1})}
(\overline m_{j-1},x)\\
&\quad=\int
\mu_j^{(d_{j+1},\ldots,d_{J+1})}
(\overline m_j,x)
\,dP(m_j\mid D=d_j,\overline M_{j-1}=\overline m_{j-1},X=x)\\
&\quad=\mathbb E\!\left[
\mu_j^{(d_{j+1},\ldots,d_{J+1})}(\overline M_j,X)
\mid D=d_j,\overline M_{j-1}=\overline m_{j-1},X=x
\right].
\end{aligned}
}

For \(j=J\), the regression on the right-hand side is understood as \(\mu_J^{(d_{J+1})}\). The superscripts list the treatment inputs of the mediators already integrated out in their causal order, followed by the outcome treatment input. The recursion successively integrates out \(M_J,M_{J-1},\ldots,M_1\) under the treatment inputs specified by \(\boldsymbol d\).

\subsection{Identification assumptions}\label{sec-2-2}

We impose the following identification assumptions, corresponding to Assumptions \(1^*\)--\(3^*\) in \citet[Section 2.2]{Zhou-SemiparametricEstimationMediators-2022}, using the treatment indicator and mediator histories defined above.

\par\medskip

\setcounter{assumption}{0}
\begin{assumption}[Consistency]

For every unit, every \(j=1,\ldots,J\), every \(d_j\in\{0,1\}\), and every fixed history \(\overline m_{j-1}\), \(D=d_j,\ \overline M_{j-1}=\overline m_{j-1}\Longrightarrow M_j=M_j(d_j,\overline m_{j-1})\), and, for every \(d_{J+1}\in\{0,1\}\) and \(\overline m_J\), \(D=d_{J+1},\ \overline M_J=\overline m_J\Longrightarrow Y=Y(d_{J+1},m_J,\ldots,m_1)\).

\end{assumption}

Consistency requires that the observed value of each mediator agrees with its potential value under the realized treatment and preceding mediator history. Similarly, the observed outcome equals its potential outcome under the realized treatment and complete mediator history.

\par\medskip

\setcounter{assumption}{1}
\begin{assumption}[Sequential conditional independence]

For every \(\boldsymbol d\in\{0,1\}^{J+1}\) and every mediator history \(\overline m_J\),
\BodyDisplay{8}{
\bigl(M_1(d_1),M_2(d_2,\overline m_1),\ldots,M_J(d_J,\overline m_{J-1}),Y(d_{J+1},m_J,\ldots,m_1)\bigr)
\perp\!\!\!\perp D\mid X.
}

For every \(j=1,\ldots,J\), the same treatment inputs, and histories \(\overline m_J\) and \(\overline m_{j-1}^{*}\),
\BodyDisplay{9}{
\bigl(M_{j+1}(d_{j+1},\overline m_j),\ldots,M_J(d_J,\overline m_{J-1}),Y(d_{J+1},m_J,\ldots,m_1)\bigr)
\perp\!\!\!\perp M_j(d_j,\overline m_{j-1}^{*})\mid X,D,\overline M_{j-1}.
}

\end{assumption}

At \(j=J\), only the potential outcome remains on the left-hand side. The first condition requires treatment assignment to be independent of the joint potential responses, conditional on pretreatment covariates. The second requires each mediator's potential response to be independent of the joint subsequent potential responses, conditional on treatment, pretreatment covariates, and the observed preceding mediator history. These restrictions rule out the forms of unmeasured confounding relevant to identification and impose cross-world independence. Remark 2.1 illustrates their implications in the single-mediator case.

\par\medskip

\setcounter{assumption}{2}
\begin{assumption}[Positivity]

There exists a constant \(c\in(0,1/2)\) such that, for every \(j=0,\ldots,J\) and \(d\in\{0,1\}\), \(c\leq \pi_j(d;\overline M_j,X)\leq1-c\) almost surely under the corresponding observed-history distribution.

\end{assumption}

Positivity requires both treatment groups to be represented at every relevant covariate value and observed mediator history. This ensures the common support needed to combine conditional mediator distributions across treatment groups and to construct the corresponding inverse-probability weights.

\par\medskip

\setcounter{remark}{0}
\begin{remark}[Interpretation of the identification assumptions for a single mediator]

When \(J=1\), write \(M=M_1\), \(a=d_2\), and \(b=d_1\). Assumption 2.1 reduces to the standard consistency conditions \(D=b\Longrightarrow M=M(b)\) and \(D=a,\ M=m\Longrightarrow Y=Y(a,m)\). Assumption 2.3 reduces to \(c\leq\pi_0(d;X)\leq1-c\) and \(c\leq\pi_1(d;M,X)\leq1-c\) for \(d\in\{0,1\}\), almost surely on the corresponding observed supports. For every \(a,b\in\{0,1\}\) and \(m\in\mathcal M\), Assumption 2.2 reduces exactly to
\BodyDisplay{10}{
\bigl(M(b),Y(a,m)\bigr)\perp\!\!\!\perp D\mid X,
\qquad
Y(a,m)\perp\!\!\!\perp M(b)\mid X,D.
}

These joint independence restrictions imply several familiar conditions used in single-mediator causal mediation analysis \citep{Pearl-DirectIndirectEffects-2001g,Imai-IdentificationInferenceEffects-2010a}. In particular, for every \(d\in\{0,1\}\) and \(m\in\mathcal M\),
\BodyDisplay{11}{
Y(d,m)\perp\!\!\!\perp D\mid X,
\qquad
M(d)\perp\!\!\!\perp D\mid X,
}

\BodyDisplay{12}{
Y(d,m)\perp\!\!\!\perp M\mid D=d,X,
\qquad
Y(d,m)\perp\!\!\!\perp M(1-d)\mid X.
}

The first two conditions express the absence of treatment--outcome and treatment--mediator confounding conditional on pretreatment covariates. The third condition expresses mediator--outcome unconfoundedness conditional on treatment and pretreatment covariates. The fourth is a cross-world independence restriction relating the potential outcome under one treatment state to the potential mediator under the opposite treatment state. Such independence may fail in the presence of a treatment-induced confounder, such as a variable \(L\) affected by treatment that subsequently influences both the mediator and the outcome, as illustrated in Figure \ref{block-fig-confounders}. If \(L\) is observed and causally precedes \(M\), it can instead be incorporated as an additional mediator by setting \(M_1=L\) and \(M_2=M\). Under the corresponding identification assumptions, the generalized mediation functionals for these two causally ordered mediators can then be identified. However, this does not generally restore identification of the original natural direct and indirect effects defined with respect to \(M\) alone.

\end{remark}

\begin{figure}[htbp]
\centering
\includegraphics[scale=1.2]{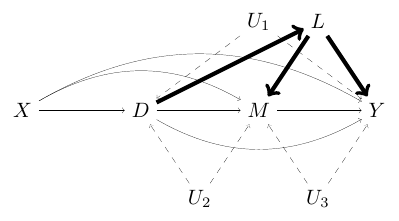}
\setcounter{figure}{0}
\caption{Potential violations of the identification assumptions. The dashed arrows associated with \(U_1,U_2,U_3\) represent possible unobserved confounding between treatment and outcome, treatment and mediator, and mediator and outcome, respectively. The variable \(L\) is affected by treatment and subsequently affects both the mediator and the outcome. These structures illustrate potential violations of the identification assumptions.}
\label{block-fig-confounders}
\end{figure}

\par\medskip

\subsection{Generalized mediation formula and efficient influence function}\label{sec-2-3}

Under Assumptions 2.1--2.3, the causal target \(\theta_{\boldsymbol d}\) can be expressed as a functional of the observed-data distribution.

\par\medskip

\setcounter{theorem}{0}
\begin{theorem}[Identification and efficient influence function of the generalized mediation functional]

Under Assumptions 2.1--2.3, the following hold for every \(\boldsymbol d=(d_{J+1},d_J,\ldots,d_1)\in\{0,1\}^{J+1}\).

(i) Regression-based identification. The causal target \(\theta_{\boldsymbol d}\) is identified by
\BodyDisplay{13}{
\begin{aligned}
\theta_{\boldsymbol d}
&=\int \mu_J^{(d_{J+1})}(\overline m_J,x)
\left\{
\prod_{r=1}^{J}
dP(m_r\mid D=d_r,\overline M_{r-1}=\overline m_{r-1},X=x)
\right\}dP_X(x)\\
&=\mathbb E\!\left[
\mu_0^{(d_1,\ldots,d_{J+1})}(X)
\right].
\end{aligned}
}

(ii) Weight-based identification. For \(s=1,\ldots,J+1\), define the population weights
\BodyDisplay{14}{
w_s^*(\overline m_{s-1},x)
=\frac{1}{\pi_0(d_1;x)}
\prod_{r=1}^{s-1}
\frac{\pi_r(d_r;\overline m_r,x)}
{\pi_r(d_{r+1};\overline m_r,x)}.
}

The empty product is defined as one, so that \(w_1^*(x)=1/\pi_0(d_1;x)\). The generalized mediation formula then admits the inverse-probability-weighting representation \(\theta_{\boldsymbol d} =\mathbb E\!\left[ \mathbf1\{D=d_{J+1}\} w_{J+1}^*(\overline M_J,X)Y \right].\)

(iii) Augmentation. Combining the backward regression functions with the sequential weights, define the uncentered score
\BodyDisplay{15}{
\begin{aligned}
\phi_{\boldsymbol d}(O)
&=\mu_0^{(d_1,\ldots,d_{J+1})}(X)\\
&\quad+\sum_{j=1}^{J}
\mathbf1\{D=d_j\}
w_j^*(\overline M_{j-1},X)
\left\{
\mu_j^{(d_{j+1},\ldots,d_{J+1})}(\overline M_j,X)
-\mu_{j-1}^{(d_j,\ldots,d_{J+1})}(\overline M_{j-1},X)
\right\}\\
&\quad+\mathbf1\{D=d_{J+1}\}
w_{J+1}^*(\overline M_J,X)
\left\{
Y-\mu_J^{(d_{J+1})}(\overline M_J,X)
\right\}.
\end{aligned}
}

Then \(\mathbb E[\phi_{\boldsymbol d}(O)] =\theta_{\boldsymbol d}.\) If \(\mathbb E[(\phi_{\boldsymbol d}(O)-\theta_{\boldsymbol d})^2]<\infty\), then the centered score \(\psi_{\boldsymbol d}(O) =\phi_{\boldsymbol d}(O)-\theta_{\boldsymbol d}\) is the efficient influence function of the generalized mediation functional.

\end{theorem}

\par\medskip

The integral in (i) is the generalized mediation formula for the ordered treatment inputs \(\boldsymbol d\), as established in \citet[Theorem 1]{Zhou-SemiparametricEstimationMediators-2022}. The product represents successive integration with respect to the conditional mediator distributions in their causal order. The second equality in (i) follows by recursively applying the definitions of the backward regression functions introduced in Section 2.1. The representations in (ii) and (iii) are particularly useful for the proposed method because they characterize the population weights that the sequential balancing procedure seeks to approximate.

The sequential weights and backward regression functions appearing in this representation provide the starting point for our proposed estimator. In the following sections, we replace the inverse-probability weights with directly estimated balancing weights while retaining the same augmented regression structure.

\FloatBarrier

\section{The augmented weighting estimator and its finite-sample bias decomposition}\label{sec-3}

This section defines an augmented weighting estimator and examines how its finite-sample bias depends on the construction of the weights. The estimator is built from the representation in Theorem 2.1, which expresses the generalized mediation functional through backward regressions and sequential weights. Section 3.1 introduces the estimator, Section 3.2 derives a bias decomposition in terms of the empirical discrepancies, and Section 3.3 characterizes the population balancing properties of the true weights.

\subsection{The augmented weighting estimator}\label{sec-3-1}

For \(s=1,\ldots,J+1\), let \(\widehat w_{s,i}\) denote a generic estimated weight for observation \(i\) at stage \(s\) and let \(\widehat\mu_j^{(d_{j+1},\ldots,d_{J+1})}\), \(j=0,\ldots,J\), denote estimators of the backward regression functions. We define the augmented weighting estimator by
\BodyDisplay{16}{
\begin{aligned}
\widehat\theta_{\boldsymbol d}^{\mathrm{AW}}
&=\frac1N\sum_{i=1}^{N}
\widehat\mu_0^{(d_1,\ldots,d_{J+1})}(X_i)\\
&\quad+\frac1N\sum_{i=1}^{N}\sum_{j=1}^{J}
\mathbf1\{D_i=d_j\}\widehat w_{j,i}
\left\{
\widehat\mu_j^{(d_{j+1},\ldots,d_{J+1})}(\overline M_{j,i},X_i)
-\widehat\mu_{j-1}^{(d_j,\ldots,d_{J+1})}(\overline M_{j-1,i},X_i)
\right\}\\
&\quad+\frac1N\sum_{i=1}^{N}
\mathbf1\{D_i=d_{J+1}\}\widehat w_{J+1,i}
\left\{Y_i-\widehat\mu_J^{(d_{J+1})}(\overline M_{J,i},X_i)\right\}.
\end{aligned}
}

One implementation substitutes estimated conditional treatment probabilities into the population weight formula, giving
\BodyDisplay{17}{
\widehat w_{s,i}^{\mathrm{PS}}
=\frac{1}{\widehat\pi_0(d_1;X_i)}
\prod_{r=1}^{s-1}
\frac{\widehat\pi_r(d_r;\overline M_{r,i},X_i)}
{\widehat\pi_r(d_{r+1};\overline M_{r,i},X_i)},
\qquad s=1,\ldots,J+1.
}

We denote the resulting estimator by \(\widehat\theta_{\boldsymbol d}^{\mathrm{AW\text{-}PS}}\).

Our proposed estimator, denoted by \(\widehat\theta_{\boldsymbol d}^{\mathrm{AW\text{-}BW}}\), instead uses weights constructed directly through the balancing algorithms in Section 4. Both estimators have the same augmented regression structure and differ only in how the weights are obtained. We next derive the finite-sample bias decomposition to determine suitable targets for direct weight construction.

\subsection{Finite-sample bias and its implications for balancing}\label{sec-3-2}

The augmented weighting estimator admits a finite-sample bias decomposition that holds for any method of estimating the weights. The only requirement is that the weights at each stage do not use information that is revealed later in the causal ordering.

\par\medskip

\setcounter{proposition}{0}
\begin{proposition}[Finite-sample bias of the augmented weighting estimator]

Suppose the observations are independent and identically distributed and the identification formula in Theorem 2.1 holds. Assume that for every observation \(i\) and stage \(s=1,\ldots,J+1\), \(\widehat w_{s,i}\) is \(\sigma((D_\ell,X_\ell,\overline M_{s-1,\ell})_{\ell=1}^{N})\)-measurable. Assume that all expectations in the bias identity below and its proof exist as finite real numbers. Then
\BodyDisplay{18}{
\begin{aligned}
&\mathbb E\!\left[
\widehat\theta_{\boldsymbol d}^{\mathrm{AW}}-\theta_{\boldsymbol d}
\right]
=\mathbb E\!\Bigg[
\frac1N\sum_{i=1}^{N}
\left\{\mathbf1\{D_i=d_1\}\widehat w_{1,i}-1\right\}
\left\{
\mu_0^{(d_1,\ldots,d_{J+1})}(X_i)
-\widehat\mu_0^{(d_1,\ldots,d_{J+1})}(X_i)
\right\}
\Bigg]\\
&\quad+\sum_{j=1}^{J}\mathbb E\!\left[
\begin{aligned}
&\frac1N\sum_{i=1}^{N}
\left\{
\mathbf1\{D_i=d_{j+1}\}\widehat w_{j+1,i}
-\mathbf1\{D_i=d_j\}\widehat w_{j,i}
\right\}\\
&\quad\times
\left\{
\mu_j^{(d_{j+1},\ldots,d_{J+1})}(\overline M_{j,i},X_i)
-\widehat\mu_j^{(d_{j+1},\ldots,d_{J+1})}(\overline M_{j,i},X_i)
\right\}
\end{aligned}
\right].
\end{aligned}
}

\end{proposition}

\par\medskip

The assumption that \(\widehat w_{s,i}\) is \(\sigma((D_\ell,X_\ell,\overline M_{s-1,\ell})_{\ell=1}^{N})\)-measurable means that the first-stage weights may use treatment and pretreatment covariates from all observations. At stage \(s\leq J\), the weights may additionally use the preceding mediator history, but not \(M_s\) or subsequent mediators. The final-stage weights may use the complete mediator history, but no stage may use outcomes. These restrictions apply to the entire weight-learning procedure, including basis selection, tuning, and the choice among solutions.

The key to the decomposition is to replace the fitted regressions with their true counterparts while keeping the estimated weights unchanged. Under the stagewise information restrictions and i.i.d. sampling, the weighted true-regression residuals have mean zero. Specifically, for each \(j=1,\ldots,J\),
\BodyDisplay{19}{
\begin{aligned}
&\mathbb E\Big[
\mathbf1\{D_i=d_j\}\widehat w_{j,i}
\big\{
\mu_j^{(d_{j+1},\ldots,d_{J+1})}(\overline M_{j,i},X_i)
-\mu_{j-1}^{(d_j,\ldots,d_{J+1})}(\overline M_{j-1,i},X_i)
\big\}
\mid (D_\ell,X_\ell,\overline M_{j-1,\ell})_{\ell=1}^N
\Big]\\
&\quad=
\mathbf1\{D_i=d_j\}\widehat w_{j,i}
\underbrace{\left\{\begin{aligned}
&\mathbb E\big[
\mu_j^{(d_{j+1},\ldots,d_{J+1})}(\overline M_{j,i},X_i)
\mid D_i=d_j,\overline M_{j-1,i},X_i
\big]\\
&-\mu_{j-1}^{(d_j,\ldots,d_{J+1})}(\overline M_{j-1,i},X_i)
\end{aligned}\right\}}_{=0}=0,
\end{aligned}
}
where the last equality follows from the definition of the backward regression. Thus, these conditional mean-zero properties hold even when the weights are estimated from the same sample.

The decomposition identifies both the variables to balance and the distributions against which they should be balanced. The first term compares the weighted \(D=d_1\) group with the full sample through their means of the error in the regression that depends only on \(X\). At each subsequent stage, the relevant comparison is between the means of the regression error that depends on \((\overline M_j,X)\) in the weighted \(D=d_{j+1}\) group and in the preceding weighted \(D=d_j\) group. The earlier weights therefore determine the target for the next balancing step. The decomposition also indicates what the balanced functions at each stage must be able to represent. Each term vanishes when the weights at the corresponding stage exactly balance a set of functions, fixed before the outcomes are used, whose linear span contains the regression estimation error in that term. Linear combinations of the balanced functions at each stage should therefore approximate the corresponding backward regression, and their linear span should contain its estimate. When the backward regressions and their estimates lie in the linear span of the balancing basis, Theorem 4.1 implies that the bias of the augmented weighting estimator is zero under exact balance. Beyond this requirement, the decomposition does not single out a particular basis, and functions that capture features predictive of the corresponding backward regressions are natural candidates for balancing.

The single-mediator balance problem raised in the Introduction is recovered as a special case of Proposition 3.1. For \(\theta_{1,0}\) with a single mediator, the balancing procedure has two stages. The controls' covariate moments are balanced against the full sample, followed by the treated group's mediator--covariate moments against the weighted controls.

\par\medskip

\subsection{Population balance}\label{sec-3-3}

Before introducing our proposed balancing procedure, we examine the population balancing properties of the inverse-probability weights defined in Theorem 2.1. Recall that the conventional propensity-score approach constructs weights by substituting estimated conditional treatment probabilities into the population weight formula, as described in Section 3.1. To understand this approach from a balancing perspective, we first characterize the population weights \(w_s^*\) through a sequence of balancing identities.

\par\medskip

\setcounter{proposition}{1}
\begin{proposition}[Population balance and characterization of the population sequential weights]

Under Assumption 2.3, the population weights defined in Theorem 2.1 satisfy, for every bounded measurable function \(h\) of the arguments appearing in the corresponding identity,
\BodyDisplay{20}{
\mathbb E\!\left[
\mathbf1\{D=d_1\}w_1^*(X)h(X)
\right]
=\mathbb E[h(X)]
}
and, for \(s=2,\ldots,J+1\),
\BodyDisplay{21}{
\begin{aligned}
&\mathbb E\!\left[
\mathbf1\{D=d_s\}
w_s^*(\overline M_{s-1},X)
h(\overline M_{s-1},X)
\right] =
\mathbb E\!\left[
\mathbf1\{D=d_{s-1}\}
w_{s-1}^*(\overline M_{s-2},X)
h(\overline M_{s-1},X)
\right].
\end{aligned}
}

Conversely, suppose measurable functions \(w_s(\overline m_{s-1},x)\) satisfy the preceding balancing identities for every bounded measurable function \(h\). Then, for each \(s=1,\ldots,J+1\), \(w_s(\overline M_{s-1},X)=w_s^*(\overline M_{s-1},X)\) almost surely.

\end{proposition}

\par\medskip

At the population level, the balancing identities uniquely characterize the weights \(w_s^*\) that appear in the inverse-probability representation. The propensity-score approach targets these weights only indirectly, through models for the conditional treatment probabilities \(\pi_0,\ldots,\pi_J\). Under appropriate consistency conditions, the resulting estimated weights converge to \(w_s^*\). However, this does not imply balance in the observed sample. Substituting estimated treatment probabilities generally leaves the empirical counterparts of the balancing identities unsatisfied. By Proposition 3.1, these empirical imbalances are exactly the terms that enter the finite-sample bias decomposition. This motivates the method proposed in Section 4, which constructs weights by directly constraining the empirical imbalances of selected basis functions.

The population characterization in Proposition 3.2 also provides a foundation for the asymptotic analysis of the proposed estimator in Section 5. As the basis dimensions \(K_s\) increase, sufficiently rich basis spaces can approximate broader classes of functions. Under suitable assumptions, the resulting balancing weights can recover the population weights characterized by Proposition 3.2.

\FloatBarrier

\section{Proposed method: sequential balancing weights}\label{sec-4}

This section turns these targets into a sequential weight-construction procedure. Section 4.1 constructs the weights, Section 4.2 describes exact balance and the tolerances introduced when exact balance cannot be attained, and Section 4.3 examines the benefit of using the augmented weighting estimator rather than the weighting estimator with these weights.

\subsection{Sequential balancing-weight construction for ordered mediators}\label{sec-4-1}

This section introduces the sequential balancing-weight construction. We begin by fixing notation. For each stage \(s=1,\ldots,J+1\), let \(B_s\) denote a collection of \(K_s\) balancing functions. Specifically, define \(B_1(X) = \bigl( b_{1,1}(X),\ldots,b_{1,K_1}(X) \bigr)^\top\) and, for \(s=2,\ldots,J+1\), \(B_s(\overline M_{s-1},X) = \bigl( b_{s,1}(\overline M_{s-1},X), \ldots, b_{s,K_s}(\overline M_{s-1},X) \bigr)^\top.\) We include a constant as the first component, so that \(b_{s,1}=1\), and impose zero tolerance on this component. Let \(\varepsilon_s=(\varepsilon_{s,1},\ldots,\varepsilon_{s,K_s})^\top\) denote the stage-specific tolerance vector, where \(\varepsilon_{s,1}=0\) and \(\varepsilon_{s,k}\geq0\) for all \(s\) and \(k\). Let \(\ell:[0,\infty)\to\mathbb R\) be a strictly convex dispersion loss. Algorithm 4.1 then constructs the sequence of balancing weights stage by stage.

\par\medskip

\begin{manuscriptalgorithm}[10]{Algorithm 4.1 Sequential balancing-weight construction for causally ordered mediators}

\begin{itemize}
\item
  \textbf{Inputs.} Fix \(\boldsymbol d=(d_{J+1},d_J,\ldots,d_1)\). For each \(s=1,\ldots,J+1\), specify the basis \(B_s\), its dimension \(K_s\), and the tolerance vector \(\varepsilon_s\). Choose a strictly convex dispersion loss \(\ell:[0,\infty)\to\mathbb R\).
\item
  \textbf{First stage.} Assign nonnegative weights to observations with \(D_i=d_1\) by solving \BodyDisplay{22}{
  \begin{aligned}
  \min_{\{w_{1,i}\geq0:D_i=d_1\}}
  &\quad
  \frac1N\sum_{i:D_i=d_1}\ell(w_{1,i})\\
  \text{subject to}
  &\quad
  \left|
  \frac1N\sum_{i:D_i=d_1}
  w_{1,i}b_{1,k}(X_i)
  -
  \frac1N\sum_{i=1}^{N}b_{1,k}(X_i)
  \right|
  \leq\varepsilon_{1,k},\\
  &\quad k=1,\ldots,K_1.
  \end{aligned}
  }

  Denote the solution by \(\widehat w_{1,i}\) for observations with \(D_i=d_1\).
\item
  \textbf{Subsequent stages.} For each \(s=2,\ldots,J+1\), hold the preceding solution \(\{\widehat w_{s-1,i}:D_i=d_{s-1}\}\) fixed. Assign nonnegative weights to observations with \(D_i=d_s\) by solving \BodyDisplay{23}{
  \begin{aligned}
  &\min_{\{w_{s,i}\geq0:D_i=d_s\}}
  \quad
  \frac1N\sum_{i:D_i=d_s}\ell(w_{s,i})\\
  &\text{subject to}
  \quad
  \left|
  \frac1N\sum_{i:D_i=d_s}
  w_{s,i}
  b_{s,k}(\overline M_{s-1,i},X_i)
  -
  \frac1N\sum_{i:D_i=d_{s-1}}
  \widehat w_{s-1,i}
  b_{s,k}(\overline M_{s-1,i},X_i)
  \right|
  \leq\varepsilon_{s,k},\\
  &\quad k=1,\ldots,K_s.
  \end{aligned}
  }

  Denote the solution by \(\widehat w_{s,i}\) for observations with \(D_i=d_s\).
\item
  \textbf{Output.} Return the stage-specific weights \(\{\widehat w_{s,i}:D_i=d_s\}\), \(s=1,\ldots,J+1\).
\end{itemize}

\end{manuscriptalgorithm}

\par\medskip

We now describe four features of Algorithm 4.1 in turn, namely the flow of information across stages, the dispersion loss, the role of the basis together with exact and approximate balance, and the normalization of the weights. First, the sequential ordering preserves the learning-information restriction used in Proposition 3.1. More precisely, every data-dependent choice affecting stage \(s\) must be measurable with respect to \(\sigma\!\left((D_\ell,X_\ell,\overline M_{s-1,\ell})_{\ell=1}^{N}\right)\). The tolerance selection of Section 4.2 and Appendix C respects the same information restriction.

Second, the convex objective controls dispersion among weights satisfying the balance restrictions. Entropy and quadratic losses both discourage unnecessary concentration because the exact constant constraint fixes the total weighted mass. The entropy loss is closely related to entropy balancing, whereas the quadratic loss directly minimizes the squared Euclidean norm of the weights. By the dual representation in Theorem 5.1, the weights are obtained by applying the derivative \((\ell^{\mathrm c})'\) of the convex conjugate of the loss to a linear combination of the basis functions, so the choice of loss can be interpreted as a choice of link function for the weights. For the entropy loss \(\ell(w)=w\log w-w\), \((\ell^{\mathrm c})'(a)=\exp(a)\), which gives exponential weights. For the quadratic loss \(\ell(w)=w^2/2\), \((\ell^{\mathrm c})'(a)=\max(a,0)\), which gives weights that are linear in the basis functions where they are positive and zero elsewhere. This interpretation does not by itself determine which link function is preferable, and a general criterion for choosing the loss remains to be developed.

Third, the basis and the tolerances determine which empirical imbalances are controlled and how tightly, which connects the constraints directly to the finite-sample bias decomposition in Proposition 3.1. The tolerances also matter for computation. With exact balance, the target moment vector must lie in the convex hull generated by the available observations in the target treatment group. This condition can fail under limited overlap, a small target group, or a large balance basis. Positive tolerances enlarge the feasible set and can substantially reduce both computational failure and weight concentration.

Finally, the constant constraints imply exact normalization at every stage. At the first stage, \(N^{-1}\sum_{i:D_i=d_1}\widehat w_{1,i}=1\). Proceeding recursively, for \(s=2,\ldots,J+1\), we obtain \(N^{-1}\sum_{i:D_i=d_s}\widehat w_{s,i}=N^{-1}\sum_{i:D_i=d_{s-1}}\widehat w_{s-1,i}=1\). Consequently, \(\sum_{i:D_i=d_s}\widehat w_{s,i}=N\) for every \(s=1,\ldots,J+1\).

\par\medskip

\subsection{Exact balance and balance tolerances}\label{sec-4-2}

The proposed method sets \(\varepsilon_s=0\) at every stage, so the weighted sample moments of every basis function equal their targets. Exact balance requires the target moment vector at each stage to lie in the convex hull generated by the target treatment group, a condition that can fail in small samples. We therefore introduce positive tolerances only at a stage whose exact balancing problem has no solution. The stages are solved in the order of Algorithm 4.1, and each stage sets \(\varepsilon_s=0\) whenever its exact problem has a solution. Otherwise, positive tolerances are selected for that stage alone, with the weights of the earlier stages held fixed. The tolerances are parameterized by one standardized scalar \(\lambda_s\geq0\) at each stage, with \(\varepsilon_{s,1}=0\) and \(\varepsilon_{s,k}=\lambda_s\widehat\sigma_{s,k}\) for \(k\geq2\), where \(\widehat\sigma_{s,k}\) is the full-sample standard deviation of \(b_{s,k}\) with divisor \(N\). Because outcome information is deliberately excluded from weight construction, \(\lambda_s\) is not chosen by minimizing an outcome-prediction loss. Instead, Algorithm C.1 in Appendix C selects \(\lambda_s\) at such a stage from the positive values of a finite grid, using a criterion that measures the stability of the weighted balance under bootstrap resampling of the observations. This criterion extends the tuning algorithm of \citet{Wang-MinimalDispersionConsiderations-2019m}. Sections 6 and 7 use the grid \(\{0,0.005,0.01,0.025,0.05,0.10\}\) with \(R=500\) bootstrap replications. Both the decision to introduce tolerances at a stage and their selection satisfy the stagewise information restriction of Section 3.2, so Proposition 3.1 applies to the whole procedure.

\par\medskip

\subsection{The role of augmentation}\label{sec-4-3}

This section compares weighting and augmented weighting estimators that use the same sequential balancing weights. We use the same stage-specific weights \(\widehat w_{s,i}\), \(s=1,\ldots,J+1\). Define the weighting estimator as
\BodyDisplay{24}{
\widehat\theta_{\boldsymbol d}^{\mathrm{W\text{-}BW}}
=
\frac1N\sum_{i=1}^{N}
\mathbf1\{D_i=d_{J+1}\}\widehat w_{J+1,i}Y_i.
}

The augmented weighting estimator used for comparison is \(\widehat\theta_{\boldsymbol d}^{\mathrm{AW\text{-}BW}}\), defined in Section 3.1 with these weights. For the comparison, we impose the following assumption.

\par\medskip

\setcounter{assumption}{0}
\begin{assumption}[Conditions for comparing weighting and augmented weighting]

Suppose that the following two conditions hold. (a) Suppose that the basis \(B_s\) at each stage is specified deterministically. For every \(s=1,\ldots,J+1\), there is a deterministic coefficient vector \(\beta_s^*\in\mathbb R^{K_s}\) such that \(\mu_{s-1}^{(d_s,\ldots,d_{J+1})} (\overline m_{s-1},x) = B_s(\overline m_{s-1},x)^\top\beta_s^*\) holds. (b) Suppose that the estimated regressions are also represented as linear combinations of the same basis, \(\widehat\mu_{s-1}^{(d_s,\ldots,d_{J+1})} (\overline m_{s-1},x) = B_s(\overline m_{s-1},x)^\top\widehat\beta_s\).

\end{assumption}

\par\medskip

The next theorem compares the bias structures of the weighting and augmented weighting estimators.

\par\medskip

\setcounter{theorem}{0}
\begin{theorem}[Equivalence under exact balance and imbalance correction under approximate balance]

Suppose that the identification formula in Theorem 2.1 holds and that the weights satisfy the stagewise information restrictions in Section 3.2, as in the preceding sections. Under Assumption 4.1, assume that all expectations in the two bias identities below and their derivation exist as finite real numbers. The following results hold.

The bias of the weighting estimator is
\BodyDisplay{25}{
\begin{aligned}
\mathbb E\!\left[
\widehat\theta_{\boldsymbol d}^{\mathrm{W\text{-}BW}}-\theta_{\boldsymbol d}
\right]
&=
\mathbb E\!\left[
(\beta_1^*)^\top
\frac1N\sum_{i=1}^{N}
\{\mathbf1\{D_i=d_1\}\widehat w_{1,i}-1\}B_1
\right]\\
&\quad+
\sum_{s=2}^{J+1}
\mathbb E\!\left[
(\beta_s^*)^\top
\frac1N\sum_{i=1}^{N}
\{\mathbf1\{D_i=d_s\}\widehat w_{s,i}
-\mathbf1\{D_i=d_{s-1}\}\widehat w_{s-1,i}\}B_s
\right].
\end{aligned}
}

In contrast, the bias of the augmented weighting estimator is
\BodyDisplay{26}{
\begin{aligned}
\mathbb E\!\left[
\widehat\theta_{\boldsymbol d}^{\mathrm{AW\text{-}BW}}-\theta_{\boldsymbol d}
\right]
&=
\mathbb E\!\left[
(\beta_1^*-\widehat\beta_1)^\top
\frac1N\sum_{i=1}^{N}
\{\mathbf1\{D_i=d_1\}\widehat w_{1,i}-1\}B_1
\right]\\
&\quad+
\sum_{s=2}^{J+1}
\mathbb E\!\left[
(\beta_s^*-\widehat\beta_s)^\top
\frac1N\sum_{i=1}^{N}
\{\mathbf1\{D_i=d_s\}\widehat w_{s,i}
-\mathbf1\{D_i=d_{s-1}\}\widehat w_{s-1,i}\}B_s
\right].
\end{aligned}
}

Consequently, under exact balance the two point estimates coincide. In a sample in which the basis is exactly balanced at every stage, \(\widehat\theta_{\boldsymbol d}^{\mathrm{AW\text{-}BW}}=\widehat\theta_{\boldsymbol d}^{\mathrm{W\text{-}BW}}\).

\end{theorem}

\par\medskip

Theorem 4.1 shows that, under exact balance, the regression correction terms are exactly zero, so weighting and augmented weighting give the same point estimate. Under approximate balance, imbalance with respect to the basis remains. Accordingly, the contribution of imbalance is bounded above by \(\sum_{s=1}^{J+1}\|\varepsilon_s\|_\infty\|\beta_s^*\|_1\) for weighting and by \(\sum_{s=1}^{J+1}\|\varepsilon_s\|_\infty\|\widehat\beta_s-\beta_s^*\|_1\) for augmented weighting. Thus, if the true backward regressions lie in the linear span of the basis under approximate balance, augmentation has the advantage that the bias arising from the remaining basis imbalance is small whenever the regression coefficients are accurately estimated.

\par\medskip

\FloatBarrier

\section{Asymptotic theory for sequential balancing weights and the augmented weighting estimator}\label{sec-5}

This section establishes the asymptotic properties of the proposed weights and of the augmented weighting estimator. Section 5.1 derives the dual representation and the convergence rates of the weights, Section 5.2 uses these rates to show that the augmented weighting estimator is consistent, asymptotically normal, and attains the semiparametric efficiency bound, and Section 5.3 shows that its asymptotic variance can be estimated consistently.

\subsection{Dual representation and convergence rates of sequential balancing weights}\label{sec-5-1}

This subsection derives the dual of the balancing problems in Algorithm 4.1 and then the convergence rates of the resulting weights.

\par\medskip

\setcounter{theorem}{0}
\begin{theorem}[Dual representations and weights at all stages]

Define the convex conjugate of the loss on its nonnegative domain by \(\ell^{\mathrm c}(a):=\sup_{w\geq0}\{aw-\ell(w)\}\). The superscript \(\mathrm c\) denotes the convex conjugate. Assume that \(\ell\), extended by \(+\infty\) to negative arguments, is proper, closed, and convex. When the primal problems are feasible and each of the following dual minimization problems has a finite solution, denote these solutions by \(\widehat\gamma_s\in\mathbb R^{K_s}\), \(s=1,\ldots,J+1\). Assume further that \(\ell^{\mathrm c}\) is finite and differentiable at \(B_s(\overline M_{s-1,i},X_i)^\top\widehat\gamma_s\) for every \(s=1,\ldots,J+1\) and \(i\) with \(D_i=d_s\), using \(B_1(X_i)\) at \(s=1\). At the first stage,
\BodyDisplay{27}{
\begin{aligned}
\widehat\gamma_1\in\operatorname*{arg\,min}_{\gamma_1\in\mathbb R^{K_1}}
\Bigg[
&\frac1N\sum_{i=1}^{N}
\left\{
\mathbf1\{D_i=d_1\}\ell^{\mathrm c}(B_1(X_i)^\top\gamma_1)
-B_1(X_i)^\top\gamma_1
\right\} +\sum_{k=2}^{K_1}\varepsilon_{1,k}|\gamma_{1,k}|
\Bigg].
\end{aligned}
}

For each \(s=2,\ldots,J+1\),
\BodyDisplay{28}{
\begin{aligned}
\widehat\gamma_s\in\operatorname*{arg\,min}_{\gamma_s\in\mathbb R^{K_s}}
\Bigg[
&\frac1N\sum_{i=1}^{N}
\Big\{
\mathbf1\{D_i=d_s\}
\ell^{\mathrm c}\!\left(B_s(\overline M_{s-1,i},X_i)^\top\gamma_s\right)\\
&\qquad-
\mathbf1\{D_i=d_{s-1}\}\widehat w_{s-1,i}
B_s(\overline M_{s-1,i},X_i)^\top\gamma_s
\Big\}+\sum_{k=2}^{K_s}\varepsilon_{s,k}|\gamma_{s,k}|
\Bigg].
\end{aligned}
}

These are minimization forms obtained by reversing the sign of the dual maximization problems. The preceding weights are held fixed at each subsequent stage. The primal solutions in the target groups are
\BodyDisplay{29}{
\widehat w_{s,i}
=(\ell^{\mathrm c})'\!\left(
B_s(\overline M_{s-1,i},X_i)^\top\widehat\gamma_s
\right),
\qquad\text{for } D_i=d_s,\quad s=1,\ldots,J+1.
}

\end{theorem}

\par\medskip

This dual representation shows that the weights are estimated directly from a linear combination of basis functions, with the derivative \((\ell^{\mathrm c})'\) of the convex conjugate as the link function. In the primal problem, \(\varepsilon_{s,k}\) is the permissible discrepancy in moment \(k\); in the dual, it is the penalty multiplier on the absolute coefficient \(|\gamma_{s,k}|\). Holding the other conditions fixed, increasing this tolerance relaxes the corresponding balance constraint and increases the penalty on the absolute coefficient.

Using these dual representations, we derive the rates at which the estimated weights converge to the weights based on the true conditional treatment probabilities. For population and full-sample error evaluation, extend the estimated weights to the common support by \(\widehat w_s(\overline m_{s-1},x)=(\ell^{\mathrm c})'(B_s(\overline m_{s-1},x)^\top\widehat\gamma_s)\), and write \(\widehat w_{s,i}=\widehat w_s(\overline M_{s-1,i},X_i)\) at all observations. We impose the following assumptions.

\par\medskip

\setcounter{assumption}{0}
\begin{assumption}[Nondegenerate bases]

For every \(s=1,\ldots,J+1\), the basis \(B_s\) and its dimension \(K_s\) are specified deterministically before observing the analysis sample.

\(\sup_{\overline m_{s-1},x}\|B_s(\overline m_{s-1},x)\|_2\leq C\sqrt{K_s}\) and \(cI\preceq\mathbb E[B_s(\overline M_{s-1},X)B_s(\overline M_{s-1},X)^\top]\preceq CI\) for every \(s=1,\ldots,J+1\).

\end{assumption}

\par\medskip

Here, \(c,C>0\) are constants independent of \(N\) and the basis dimensions. The first condition controls the magnitude of the basis vector, allowing its Euclidean norm to grow at most at rate \(\sqrt{K_s}\). The second requires the eigenvalues of the population Gram matrix to be uniformly bounded away from zero and infinity. Together, these conditions ensure that the basis functions remain sufficiently well-conditioned as their dimensions increase, ruling out redundant or nearly collinear basis functions.

\par\medskip

\setcounter{assumption}{1}
\begin{assumption}[Approximation coefficients]

{\tolerance=9999\relax For every \(s=1,\ldots,J+1\), there exist \(r_{w,s}>1\) and deterministic approximation coefficients \(\gamma_s^*\in\mathbb R^{K_s}\) such that \(\sup_{\overline m_{s-1},x}|(\ell^{\mathrm c})'(B_s(\overline m_{s-1},x)^\top\gamma_s^*)-w_s^*(\overline m_{s-1},x)|=O(K_s^{-r_{w,s}})\).\par}

\end{assumption}

\par\medskip

This condition requires the true population weight \(w_s^*\) to be uniformly approximated by applying the link function \((\ell^{\mathrm c})'\) to a linear combination of basis functions. The form of this approximation is motivated by the population balancing characterization in Proposition 3.2. The exponent \(r_{w,s}>1\) determines how rapidly the approximation error decreases as the basis dimension \(K_s\) increases. Under appropriate normalization and smoothness conditions, commonly used series bases, including regression splines, trigonometric polynomials, and wavelets, can satisfy the basis norm, Gram-matrix, and approximation requirements \citep{Newey-ConvergenceRatesEstimators-1997k,Chen-Chapter76Models-2007p,ChenChristensen-OptimalUniformRates-2015}.

\par\medskip

\setcounter{assumption}{2}
\begin{assumption}[Local curvature]

For every \(s=1,\ldots,J+1\), let \(\Gamma_s\) be a local neighborhood of the approximation coefficients \(\gamma_s^*\) defined in Assumption 5.2. For a sufficiently small positive constant \(c\) independent of \(N\), \(\{\gamma_s\in\mathbb R^{K_s}:\sup_{\overline m_{s-1},x}|B_s(\overline m_{s-1},x)^\top(\gamma_s-\gamma_s^*)|\leq c\}\subseteq\Gamma_s\). The bounds \(c\leq(\ell^{\mathrm c})'(B_s(\overline m_{s-1},x)^\top\gamma_s)\leq C\) and \(c\leq(\ell^{\mathrm c})''(B_s(\overline m_{s-1},x)^\top\gamma_s)\leq C\) are imposed for every \(\gamma_s\in\Gamma_s\) and every argument on the support.

\end{assumption}

\par\medskip

These conditions require the link function and its derivative to be uniformly bounded above and away from zero in a neighborhood of the approximation coefficients. The lower bound on \((\ell^{\mathrm c})'\) ensures that the corresponding weights remain positive, while its upper bound prevents them from becoming arbitrarily large. The bounds on \((\ell^{\mathrm c})''\) ensure that the dual objective has nondegenerate local curvature.

\par\medskip

\setcounter{assumption}{3}
\begin{assumption}[Growth of basis dimensions]

For every \(s=1,\ldots,J+1\), let \(K_s\to\infty\) and assume \(\sqrt{K_s}\sum_{r=1}^{s}\{\sqrt{K_r/N}+K_r^{-r_{w,r}}\}\longrightarrow0\).

\end{assumption}

\par\medskip

This condition is sufficient for localization of the dual solution and uniform consistency of the weights after errors from preceding stages propagate to stage \(s\). The term involving the stage's own sampling error implies \(K_s^2/N\to0\). The cross-stage sampling terms \(\sqrt{K_sK_r/N}\) converge to zero under the corresponding squared-dimension conditions. The remaining terms control the propagation of approximation errors. One sufficient choice of dimensions for consistency is \(K_s\asymp N^{1/3}\) for every \(s=1,\ldots,J+1\), with all \(r_{w,s}>1\). Asymptotic normality also requires regression approximation and product conditions, which we verify below for the same dimension choice.

\par\medskip

\setcounter{assumption}{4}
\begin{assumption}[Tolerances]

For every \(s=1,\ldots,J+1\), let \(\varepsilon_{s,k}\geq0\) and \(\varepsilon_{s,1}=0\), and assume \(\|\varepsilon_s\|_\infty=o_p((NK_s)^{-1/2})\). Consequently, \(\|\varepsilon_s\|_2=o_p(N^{-1/2})\). Exact balance is the case in which every component is zero.

\end{assumption}

\par\medskip

This assumption restricts the magnitude of the tolerances to prevent excessive empirical imbalance. If the tolerances are too large, the resulting imbalance may contribute to the first-order estimation error and affect the asymptotic distribution.

\par\medskip

\setcounter{theorem}{1}
\begin{theorem}[Stagewise convergence rates for sequential balancing weights]

Suppose the i.i.d. sample satisfies Assumption 2.3, the loss and learning-information structure of Algorithm 4.1, and Assumptions 5.1--5.5. Then, on events whose probability tends to one, all \(J+1\) dual problems have finite, unique solutions, and the resulting target-group weights are the unique primal solutions.

For every \(s=1,\ldots,J+1\), the dual-coefficient errors satisfy
\BodyDisplay{30}{
\|\widehat\gamma_s-\gamma_s^*\|_2=O_p(\sum_{r=1}^{s}\{\sqrt{K_r/N}+K_r^{-r_{w,r}}\})=o_p(1).
}

The root-mean-square weight errors, both in the population and over the learning sample, satisfy the same rate bound as the dual-coefficient errors, as shown in the proof of Theorem 5.2 in Appendix A. Finally, the uniform errors satisfy
\BodyDisplay{31}{
\|\widehat w_s-w_s^*\|_\infty
=O_p\!\left(
\sqrt{K_s}
\sum_{r=1}^{s}
\left\{\sqrt{\frac{K_r}{N}}+K_r^{-r_{w,r}}\right\}
\right)
=o_p(1).
}

\end{theorem}

\par\medskip

Theorem 5.2 establishes the consistency of the sequential balancing weights at every stage. Specifically, the estimated dual coefficients converge to their approximation counterparts, while the estimated weights converge to the true population weights in both population and sample root-mean-square norms, as well as uniformly over the support. In these rates, the errors from the current and preceding stages enter additively and are not amplified at successive stages.

\par\medskip

\subsection{Asymptotic normality and semiparametric efficiency}\label{sec-5-2}

This subsection derives the asymptotic distribution of the augmented weighting estimator and shows that it attains the semiparametric efficiency bound.

\par\medskip

\setcounter{assumption}{5}
\begin{assumption}[Moments]

{\tolerance=9999\relax Assume \(\mathbb E[Y^2]<\infty\). For every \(j=1,\ldots,J\), \(\mathbb E[\{\mu_j^{(d_{j+1},\ldots,d_{J+1})}(\overline M_j,X)-\mu_{j-1}^{(d_j,\ldots,d_{J+1})}(\overline M_{j-1},X)\}^2\mid D=d_j,\overline M_{j-1},X]\leq C\). At the final stage, \(\mathbb E[\{Y-\mu_J^{(d_{J+1})}(\overline M_J,X)\}^2\mid D=d_{J+1},\overline M_J,X]\leq C\).\par}

\end{assumption}

\par\medskip

These conditions require the true regression residuals appearing in the augmented weighting estimator to have uniformly bounded conditional second moments. Together with the positivity condition in Assumption 2.3, these moment bounds ensure that the weighted residual terms have finite second moments.

\par\medskip

\setcounter{assumption}{6}
\begin{assumption}[Regression approximation]

For every \(s=1,\ldots,J+1\), there exist \(r_{\mu,s-1}>1\) and deterministic coefficients \(\beta_s^*\in\mathbb R^{K_s}\) such that \(\sup_{\overline m_{s-1},x}|\mu_{s-1}^{(d_s,\ldots,d_{J+1})}(\overline m_{s-1},x)-B_s(\overline m_{s-1},x)^\top\beta_s^*|=O(K_s^{-r_{\mu,s-1}})\).

\end{assumption}

\par\medskip

This assumption requires each true backward regression to be uniformly approximated by a linear combination of the basis functions used for balancing at the corresponding stage.

\par\medskip

\setcounter{assumption}{7}
\begin{assumption}[Product conditions for the asymptotic distribution]

Impose the two deterministic conditions \(\sqrt N K_{J+1}^{-r_{\mu,J}}\sum_{r=1}^{J+1}\{\sqrt{K_r/N}+K_r^{-r_{w,r}}\}\longrightarrow0\) and \(\sqrt N\sum_{s=1}^{J}(K_s^{-r_{\mu,s-1}}+K_{s+1}^{-r_{\mu,s}})\sum_{r=1}^{s}\{\sqrt{K_r/N}+K_r^{-r_{w,r}}\}\longrightarrow0\).

\end{assumption}

\par\medskip

These conditions control the interaction between weight estimation errors and regression approximation errors. The first condition concerns the terminal outcome regression, whereas the second accounts for the approximation errors of adjacent backward regressions at each preceding stage. A sufficient choice of basis dimensions satisfying the two product conditions is \(K_s\asymp N^{1/3}\) for every \(s=1,\ldots,J+1\), provided that all approximation exponents \(r_{w,s}\) and \(r_{\mu,s-1}\) exceed one.

\par\medskip

\setcounter{assumption}{8}
\begin{assumption}[Balancing-basis control of estimated backward regressions]

For every \(s=1,\ldots,J+1\), suppose that the estimated backward regression belongs to the linear span of the corresponding balancing basis. Specifically, there exists a data-dependent coefficient vector \(\widehat\beta_s\in\mathbb R^{K_s}\) such that \(\widehat\mu_{s-1}^{(d_s,\ldots,d_{J+1})}(\overline m_{s-1},x)=B_s(\overline m_{s-1},x)^\top\widehat\beta_s\). Furthermore, assume that \(\sqrt N\|\widehat\beta_s\|_1\|\varepsilon_s\|_\infty=o_p(1)\).

\end{assumption}

\par\medskip

Assumption 5.9 provides a sufficient condition for controlling the empirical balance residuals involving estimated backward regressions. Because Algorithm 4.1 directly constrains the imbalance of each basis function, the same constraints also control any linear combination of those functions. Regression estimators such as ordinary least squares, ridge regression, lasso, and elastic net satisfy the linear-span requirement when they are fitted using the corresponding balancing basis.

\par\medskip

\setcounter{remark}{0}
\begin{remark}[Extension beyond the balancing-basis control by empirical process theory]

Assumption 5.9 is sufficient but not necessary for the asymptotic linearity and normality in Theorem 5.3, and the requirement that the estimated backward regressions lie in the linear span of the balancing basis can be relaxed. The empirical balance residual for the regression estimation error decomposes into a population-weight component and a weight-estimation-error component. A Donsker condition \citep{vanDerVaart-Wellner-WeakConvergence-2023} on the corresponding true-weighted class of regression functions and population root-mean-square consistency of the estimated regressions control the first component. Theorem 5.2's sample root-mean-square weight rates, together with product-rate conditions on the estimated regressions, control the second. Appendix B gives these alternative conditions and their proof.

\end{remark}

\par\medskip

The preceding assumptions lead to our main theorem, which establishes the asymptotic linearity, asymptotic normality, and semiparametric efficiency of the augmented weighting estimator.

\par\medskip

\setcounter{theorem}{2}
\begin{theorem}[Convergence rate and asymptotic distribution of the augmented weighting estimator]

Suppose the identification formula in Theorem 2.1, Assumption 2.3, the learning-information restrictions in Algorithm 4.1, Assumptions 5.1--5.7, and Assumption 5.9 hold. Then
\BodyDisplay{32}{
\begin{aligned}
\widehat\theta_{\boldsymbol d}^{\mathrm{AW\text{-}BW}}-\theta_{\boldsymbol d}
&=O_p(N^{-1/2}) +
O_p\!\left(
\sum_{s=1}^{J}
\left(K_s^{-r_{\mu,s-1}}+K_{s+1}^{-r_{\mu,s}}\right)
\sum_{r=1}^{s}
\left\{\sqrt{\frac{K_r}{N}}+K_r^{-r_{w,r}}\right\}
\right)\\
&\quad+
O_p\!\left(
K_{J+1}^{-r_{\mu,J}}
\sum_{r=1}^{J+1}
\left\{\sqrt{\frac{K_r}{N}}+K_r^{-r_{w,r}}\right\}
\right)
=o_p(1).
\end{aligned}
}

Recall the true centered score \(\psi_{\boldsymbol d}(O)=\phi_{\boldsymbol d}(O)-\theta_{\boldsymbol d}\) from Theorem 2.1(iii), which uses the population weights, with \(O=(Y,D,\overline M_J,X)\). If Assumption 5.8 also holds, then
\BodyDisplay{33}{
\begin{aligned}
\sqrt N\left(
\widehat\theta_{\boldsymbol d}^{\mathrm{AW\text{-}BW}}-\theta_{\boldsymbol d}
\right)
&=\frac1{\sqrt N}\sum_{i=1}^{N}
\psi_{\boldsymbol d}(Y_i,D_i,\overline M_{J,i},X_i)
+o_p(1)\\
&\xrightarrow{d}\mathcal N(0,V_{\boldsymbol d}),
\qquad V_{\boldsymbol d}=\mathbb E[\psi_{\boldsymbol d}^2]<\infty.
\end{aligned}
}

Under these conditions, \(V_{\boldsymbol d}\) equals the semiparametric efficiency bound for \(\theta_{\boldsymbol d}\), because Theorem 2.1(iii) shows that \(\psi_{\boldsymbol d}\) is the efficient influence function.

\end{theorem}

\par\medskip

Theorem 5.3 shows that the convergence rate depends on the interaction between outcome-regression approximation errors and sequential weight-estimation errors. Under Assumption 5.8, these errors are asymptotically negligible at the root-\(N\) scale, yielding asymptotic normality and semiparametric efficiency.

\subsection{Consistency of the variance estimator}\label{sec-5-3}

Finally, we show that the asymptotic variance \(V_{\boldsymbol d}\) in Theorem 5.3 can be consistently estimated, which requires one additional condition on the regression errors.

\par\medskip

\setcounter{assumption}{9}
\begin{assumption}[Regression errors required for variance estimation]

{\tolerance=9999\relax Suppose that, at all evaluation points of the same full sample used for learning, \(\{N^{-1}\sum_{i=1}^{N}|\widehat\mu_{s-1}^{(d_s,\ldots,d_{J+1})}(\overline M_{s-1,i},X_i)-\mu_{s-1}^{(d_s,\ldots,d_{J+1})}(\overline M_{s-1,i},X_i)|^2\}^{1/2}=o_p(1)\) for every \(s=1,\ldots,J+1\).\par}

\end{assumption}

\par\medskip

These conditions include errors from evaluating regressions learned in one treatment group on observations in the other group. They differ from \(L_2\) errors integrated over a population distribution. As a sufficient example, consider backward linear regression on the stage-specific bases. Under Assumption 2.3 and Assumptions 5.1--5.7, \(\{N^{-1}\sum_{i=1}^{N}|\widehat\mu_{s-1}^{(d_s,\ldots,d_{J+1})}(\overline M_{s-1,i},X_i)-\mu_{s-1}^{(d_s,\ldots,d_{J+1})}(\overline M_{s-1,i},X_i)|^2\}^{1/2}=O_p(\sum_{r=s}^{J+1}\{\sqrt{K_r/N}+K_r^{-r_{\mu,r-1}}\})\) for \(s=1,\ldots,J+1\). Each bound retains the errors from subsequent regressions used in the backward recursion. If all approximation exponents exceed one and \(K_s\asymp N^{1/3}\) at every stage, all these errors are \(O_p(N^{-1/3})=o_p(1)\).

The next theorem shows that the plug-in estimator of \(V_{\boldsymbol d}\), which replaces each component of the influence function with the estimates used to compute \(\widehat\theta_{\boldsymbol d}^{\mathrm{AW\text{-}BW}}\), is consistent.

\par\medskip

\setcounter{theorem}{3}
\begin{theorem}[Consistency of the variance estimator]

Suppose that Assumption 5.10 holds in addition to the asymptotic-normality conditions of Theorem 5.3. Then the following variance estimator converges in probability to \(V_{\boldsymbol d}\):
\BodyDisplay{34}{
\begin{aligned}
&\widehat V_{\boldsymbol d}^{\mathrm{AW\text{-}BW}}
=\frac1N\sum_{i=1}^{N}
\Bigg[
\widehat\mu_0^{(d_1,\ldots,d_{J+1})}(X_i)
-\widehat\theta_{\boldsymbol d}^{\mathrm{AW\text{-}BW}}\\
&\qquad+
\sum_{j=1}^{J}
\mathbf1\{D_i=d_j\}\widehat w_{j,i}
\Big\{
\widehat\mu_j^{(d_{j+1},\ldots,d_{J+1})}(\overline M_{j,i},X_i) -
\widehat\mu_{j-1}^{(d_j,\ldots,d_{J+1})}(\overline M_{j-1,i},X_i)
\Big\}\\
&\qquad+
\mathbf1\{D_i=d_{J+1}\}\widehat w_{J+1,i}
\left\{Y_i-\widehat\mu_J^{(d_{J+1})}(\overline M_{J,i},X_i)\right\}
\Bigg]^2\xrightarrow{p}V_{\boldsymbol d}.
\end{aligned}
}

\end{theorem}

\par\medskip

When \(V_{\boldsymbol d}>0\), Theorem 5.4 combined with Theorem 5.3 and Slutsky's lemma yields asymptotically valid Wald-type confidence intervals. The same construction applies to contrasts of two generalized mediation functionals, such as the NDE and NIE, when the variance of the difference of their scores is positive.

\FloatBarrier

\section{Simulation}\label{sec-6}

This simulation compares sequential balancing weights for two ordered mediators under nonlinear treatment and mediator generation and weak overlap. The proposed method estimates generalized mediation functionals (GMFs). With binary treatment and two mediators, the combinations of treatment inputs to the outcome and the two mediators give eight GMFs. We consider the contrast \(\theta_{(1,1,1)}-\theta_{(1,0,1)}\). For the latter mean, the target group, whose weights are estimated at each stage, switches as \(1\to0\to1\), involving the maximum of two group switches possible across the three stages with two mediators. The design therefore shows whether the weights remain stable and available when they repeatedly balance one treatment group against the weighted other group.

\subsection{Data-generating process}\label{sec-6-1}

We first introduce the data-generating process, which extends the simulation design of \citet[Section 6]{TchetgenTchetgen-SemiparametricTheoryAnalysis-2012y}. The independent and identically distributed observations consist of covariates \(Z=(Z_1,Z_2,Z_3,Z_4)\), binary treatment \(D\), binary mediators \(M_1,M_2\), and a real-valued outcome \(Y\). Thus \(J=2\), and \(Z\) takes the role of \(X\) in Sections 2--5. The potential mediators are \(M_1(d)\) and \(M_2(d,m_1)\), and the potential outcome is \(Y(d,m_2,m_1)\). As in Section 2.1, the outcome input comes first in \(\theta_{(d_3,d_2,d_1)}=\mathbb E[Y(d_3,M_2\{d_2,M_1(d_1)\},M_1(d_1))]\).

Let \(Z_1,\ldots,Z_4\overset{\mathrm{i.i.d.}}{\sim}\mathcal N(0,1)\) and \(U_D,U_{M_1},U_{M_2}\overset{\mathrm{i.i.d.}}{\sim}\operatorname{Unif}(0,1)\), where the uniform variables are independent of \((Z,U_Y)\). Write \(\operatorname{expit}(t)=1/(1+\exp(-t))\). Treatment is \(D=\mathbf 1\{U_D\leq\operatorname{expit}(-Z_1+0.5Z_2-0.25Z_3-0.1Z_4)\}\), and the potential mediators are
\BodyDisplay{35}{
\begin{aligned}
M_1(d)&=\mathbf 1\{U_{M_1}\leq\operatorname{expit}(0.5-Z_1+0.5Z_2-0.9Z_3+Z_4-1.5d)\},\\
M_2(d,m_1)&=\mathbf 1\{U_{M_2}\leq\operatorname{expit}(0.5-Z_1+0.5Z_2-0.9Z_3+Z_4-1.5d+0.75m_1)\}.
\end{aligned}
}

The observed mediators are \(M_1=M_1(D)\) and \(M_2=M_2(D,M_1(D))\). With an outcome error \(U_Y\sim\mathcal N(0,1)\) that is independent of \(Z\) and the uniform variables, the potential outcome is
\BodyDisplay{36}{
Y(d,m_2,m_1)=210+27.4Z_1+13.7Z_3+13.7Z_4+m_1+m_2+d+5m_2Z_1^2+U_Y,
}
and the observed outcome is \(Y=Y(D,M_2,M_1)\).

The treatment and first-mediator probabilities are those of \citet[Section 6]{TchetgenTchetgen-SemiparametricTheoryAnalysis-2012y}. The second mediator, the term \(0.75m_1\), the mediator terms \(m_1+m_2\), and the term \(5m_2Z_1^2\) are our extensions. The effect \(\theta_{(1,1,1)}-\theta_{(1,0,1)}\) fixes the treatment inputs to the outcome and the first mediator at 1 and changes the input to the second mediator from 0 to 1. Because \(Y(1,m_2,m_1)\) depends on \(m_2\) only through \(m_2(1+5Z_1^2)\),
\BodyDisplay{37}{
\theta_{(1,1,1)}-\theta_{(1,0,1)}
=\mathbb E\bigl[\bigl(M_2\{1,M_1(1)\}-M_2\{0,M_1(1)\}\bigr)(1+5Z_1^2)\bigr].
}

Since \(M_2(1,m_1)\leq M_2(0,m_1)\) for each \(m_1\), the effect is negative. Its true value is approximately \(-1.133\).

\subsection{Estimation methods}\label{sec-6-2}

This subsection specifies the bases, the three weighting methods, and the outcome regression. Every weighting method is combined with the same misspecified outcome regression, so differences between the weighting estimates arise from the weights alone. The stage-specific bases are built from \(Z\). The first-stage basis \(B_1(Z)\in\mathbb R^{15}\) consists of the constant, \(Z_1,\ldots,Z_4\), \(Z_1^2,\ldots,Z_4^2\), and the six products \(Z_jZ_k\) with \(j<k\). The later stages add each new mediator and its interactions with \(Z\), and the third stage also adds \(M_1M_2\), so that \(B_2(M_1,Z)=(B_1(Z)^\top,M_1,M_1Z_1,\ldots,M_1Z_4)^\top\), \(B_3(M_1,M_2,Z)=(B_2(M_1,Z)^\top,M_2,M_2Z_1,\ldots,M_2Z_4,M_1M_2)^\top\), and \((K_1,K_2,K_3)=(15,20,26)\). Squares of the binary mediators duplicate the mediators and are omitted. Without \(5M_2Z_1^2\), the outcome mean in each treatment group would lie in the linear span of \(B_3\). The OLS outcome regression on \(B_3\) is therefore misspecified, and the stage-3 balance constraints, which use the same basis, do not include \(M_2Z_1^2\) either.

We compare three weighting methods. For \(\theta_{(1,1,1)}\), the \(D=1\) group receives the weights at every stage. For \(\theta_{(1,0,1)}\), the \(D=1\), \(D=0\), and \(D=1\) groups receive them at stages 1, 2, and 3, and each later stage balances its basis against the preceding weighted group. PS estimates \(\pi_0\), \(\pi_1\), and \(\pi_2\) by logistic regressions of \(D\) on \(B_1\), \(B_2\), and \(B_3\), respectively, substitutes the fitted probabilities into \(\widehat w_{s,i}^{\mathrm{PS}}\) of Section 3.1, and normalizes the weights at each stage to sum to \(N\) in the target group. CBPS forms and normalizes the weights as PS does and carries the preceding weights forward when consecutive stages share the target group, but estimates the same logistic models with the R package CBPS \citep{R-CBPS}, which combines the logistic score with balance moments by GMM. At stage 1 these moments balance the inverse-probability-weighted treated and control groups against each other, and at later stages they balance the target group against the preceding weighted group (Appendix C). We use the default settings of the package and the fitted probabilities that it returns. BW constructs the weights by Algorithm 4.1 with the entropy dispersion \(\ell(w)=w\log w-w\) and every tolerance component set to 0. A stage is accepted when the KKT, constraint, and duality-gap residuals of the entropy dual problem fall below the thresholds given in Appendix C. If a stage is not accepted, tolerances are introduced at that stage as in Section 4.2, with \(\lambda_s\) selected by Algorithm C.1 from the positive values of the grid and with the number of bootstrap replications stated there.

The outcome regression is fitted by OLS on \(B_3\) in the \(D=d_3\) group. Its predictions are regressed by OLS on \(B_2\) in the \(D=d_2\) group, and the resulting predictions on \(B_1\) in the \(D=d_1\) group. The three weighting methods use the same predictions in the augmented weighting estimator \(\widehat\theta_{\boldsymbol d}^{\mathrm{AW}}\) of Section 3.1 and report \(\widehat\theta_{(1,1,1)}^{\mathrm{AW}}-\widehat\theta_{(1,0,1)}^{\mathrm{AW}}\). Regression-only OLS reports \(N^{-1}\sum_{i=1}^{N}\{\widehat\mu_0^{(1,1,1)}(Z_i)-\widehat\mu_0^{(1,0,1)}(Z_i)\}\). For the weighting methods, the standard error is the plug-in estimator of Theorem 5.4 applied to the contrasts. Regression-only OLS uses a sandwich standard error from the stacked OLS estimating equations. Nominal 95\% Wald intervals are the estimate plus or minus \(z_{0.975}\) times its standard error.

\subsection{Results}\label{sec-6-3}

We generate 3,000 Monte Carlo repetitions at each of \(N=265\), \(500\), and \(1{,}000\), where \(N=265\) equals the sample size of the empirical application in Section 7. All methods use the same simulated sample in each repetition and return an estimate in every repetition.

{
\small
\setcounter{table}{0}
\setlength{\LTcapwidth}{\linewidth}
\begin{longtable}[]{@{}
  >{\raggedleft\arraybackslash}p{(\linewidth - 12\tabcolsep) * \real{0.1017}}
  >{\raggedright\arraybackslash}p{(\linewidth - 12\tabcolsep) * \real{0.3559}}
  >{\raggedleft\arraybackslash}p{(\linewidth - 12\tabcolsep) * \real{0.1017}}
  >{\raggedleft\arraybackslash}p{(\linewidth - 12\tabcolsep) * \real{0.1356}}
  >{\raggedleft\arraybackslash}p{(\linewidth - 12\tabcolsep) * \real{0.1017}}
  >{\raggedleft\arraybackslash}p{(\linewidth - 12\tabcolsep) * \real{0.1017}}
  >{\raggedleft\arraybackslash}p{(\linewidth - 12\tabcolsep) * \real{0.1017}}@{}}
\toprule\noalign{}
\begin{minipage}[b]{\linewidth}\raggedleft
\(N\)
\end{minipage} & \begin{minipage}[b]{\linewidth}\raggedright
Method
\end{minipage} & \begin{minipage}[b]{\linewidth}\raggedleft
Bias
\end{minipage} & \begin{minipage}[b]{\linewidth}\raggedleft
Variance
\end{minipage} & \begin{minipage}[b]{\linewidth}\raggedleft
RMSE
\end{minipage} & \begin{minipage}[b]{\linewidth}\raggedleft
Mean SE
\end{minipage} & \begin{minipage}[b]{\linewidth}\raggedleft
Cov.
\end{minipage} \\
\midrule\noalign{}
\endfirsthead
\toprule\noalign{}
\begin{minipage}[b]{\linewidth}\raggedleft
\(N\)
\end{minipage} & \begin{minipage}[b]{\linewidth}\raggedright
Method
\end{minipage} & \begin{minipage}[b]{\linewidth}\raggedleft
Bias
\end{minipage} & \begin{minipage}[b]{\linewidth}\raggedleft
Variance
\end{minipage} & \begin{minipage}[b]{\linewidth}\raggedleft
RMSE
\end{minipage} & \begin{minipage}[b]{\linewidth}\raggedleft
Mean SE
\end{minipage} & \begin{minipage}[b]{\linewidth}\raggedleft
Cov.
\end{minipage} \\
\midrule\noalign{}
\endhead
1,000 & PS & 0.210 & 0.086 & 0.361 & 0.268 & 81.6\% \\
1,000 & CBPS & 0.198 & 0.040 & 0.281 & 0.240 & 86.9\% \\
1,000 & BW & 0.155 & 0.036 & 0.245 & 0.258 & 92.8\% \\
1,000 & Regression-only (OLS) & 0.400 & 0.050 & 0.458 & 0.207 & 49.5\% \\
\bottomrule\noalign{}
\noalign{\vskip\abovecaptionskip}
\caption{Estimates of \(\theta_{(1,1,1)}-\theta_{(1,0,1)}\) in the two-mediator extension. The true value is \(-1.133\). At \(N=265\), exact balance is not attained in 256 repetitions, in which BW introduces the tolerances of Section 4.2. Exact balance is attained in every repetition at \(N=500\) and \(1{,}000\). Bias is the average of the estimate minus the true value, Variance is the sample variance of the estimates with divisor 2,999, and RMSE is the square root of the average squared estimation error, each over the 3,000 repetitions. Mean SE is the average estimated standard error, and Cov. is the proportion of the 3,000 repetitions in which the nominal 95\% interval covers the true value.}\tabularnewline
\endlastfoot
265 & PS & 0.370 & 0.269 & 0.637 & 0.426 & 77.2\% \\*
265 & CBPS & 0.393 & 0.178 & 0.576 & 0.402 & 79.4\% \\*
265 & BW & 0.321 & 0.208 & 0.557 & 0.468 & 86.3\% \\*
265 & Regression-only (OLS) & 0.477 & 0.219 & 0.668 & 0.402 & 73.2\% \\
500 & PS & 0.273 & 0.145 & 0.468 & 0.343 & 81.2\% \\*
500 & CBPS & 0.281 & 0.076 & 0.393 & 0.309 & 83.9\% \\*
500 & BW & 0.214 & 0.079 & 0.353 & 0.346 & 91.6\% \\*
500 & Regression-only (OLS) & 0.421 & 0.101 & 0.527 & 0.289 & 65.4\% \\
\end{longtable}
}

Table 6.1 reports the bias, variance, RMSE, mean standard error, and coverage of the four methods. All methods have positive bias and thus understate the magnitude of the negative effect on average. Regression-only OLS has the largest bias at every sample size because its outcome regression is misspecified, and each weighting method reduces this bias. In this design, BW achieves the smallest bias and RMSE and the highest coverage among the four methods at every sample size. Among the weighting methods, CBPS has the second smallest RMSE and PS has the largest variance.

Relative to PS, BW has a smaller bias, variance, and RMSE at every sample size, and its RMSE is about 12\% to 32\% smaller. The coverage of its intervals is 86.3\%, 91.6\%, and 92.8\% at \(N=265\), \(500\), and \(1{,}000\), compared with 77.2\%, 81.2\%, and 81.6\% for PS, and it also exceeds that of CBPS. Although the bias reduction may not look large in absolute terms, the smaller bias helps explain why BW achieves higher coverage despite having a smaller variance.

We next compare the methods separately on the repetitions with and without exact balance. Table 6.1 does not report these subsample results. At \(N=265\), exact balance is attained in 2,744 of the 3,000 repetitions. On these repetitions, BW has a bias of 0.313 and an RMSE of 0.533, compared with 0.358 and 0.609 for PS and 0.380 and 0.555 for CBPS, and its coverage is 87.5\%, compared with 78.1\% for PS and 80.5\% for CBPS. In the remaining 256 repetitions, which are difficult for every method, BW introduces the tolerances of Section 4.2. In these repetitions, BW has a bias of 0.405, compared with 0.492 for PS and 0.528 for CBPS, and its RMSE of 0.774 is smaller than the 0.880 of PS and close to the 0.770 of CBPS. Exact balance is attained in every repetition at \(N=500\) and \(1{,}000\), so the tolerances matter only in the smallest samples.

Appendix E repeats this simulation in two other settings. With the elastic-net backward regressions of Section 7.2, the bias and RMSE of BW are comparable to those of the other methods at \(N=265\) and the smallest at \(N=500\) and \(1{,}000\), and its coverage is the highest among the three weighting methods at every sample size. When only the treatment probability is changed to strengthen or weaken overlap, BW keeps the highest coverage among the four methods at every sample size, and its RMSE is 11\% to 22\% smaller than that of PS with stronger overlap and 13\% to 39\% smaller with weaker overlap.

\FloatBarrier

\section{Empirical application: the effect of media framing on immigration attitudes}\label{sec-7}

This section applies the proposed sequential balancing weights to a media-framing experiment with two causally ordered mediators. Section 7.1 describes the data and defines the effects, Section 7.2 specifies the bases and the three weighting methods, Section 7.3 compares the balance and concentration of their weights at each stage, and Section 7.4 reports the estimates.

\subsection{Data, ordered mediators, and estimands}\label{sec-7-1}

We use the experimental data of \citet{brader2008triggers}, which are available as the \texttt{framing} data in the R package \texttt{mediation} \citep{R-mediation}. In the experiment, participants were randomly assigned to read one of four versions of a news story about immigration, which varied the tone of the story and the ethnicity of the featured immigrant, and then answered questions from which the mediators and the outcome are measured. The sample contains 265 participants, and the variables used have no missing values. The treatment \(D\) equals 1 for the story with a negative tone and a Latino immigrant and 0 for the other three versions, so the comparison pools the remaining three cells of the \(2\times2\) design. The sample thus has 68 treated and 197 control participants. The outcome \(Y\) is the participant's attitude toward increased immigration on a four-point scale, with larger values indicating a more negative attitude. The covariates \(X\) are age, education, gender, and income, denoted by \(X_1\), \(X_2\), \(X_3\), and \(X_4\) in this order.

The two mediators are perceived harm from increased immigration, \(M_1\), measured on a 2--8 scale, and anxiety, \(M_2\), a measure of negative emotions during the experiment that sums three items on feeling anxious, worried, and angry and ranges from 3 to 12. Following \citet{ImaiYamamoto-MultipleCausalMechanisms-2013} and \citet{Tingley-MediationRPackage-2014}, we assume that perceived harm may affect anxiety but not the reverse. As in Remark 2.1, we therefore enter perceived harm as the earlier mediator. These two studies also take the attitude toward increased immigration as the outcome, and we follow them in the choice of the two mediators, their order, and the outcome.

\begingroup\tolerance=9999\relax With \(J=2\), the generalized mediation functional of Section 2.1 is \(\theta_{(d_3,d_2,d_1)}=\mathbb E[Y(d_3,M_2\{d_2,M_1(d_1)\},M_1(d_1))]\), where \(d_1\), \(d_2\), and \(d_3\) are the treatment inputs to perceived harm, anxiety, and the outcome. Following \citet{Zhou-SemiparametricEstimationMediators-2022}, we decompose the ATE by switching these inputs from 0 to 1 in the order \(d_3\), \(d_2\), \(d_1\),
\BodyDisplay{38}{
\begin{aligned}
\mathrm{ATE}&=\theta_{(1,1,1)}-\theta_{(0,0,0)}\\
&=\underbrace{\theta_{(1,0,0)}-\theta_{(0,0,0)}}_{\text{(i)}}
+\underbrace{\theta_{(1,1,0)}-\theta_{(1,0,0)}}_{\text{(ii)}}
+\underbrace{\theta_{(1,1,1)}-\theta_{(1,1,0)}}_{\text{(iii)}}.
\end{aligned}
}\endgroup

Term (i) changes only the input to the outcome while both mediators keep their potential values under control, so it is the natural direct effect with respect to the two mediators. Term (ii) changes the input to anxiety and is the path-specific effect through anxiety alone. Term (iii) changes the input to perceived harm and is the path-specific effect through perceived harm, including the path from perceived harm through anxiety.

\subsection{Estimation methods}\label{sec-7-2}

We use the same bases for balancing and regression augmentation. \(B_1(X)\) contains a constant, the covariate main effects, squared and cubic terms for age, education, and income, and all two-way and three-way interactions among the four covariates. Because gender is coded 0 for female and 1 for male, its squared and cubic terms would duplicate \(X_3\) and are excluded. \(B_2(M_1,X)\) adds perceived harm and its products with the four covariates, and \(B_3(M_1,M_2,X)\) further adds anxiety, its products with the four covariates, and the product of the two mediators. Including the constant, the dimensions are \(K_1=21\), \(K_2=26\), and \(K_3=32\), so balance is diagnosed for 20, 25, and 31 terms. The regression models and predictions are common to the three weighting methods, so differences between the estimates arise only from the weights.

Regression augmentation uses the elastic net. Each regression is fitted in its training group on the nonconstant basis terms, standardized, with an unpenalized intercept, and its coefficient vector \(\beta\) minimizes half the average squared residual plus the penalty \(\rho\{(1-\alpha)\|\beta\|_2^2/2+\alpha\|\beta\|_1\}\). The mixing parameter \(\alpha\in[0,1]\) gives ridge regression at \(\alpha=0\) and the lasso at \(\alpha=1\), and \(\rho\geq0\) sets the strength of the penalty. For each nonconstant response, unpenalized linear regression and the elastic net with ten equally spaced values of \(\alpha\) from 0 to 1 are compared by out-of-fold mean squared error on a common 3-fold split of the training group. For each \(\alpha\), \(\rho\) is selected from a path of ten values, and the selected model is fitted to the whole training group. A constant response is fitted by its constant value without cross-validation. When consecutive treatment inputs coincide, the population weight ratio of the later stage equals one, so the weights of the earlier stage are kept and the regressions of the two stages are combined into one regression on the basis of the earlier stage. For example, \(\theta_{(1,0,0)}\) uses first-stage weights for the controls and third-stage weights for the treated observations, and \(\theta_{(0,0,0)}\) and \(\theta_{(1,1,1)}\) use only first-stage weights and a regression of the outcome on \(B_1\) in the corresponding treatment group.

We compare PS, CBPS, and BW and report augmented weighting estimates for all three. PS uses probability ratios estimated by stage-specific logistic regressions of \(D\) on \(B_1\), \(B_2\), and \(B_3\) and normalizes the weights at each stage to sum to \(N\) in the weighted group. CBPS estimates the same logistic models with the R package CBPS \citep{R-CBPS}, as in Section 6, using its default settings and the fitted probabilities that it returns (Appendix C), and normalizes the weights as PS does. BW constructs the weights by Algorithm 4.1 with the entropy dispersion \(\ell(w)=w\log w-w\), imposes nonnegative weights, and requires exact balance for every basis term. Among the stage problems that the reported effects use, exact balance is attained in all but the third stage of \(\theta_{(1,0,0)}\), where the 68 treated observations are weighted toward the weighted controls on the 32 terms of \(B_3\). At this stage alone, BW introduces the tolerances of Section 4.2. Algorithm C.1 selects \(\lambda_3\) from the positive values of the grid in Section 4.2 with \(R=500\) bootstrap replications, and the selected value is \(\lambda_3=0.05\). Terms (i) and (ii) use these weights, whereas the ATE and term (iii) are estimated with exact balance at every stage. For each effect, the standard error is the plug-in estimator of Theorem 5.4 applied to the contrast, and nominal 95\% Wald intervals are the estimate plus or minus \(z_{0.975}\) times its standard error.

\subsection{Balance and weight diagnostics}\label{sec-7-3}

This subsection compares the balance and the concentration of the weights in the four stage problems that the effects use. For a stage that weights the \(D=d\) group, the target absolute standardized mean difference (TASMD) of \citet{Chattopadhyay-BalancingVsPractice-2020m} for a nonconstant basis term is the absolute difference between its weighted mean in that group and its target mean, divided by its unweighted sample standard deviation in that group. The target mean is the full-sample mean at the first stage and the weighted mean in the preceding weighted group at later stages. The balancing constraints of BW restrict exactly these differences between weighted and target means, so the TASMD measures the imbalance that BW targets directly and is numerically zero for BW at every stage where exact balance is attained. We report the largest TASMD over the nonconstant terms of each stage, the maximum share, which is the largest share of one observation in the total weight of the weighted group, and the top-10\% share, which is the share of the largest 10\% of weights, rounded up to 7 of the 68 treated observations and 20 of the 197 controls. Appendix D gives the formulas.

{
\footnotesize
\setcounter{table}{0}
\setlength{\LTcapwidth}{\linewidth}
\begin{table}[!htbp]
\centering
\footnotesize
\begin{tabular}{@{}
  >{\raggedleft\arraybackslash}p{(\linewidth - 14\tabcolsep) * \real{0.0968}}
  >{\raggedright\arraybackslash}p{(\linewidth - 14\tabcolsep) * \real{0.1290}}
  >{\raggedright\arraybackslash}p{(\linewidth - 14\tabcolsep) * \real{0.2742}}
  >{\raggedleft\arraybackslash}p{(\linewidth - 14\tabcolsep) * \real{0.0968}}
  >{\raggedright\arraybackslash}p{(\linewidth - 14\tabcolsep) * \real{0.0968}}
  >{\raggedleft\arraybackslash}p{(\linewidth - 14\tabcolsep) * \real{0.0968}}
  >{\raggedleft\arraybackslash}p{(\linewidth - 14\tabcolsep) * \real{0.0968}}
  >{\raggedleft\arraybackslash}p{(\linewidth - 14\tabcolsep) * \real{0.1129}}@{}}
\toprule\noalign{}
\begin{minipage}[b]{\linewidth}\raggedleft
Stage
\end{minipage} & \begin{minipage}[b]{\linewidth}\raggedright
Weighted group
\end{minipage} & \begin{minipage}[b]{\linewidth}\raggedright
Target
\end{minipage} & \begin{minipage}[b]{\linewidth}\raggedleft
\(K_s\)
\end{minipage} & \begin{minipage}[b]{\linewidth}\raggedright
Method
\end{minipage} & \begin{minipage}[b]{\linewidth}\raggedleft
Max TASMD
\end{minipage} & \begin{minipage}[b]{\linewidth}\raggedleft
Max share
\end{minipage} & \begin{minipage}[b]{\linewidth}\raggedleft
Top-10\% share
\end{minipage} \\
\midrule\noalign{}
1 & Controls & Full sample & 21 & PS & 0.011 & 0.011 & 0.148 \\
1 & Controls & Full sample & 21 & CBPS & 0.055 & 0.018 & 0.157 \\
1 & Controls & Full sample & 21 & BW & \(\approx 0\) & 0.009 & 0.141 \\
1 & Treated & Full sample & 21 & PS & 0.134 & 0.046 & 0.230 \\
1 & Treated & Full sample & 21 & CBPS & 0.183 & 0.035 & 0.201 \\
1 & Treated & Full sample & 21 & BW & \(\approx 0\) & 0.066 & 0.311 \\
2 & Treated & Weighted controls & 26 & PS & 0.308 & 0.121 & 0.395 \\
2 & Treated & Weighted controls & 26 & CBPS & 0.137 & 0.083 & 0.275 \\
2 & Treated & Weighted controls & 26 & BW & \(\approx 0\) & 0.072 & 0.340 \\
3 & Treated & Weighted controls & 32 & PS & 0.311 & 0.122 & 0.408 \\
3 & Treated & Weighted controls & 32 & CBPS & 0.138 & 0.069 & 0.289 \\
3 & Treated & Weighted controls & 32 & BW & 0.068 & 0.045 & 0.221 \\
\bottomrule
\end{tabular}
\caption{Balance and weight concentration at each stage. Each group of three rows is one stage problem, and \(K_s\) counts its basis terms including the constant. The first-stage problem for the controls enters \(\theta_{(0,0,0)}\), \(\theta_{(1,0,0)}\), and \(\theta_{(1,1,0)}\), and the one for the treated observations enters \(\theta_{(1,1,1)}\). The second-stage problem enters \(\theta_{(1,1,0)}\). The third-stage problem enters \(\theta_{(1,0,0)}\) and takes the first-stage weighted controls as its target. Max TASMD is the largest TASMD over the nonconstant terms, as defined in Section 7.3. Max share and Top-10\% share are the fractions of the total weight of the weighted group carried by its largest weight and by its largest 10\% of weights. Values are rounded to three decimal places, and \(\approx 0\) marks a BW maximum TASMD that is numerically zero. Exact balance is not attained at the third stage for the treated observations, where BW introduces the tolerances of Section 4.2 with \(\lambda_3=0.05\). Because the tolerance is set in units of the full-sample standard deviation (Appendix C) and the TASMD divides by the standard deviation in the weighted group, the BW maximum TASMD at this stage can exceed \(\lambda_3\).}
\end{table}
\FloatBarrier
}

\par\medskip

Table 7.1 shows that BW attains exact balance in three of the four stage problems, whereas PS leaves a maximum TASMD between 0.134 and 0.311 in every problem except the first stage for the controls. CBPS lowers the maximum TASMD of PS at the second and third stages to 0.137 and 0.138 but does not remove the imbalance, and at the first stage it leaves maximum TASMDs of 0.055 for the controls and 0.183 for the treated observations, which exceed those of PS. The largest difference in balance arises at the second stage, where the 68 treated observations are weighted toward the weighted controls on the functions of perceived harm and the covariates. PS leaves a maximum TASMD of 0.308 there and places 0.121 of the total weight on one observation, whereas BW removes the imbalance with a maximum share of 0.072. At the third stage for the treated observations, exact balance cannot be attained, and the selected tolerance lowers the maximum TASMD from 0.311 under PS to 0.068 and the top-10\% share from 0.408 to 0.221. For the treated observations weighted to the full sample at the first stage, exact balance requires more concentrated weights than PS, with a top-10\% share of 0.311 compared with 0.230.

\subsection{Results}\label{sec-7-4}

Table 7.2 reports the augmented weighting estimates of the ATE and of the terms (i)--(iii) of the decomposition in Section 7.1.

{
\footnotesize
\setcounter{table}{1}
\setlength{\LTcapwidth}{\linewidth}
\begin{longtable}[]{@{}llrrr@{}}
\toprule\noalign{}
Effect & Contrast & PS & CBPS & BW \\
\midrule\noalign{}
\endfirsthead
\toprule\noalign{}
Effect & Contrast & PS & CBPS & BW \\
\midrule\noalign{}
\endhead
(iii) Through perceived harm & \(\theta_{(1,1,1)}-\theta_{(1,1,0)}\) & 0.148 (0.073) & 0.080 (0.055) & 0.029 (0.063) \\*
& & {[}0.006, 0.291{]} & {[}-0.027, 0.188{]} & {[}-0.095, 0.153{]} \\
\bottomrule\noalign{}
\noalign{\vskip\abovecaptionskip}
\caption{Estimates of the ATE and its decomposition. For each effect, the PS, CBPS, and BW columns report the point estimate with its standard error in parentheses in the first row and its nominal 95\% confidence interval in the second row. Values are rounded to three decimal places. For each effect, the standard error is obtained by dividing the centered empirical second moment of the observation-level score difference between its two component functionals by \(N=265\) and taking the square root, so the covariance between the two functionals is included. The confidence intervals add and subtract the product of the standard normal 97.5\% quantile and the standard error from the point estimate. Effects (i) and (ii) use the BW weights with tolerances at the third stage of \(\theta_{(1,0,0)}\), and the other BW estimates use exact balance at every stage.}\tabularnewline
\endlastfoot
ATE & \(\theta_{(1,1,1)}-\theta_{(0,0,0)}\) & 0.414 (0.122) & 0.405 (0.123) & 0.423 (0.116) \\*
& & {[}0.174, 0.654{]} & {[}0.164, 0.646{]} & {[}0.195, 0.651{]} \\
(i) Natural direct effect & \(\theta_{(1,0,0)}-\theta_{(0,0,0)}\) & 0.296 (0.108) & 0.349 (0.100) & 0.266 (0.114) \\*
& & {[}0.084, 0.507{]} & {[}0.154, 0.544{]} & {[}0.042, 0.489{]} \\
(ii) Through anxiety alone & \(\theta_{(1,1,0)}-\theta_{(1,0,0)}\) & -0.030 (0.043) & -0.024 (0.035) & 0.128 (0.071) \\*
& & {[}-0.115, 0.055{]} & {[}-0.092, 0.044{]} & {[}-0.012, 0.268{]} \\
\end{longtable}
}

\par\medskip

Table 7.2 shows that the three methods give similar estimates of the ATE, between 0.405 and 0.423, and natural direct effects between 0.266 and 0.349, and all six of these intervals exclude zero. The ATE estimates are also close to the total effects of 0.420 in \citet[Section 6.3]{ImaiYamamoto-MultipleCausalMechanisms-2013} and 0.418 in \citet[Section 6.2]{Tingley-MediationRPackage-2014}, which use the same treatment and outcome. The methods split the remaining effect differently between the two paths. PS attributes it to perceived harm, with an estimate of 0.148 whose interval excludes zero, and gives \(-0.030\) for anxiety alone. CBPS gives the same signs with smaller magnitudes, 0.080 for perceived harm and \(-0.024\) for anxiety alone, and neither of its intervals excludes zero. BW gives 0.128 for anxiety alone and 0.029 for perceived harm, and neither of these BW intervals excludes zero either. Both terms involve \(\theta_{(1,1,0)}\), whose second stage is the problem in which PS and CBPS leave maximum TASMDs of 0.308 and 0.137 and BW attains exact balance (Table 7.1). Recall that larger values of \(Y\) indicate a more negative attitude toward increased immigration, so a positive path-specific effect means that the treatment makes this attitude more negative through that path. Both perceived harm and anxiety are expected to make attitudes toward immigration more negative, and PS and CBPS leave imbalance at the stage that the two terms share. We therefore regard the BW estimates, which are positive for both paths, as the more plausible split of the indirect effect.

\par\medskip

\FloatBarrier

\section{Conclusion}\label{sec-8}

This paper studied the estimation of natural direct and indirect effects under the finite-sample instability and covariate imbalance that arise when the weights of the EIF-based estimator are obtained by inverting estimated propensity scores. We proposed sequential balancing weights for this problem when the mediators follow a prespecified causal order. Under the stated conditions, the resulting augmented weighting estimator is asymptotically normal and semiparametrically efficient, and its asymptotic variance can be estimated consistently. In the simulation design of Section 6, BW attained the smallest bias and RMSE at every sample size, and its confidence intervals had higher coverage than those of PS and CBPS. In the empirical application with two ordered mediators, BW attained exact balance in three of the four stage problems and reduced the imbalance and weight concentration of PS in the remaining one, and it split the indirect effect between the two mediators differently from PS and CBPS. Open questions remain about the choice of dispersion loss, the rule for introducing tolerances, and the tolerance grid, and the greedy stagewise selection of tolerances need not minimize a single global criterion.

As a direction for future research, we note that causal mediation analysis is closely connected to the study of long-term treatment effects and statistical surrogacy, since mediators are often interpreted as short-term outcomes or surrogate endpoints \citep{Chen-SemiparametricEstimationEffects-2023s,Imbens-Long-termCausalCombination-2024k,athey2024estimatingtreatmenteffectsusing}. We expect that estimators analogous to those proposed in this paper can also be developed in this setting.

\section*{Declaration of generative AI and AI-assisted technologies in the manuscript preparation process}

During the preparation of this work, the author used Claude Code (Anthropic) and Codex (OpenAI) in order to draft parts of the text and edit its language and organization, check derivations and citations, and write and check the R code for the simulation and the empirical application. After using these tools, the author reviewed and edited the content as needed and takes full responsibility for the content of the published article.

\clearpage
\bibliographystyle{elsarticle-harv}
\bibliography{references}

\clearpage
\appendix
\AppendixNumbering

\section{Proofs of the main results}\label{app-A}

\subsection{Proof of Theorem 2.1}\label{app-A-1}

For the identification formula in part (i), see \citet[Theorem~1]{Zhou-SemiparametricEstimationMediators-2022}. The weighting representation in part (ii) follows by Bayes' rule. For the efficient influence function in part (iii), see \citet[Theorem~2]{Zhou-SemiparametricEstimationMediators-2022}.

\subsection{Proof of Proposition 3.1}\label{app-A-2}

The proof expands the augmented weighting estimator around the true backward regressions. Because each weight depends only on the information available at its stage, the weighted true residuals have mean zero, and the bias consists only of the weight imbalances multiplied by the regression errors.

The weights may be estimated by any method satisfying \AppendixDisplay{1}{\widehat w_{s,i}
\text{ is }
\sigma\!\left((D_\ell,X_\ell,\overline M_{s-1,\ell})_{\ell=1}^{N}\right)
\text{-measurable},
\qquad s=1,\ldots,J+1
} for every \(i=1,\ldots,N\). The estimator is \AppendixDisplay{2}{\begin{aligned}
\widehat\theta_{\boldsymbol d}^{\mathrm{AW}}
&=\frac1N\sum_{i=1}^{N}
\widehat\mu_0^{(d_1,\ldots,d_{J+1})}(X_i)\\
&\quad+\frac1N\sum_{i=1}^{N}\sum_{j=1}^{J}
\mathbf1\{D_i=d_j\}\widehat w_{j,i}
\left\{
\widehat\mu_j^{(d_{j+1},\ldots,d_{J+1})}(\overline M_{j,i},X_i)
-\widehat\mu_{j-1}^{(d_j,\ldots,d_{J+1})}(\overline M_{j-1,i},X_i)
\right\}\\
&\quad+\frac1N\sum_{i=1}^{N}
\mathbf1\{D_i=d_{J+1}\}\widehat w_{J+1,i}
\{Y_i-\widehat\mu_J^{(d_{J+1})}(\overline M_{J,i},X_i)\}.
\end{aligned}
}

In the following sample sums, an omitted regression argument means that the regression is evaluated at \((\overline M_{j,i},X_i)\) for observation \(i\); treatment-input superscripts remain explicit. Split the terminal residual as \(Y_i-\widehat\mu_J=(Y_i-\mu_J)+(\mu_J-\widehat\mu_J)\). For each adjacent fitted-regression difference, \AppendixDisplay{3}{\begin{aligned}
\widehat\mu_j^{(d_{j+1},\ldots,d_{J+1})}
-\widehat\mu_{j-1}^{(d_j,\ldots,d_{J+1})}
&=\mu_j^{(d_{j+1},\ldots,d_{J+1})}
-\mu_{j-1}^{(d_j,\ldots,d_{J+1})}\\
&\qquad-\{\mu_j^{(d_{j+1},\ldots,d_{J+1})}
-\widehat\mu_j^{(d_{j+1},\ldots,d_{J+1})}\}
+\{\mu_{j-1}^{(d_j,\ldots,d_{J+1})}
-\widehat\mu_{j-1}^{(d_j,\ldots,d_{J+1})}\}
\end{aligned}
} and at the initial stage use \(\widehat\mu_0=\mu_0-(\mu_0-\widehat\mu_0)\). Collecting the coefficients of each \(\mu_j-\widehat\mu_j\) yields the sample identity \AppendixDisplay{4}{\begin{aligned}
\widehat\theta_{\boldsymbol d}^{\mathrm{AW}}-\theta_{\boldsymbol d}
&=\underbrace{\frac1N\sum_{i=1}^{N}
\mu_0^{(d_1,\ldots,d_{J+1})}(X_i)-\theta_{\boldsymbol d}}_{\text{(i)}}\\
&+\underbrace{\frac1N\sum_{i=1}^{N}\sum_{j=1}^{J}
\mathbf1\{D_i=d_j\}\widehat w_{j,i}
\{\mu_j^{(d_{j+1},\ldots,d_{J+1})}
-\mu_{j-1}^{(d_j,\ldots,d_{J+1})}\}}_{\text{(ii)}}\\
&+\underbrace{\frac1N\sum_{i=1}^{N}
\mathbf1\{D_i=d_{J+1}\}\widehat w_{J+1,i}
\{Y_i-\mu_J^{(d_{J+1})}\}}_{\text{(iii)}}\\
&+\underbrace{\frac1N\sum_{i=1}^{N}
\{\mathbf1\{D_i=d_1\}\widehat w_{1,i}-1\}
\{\mu_0^{(d_1,\ldots,d_{J+1})}
-\widehat\mu_0^{(d_1,\ldots,d_{J+1})}\}}_{\text{(iv)}}\\
&+\underbrace{\frac1N\sum_{i=1}^{N}\sum_{j=1}^{J}
\{\mathbf1\{D_i=d_{j+1}\}\widehat w_{j+1,i}
-\mathbf1\{D_i=d_j\}\widehat w_{j,i}\}
\{\mu_j^{(d_{j+1},\ldots,d_{J+1})}
-\widehat\mu_j^{(d_{j+1},\ldots,d_{J+1})}\}}_{\text{(v)}}.
\end{aligned}
} Here (i) is the baseline term, (ii) collects the mediator residuals, (iii) is the outcome residual, (iv) is the first-stage imbalance, and (v) is the later-stage imbalance.

Fix \(j\in\{1,\ldots,J\}\) and condition on \((D_\ell,X_\ell,\overline M_{j-1,\ell})_{\ell=1}^{N}\). The weight \(\widehat w_{j,i}\) is measurable with respect to this information. Since observations are independent, adding the histories of the other observations does not change the conditional mean of observation \(i\)'s true residual. Therefore, \AppendixDisplay{5}{\begin{aligned}
&\mathbb E\!\left[
\mathbf1\{D_i=d_j\}\widehat w_{j,i}
\{\mu_j^{(d_{j+1},\ldots,d_{J+1})}
-\mu_{j-1}^{(d_j,\ldots,d_{J+1})}\}
\mid(D_\ell,X_\ell,\overline M_{j-1,\ell})_{\ell=1}^{N}
\right]\\
&\quad=\widehat w_{j,i}
\mathbb E\!\left[
\mathbf1\{D_i=d_j\}
\{\mu_j^{(d_{j+1},\ldots,d_{J+1})}
-\mu_{j-1}^{(d_j,\ldots,d_{J+1})}\}
\mid D_i,X_i,\overline M_{j-1,i}
\right]\\
&\quad=0.
\end{aligned}
} For \(D_i\neq d_j\), the indicator makes the inner expectation zero. For \(D_i=d_j\), the backward-regression definition gives \AppendixDisplay{6}{\begin{aligned}
&\mathbb E[
\mu_j^{(d_{j+1},\ldots,d_{J+1})}(\overline M_{j,i},X_i)
-\mu_{j-1}^{(d_j,\ldots,d_{J+1})}(\overline M_{j-1,i},X_i)
\mid D_i=d_j,X_i,\overline M_{j-1,i}]\\
&\quad=\mu_{j-1}^{(d_j,\ldots,d_{J+1})}(\overline M_{j-1,i},X_i)
-\mu_{j-1}^{(d_j,\ldots,d_{J+1})}(\overline M_{j-1,i},X_i)=0.
\end{aligned}
} Thus the weighted mediator residual has unconditional mean zero for every \(i\) and \(j\). At the terminal stage, condition on \((D_\ell,X_\ell,\overline M_{J,\ell})_{\ell=1}^{N}\), which leaves the outcomes unfixed. The last-stage weight is measurable with respect to this information, and the definition of the true outcome regression gives \AppendixDisplay{7}{\begin{aligned}
&\mathbb E\!\left[
\mathbf1\{D_i=d_{J+1}\}\widehat w_{J+1,i}
\{Y_i-\mu_J^{(d_{J+1})}(\overline M_{J,i},X_i)\}
\mid(D_\ell,X_\ell,\overline M_{J,\ell})_{\ell=1}^{N}
\right]\\
&\quad=\widehat w_{J+1,i}
\mathbb E\!\left[
\mathbf1\{D_i=d_{J+1}\}
\{Y_i-\mu_J^{(d_{J+1})}(\overline M_{J,i},X_i)\}
\mid D_i,X_i,\overline M_{J,i}
\right]\\
&\quad=0.
\end{aligned}
} so the weighted outcome residual also has unconditional mean zero. Finally, Theorem 2.1's identification formula and identical distribution imply \AppendixDisplay{8}{\mathbb E\!\left[
\frac1N\sum_{i=1}^{N}\mu_0^{(d_1,\ldots,d_{J+1})}(X_i)
-\theta_{\boldsymbol d}
\right]=0.
}

Taking expectations in the sample identity and using these three zero-mean relations leaves \AppendixDisplay{9}{\begin{aligned}
\mathbb E[\widehat\theta_{\boldsymbol d}^{\mathrm{AW}}-\theta_{\boldsymbol d}]
&=\mathbb E\!\Bigg[
\frac1N\sum_{i=1}^{N}
\{\mathbf1\{D_i=d_1\}\widehat w_{1,i}-1\}
\{\mu_0^{(d_1,\ldots,d_{J+1})}(X_i)
-\widehat\mu_0^{(d_1,\ldots,d_{J+1})}(X_i)\}
\Bigg]\\
&\quad+\sum_{j=1}^{J}\mathbb E\!\Bigg[
\frac1N\sum_{i=1}^{N}
\{\mathbf1\{D_i=d_{j+1}\}\widehat w_{j+1,i}
-\mathbf1\{D_i=d_j\}\widehat w_{j,i}\}\\
&\qquad\hspace{7em}\times
\{\mu_j^{(d_{j+1},\ldots,d_{J+1})}(\overline M_{j,i},X_i)
-\widehat\mu_j^{(d_{j+1},\ldots,d_{J+1})}(\overline M_{j,i},X_i)\}
\Bigg]
\end{aligned}
} which is the bias identity in Proposition 3.1. \AppendixQED

\subsection{Proof of Proposition 3.2}\label{app-A-3}

We first derive the balancing identities from the population weight recursion and then prove that the identities for every bounded measurable function characterize the weights. Here \(\pi_0(d;X)=P(D=d\mid X)\) and \(\pi_{s-1}(d;\overline M_{s-1},X)=P(D=d\mid\overline M_{s-1},X)\) for \(s=2,\ldots,J+1\). Assumption 2.3 gives \(c\leq\pi_j(d;\overline M_j,X)\leq1-c\) for \(j=0,\ldots,J\) and \(d\in\{0,1\}\), almost surely, for some \(c>0\). The population weights obey the recursion \AppendixDisplay{10}{w_1^*(x)=\frac1{\pi_0(d_1;x)},
\qquad
w_s^*(\overline m_{s-1},x)
=w_{s-1}^*(\overline m_{s-2},x)
\frac{\pi_{s-1}(d_{s-1};\overline m_{s-1},x)}
{\pi_{s-1}(d_s;\overline m_{s-1},x)},
\quad s\geq2
} For fixed \(J\), positivity also implies that each population weight is bounded.

For any bounded measurable function \(h(X)\), conditioning on \(X\) gives \AppendixDisplay{11}{\begin{aligned}
\mathbb E[\mathbf1\{D=d_1\}w_1^*(X)h(X)]
&=\mathbb E\!\left[
\mathbb E[\mathbf1\{D=d_1\}\mid X]w_1^*(X)h(X)
\right]\\
&=\mathbb E[\pi_0(d_1;X)w_1^*(X)h(X)]\\
&=\mathbb E[h(X)].
\end{aligned}
} The last equality follows from \(\pi_0(d_1;X)w_1^*(X)=1\).

Fix \(s\in\{2,\ldots,J+1\}\). The weight recursion gives \AppendixDisplay{12}{\pi_{s-1}(d_s;\overline M_{s-1},X)w_s^*(\overline M_{s-1},X)
=\pi_{s-1}(d_{s-1};\overline M_{s-1},X)w_{s-1}^*(\overline M_{s-2},X).
} For any bounded measurable function \(h(\overline M_{s-1},X)\), \AppendixDisplay{13}{\begin{aligned}
&\mathbb E[\mathbf1\{D=d_s\}w_s^*(\overline M_{s-1},X)h(\overline M_{s-1},X)]\\
&\quad=\mathbb E[
\pi_{s-1}(d_s;\overline M_{s-1},X)w_s^*(\overline M_{s-1},X)h(\overline M_{s-1},X)]\\
&\quad=\mathbb E[
\pi_{s-1}(d_{s-1};\overline M_{s-1},X)w_{s-1}^*(\overline M_{s-2},X)h(\overline M_{s-1},X)]\\
&\quad=\mathbb E[
\mathbf1\{D=d_{s-1}\}w_{s-1}^*(\overline M_{s-2},X)h(\overline M_{s-1},X)].
\end{aligned}
} The first and last equalities follow by conditioning on \((\overline M_{s-1},X)\). The preceding weight is a function of the shorter history \((\overline M_{s-2},X)\) and is therefore measurable with respect to \((\overline M_{s-1},X)\). The second equality uses the weight recursion. This proves the forward implication in Proposition 3.2.

Conversely, suppose that measurable functions \(w_s(\overline m_{s-1},x)\) satisfy the balancing identities in Proposition 3.2 for every bounded measurable function. Conditioning the first-stage identity on \(X\), we obtain, for every bounded measurable function \(h(X)\), \AppendixDisplay{14}{\mathbb E[\{\pi_0(d_1;X)w_1(X)-1\}h(X)]=0.
} Choose \(h(X)=\operatorname{sgn}\{\pi_0(d_1;X)w_1(X)-1\}\). This \(h\) is a bounded measurable function of \(X\), so the preceding identity yields \AppendixDisplay{15}{\mathbb E[|\pi_0(d_1;X)w_1(X)-1|]=0.
} The integrand is nonnegative, hence \(\pi_0(d_1;X)w_1(X)=1\) almost surely. Since positivity gives \(\pi_0(d_1;X)>0\), \AppendixDisplay{16}{w_1(X)=\frac1{\pi_0(d_1;X)}=w_1^*(X)
\qquad\text{almost surely}.
}

{\AppendixRaggedParagraph For \(s\geq2\), condition the balancing identity on \((\overline M_{s-1},X)\). For every bounded measurable function \(h\), \AppendixDisplay{17}{\begin{aligned}
\mathbb E\!\Big[
\big\{&\pi_{s-1}(d_s;\overline M_{s-1},X)w_s(\overline M_{s-1},X)\\
&-\pi_{s-1}(d_{s-1};\overline M_{s-1},X)w_{s-1}(\overline M_{s-2},X)\big\}
 h(\overline M_{s-1},X)
\Big]=0.
\end{aligned}
} Choose \(h(\overline M_{s-1},X)=\operatorname{sgn}\{\pi_{s-1}(d_s;\overline M_{s-1},X)w_s(\overline M_{s-1},X)-\pi_{s-1}(d_{s-1};\overline M_{s-1},X)w_{s-1}(\overline M_{s-2},X)\}\). This function is measurable with respect to \((\overline M_{s-1},X)\) and bounded by one. Substitution into the preceding identity gives \AppendixDisplay{18}{\begin{aligned}
\mathbb E\Big|\pi_{s-1}(d_s;\overline M_{s-1},X)w_s(\overline M_{s-1},X)
-\pi_{s-1}(d_{s-1};\overline M_{s-1},X)w_{s-1}(\overline M_{s-2},X)\Big|=0.
\end{aligned}
} Therefore, \AppendixDisplay{19}{\pi_{s-1}(d_s;\overline M_{s-1},X)w_s(\overline M_{s-1},X)
=\pi_{s-1}(d_{s-1};\overline M_{s-1},X)w_{s-1}(\overline M_{s-2},X)
} almost surely. Positivity allows us to divide by the treatment probability on the left, giving \AppendixDisplay{20}{w_s(\overline M_{s-1},X)
=\frac{\pi_{s-1}(d_{s-1};\overline M_{s-1},X)}
{\pi_{s-1}(d_s;\overline M_{s-1},X)}
w_{s-1}(\overline M_{s-2},X).
}\par}

The first-stage result gives \(w_1=w_1^*\). At stage \(s-1\), suppose \(w_{s-1}=w_{s-1}^*\). Substitution into the recursion then gives \(w_s=w_s^*\). The equality at the preceding stage holds almost surely under the marginal distribution of the shorter history and therefore also under the joint distribution of the longer history. By induction, for every \(s=1,\ldots,J+1\), \AppendixDisplay{21}{w_s(\overline M_{s-1},X)
=\frac1{\pi_0(d_1;X)}
\prod_{r=1}^{s-1}
\frac{\pi_r(d_r;\overline M_r,X)}
{\pi_r(d_{r+1};\overline M_r,X)}
=w_s^*(\overline M_{s-1},X)
\qquad\text{almost surely}
} This proves the converse in Proposition 3.2. \AppendixQED

\subsection{Proof of Theorem 4.1}\label{app-A-4}

We first identify the sample correction made by augmentation. We then apply Proposition 3.1 to the weighting and augmented weighting estimators to obtain their two bias expressions. Within each sample sum, \(B_1=B_1(X_i)\) and \(B_s=B_s(\overline M_{s-1,i},X_i)\) for \(s=2,\ldots,J+1\), with \(\beta_s^*,\widehat\beta_s\in\mathbb R^{K_s}\).

The weighting estimator is \(\widehat\theta_{\boldsymbol d}^{\mathrm{W\text{-}BW}}=N^{-1}\sum_{i=1}^{N}\mathbf1\{D_i=d_{J+1}\}\widehat w_{J+1,i}Y_i\). Subtracting it from the augmented estimator of Section 3.1 leaves \AppendixDisplay{22}{\begin{aligned}
\widehat\theta_{\boldsymbol d}^{\mathrm{AW\text{-}BW}}-\widehat\theta_{\boldsymbol d}^{\mathrm{W\text{-}BW}}
&=\frac1N\sum_{i=1}^{N}\widehat\mu_0^{(d_1,\ldots,d_{J+1})}(X_i)\\
&\quad+\frac1N\sum_{i=1}^{N}\sum_{j=1}^{J}\mathbf1\{D_i=d_j\}\widehat w_{j,i}
\{\widehat\mu_j^{(d_{j+1},\ldots,d_{J+1})}-\widehat\mu_{j-1}^{(d_j,\ldots,d_{J+1})}\}\\
&\quad-\frac1N\sum_{i=1}^{N}\mathbf1\{D_i=d_{J+1}\}\widehat w_{J+1,i}\widehat\mu_J^{(d_{J+1})}.
\end{aligned}
} Here each omitted regression argument is its observed history and \(X_i\). Assumption 4.1(b) gives \(\widehat\mu_{s-1}^{(d_s,\ldots,d_{J+1})}=B_s^\top\widehat\beta_s\). Collecting coefficients in the preceding display gives \AppendixDisplay{23}{\begin{aligned}
\widehat\theta_{\boldsymbol d}^{\mathrm{AW\text{-}BW}}
&=\widehat\theta_{\boldsymbol d}^{\mathrm{W\text{-}BW}}
-\widehat\beta_1^\top
\frac1N\sum_{i=1}^{N}
\{\mathbf1\{D_i=d_1\}\widehat w_{1,i}-1\}B_1\\
&\quad-\sum_{s=2}^{J+1}
\widehat\beta_s^\top
\frac1N\sum_{i=1}^{N}
\{\mathbf1\{D_i=d_s\}\widehat w_{s,i}
-\mathbf1\{D_i=d_{s-1}\}\widehat w_{s-1,i}\}B_s.
\end{aligned}
} Exact balance means that the first-stage sample imbalance is zero and that each subsequent imbalance is zero for \(s=2,\ldots,J+1\). Hence \(\widehat\theta_{\boldsymbol d}^{\mathrm{AW\text{-}BW}}=\widehat\theta_{\boldsymbol d}^{\mathrm{W\text{-}BW}}\) for that sample. For each \(s\) and \(i\), the weight \(\widehat w_{s,i}\) is \(\sigma((D_\ell,X_\ell,\overline M_{s-1,\ell})_{\ell=1}^{N})\)-measurable, as required by Proposition 3.1.

To obtain the weighting bias, apply this identity with every fitted regression set identically to zero. The resulting augmented estimator equals \(\widehat\theta_{\boldsymbol d}^{\mathrm{W\text{-}BW}}\). Under Assumption 4.1(a), the true regressions satisfy \(\mu_{s-1}^{(d_s,\ldots,d_{J+1})}=B_s^\top\beta_s^*\), and the two remaining regression-error terms in Proposition 3.1 become \AppendixDisplay{24}{\begin{aligned}
\mathbb E\!\left[
\widehat\theta_{\boldsymbol d}^{\mathrm{W\text{-}BW}}-\theta_{\boldsymbol d}
\right]
&=\mathbb E\!\left[
(\beta_1^*)^\top
\frac1N\sum_{i=1}^{N}
\{\mathbf1\{D_i=d_1\}\widehat w_{1,i}-1\}B_1
\right]\\
&\quad+\sum_{s=2}^{J+1}
\mathbb E\!\left[
(\beta_s^*)^\top
\frac1N\sum_{i=1}^{N}
\{\mathbf1\{D_i=d_s\}\widehat w_{s,i}
-\mathbf1\{D_i=d_{s-1}\}\widehat w_{s-1,i}\}B_s
\right].
\end{aligned}
} For the actual fitted regressions, Assumption 4.1(b) gives \(\widehat\mu_{s-1}^{(d_s,\ldots,d_{J+1})}=B_s^\top\widehat\beta_s\). Applying Proposition 3.1 again, now with the regression errors \(\mu_{s-1}-\widehat\mu_{s-1}=B_s^\top(\beta_s^*-\widehat\beta_s)\), gives \AppendixDisplay{25}{\begin{aligned}
\mathbb E\!\left[
\widehat\theta_{\boldsymbol d}^{\mathrm{AW\text{-}BW}}-\theta_{\boldsymbol d}
\right]
&=\mathbb E\!\left[
(\beta_1^*-\widehat\beta_1)^\top
\frac1N\sum_{i=1}^{N}
\{\mathbf1\{D_i=d_1\}\widehat w_{1,i}-1\}B_1
\right]\\
&\quad+\sum_{s=2}^{J+1}
\mathbb E\!\left[
(\beta_s^*-\widehat\beta_s)^\top
\frac1N\sum_{i=1}^{N}
\{\mathbf1\{D_i=d_s\}\widehat w_{s,i}
-\mathbf1\{D_i=d_{s-1}\}\widehat w_{s-1,i}\}B_s
\right].
\end{aligned}
} The fitted coefficients remain inside the expectations. \AppendixQED

\subsection{Proof of Theorem 5.1}\label{app-A-5}

Assume, as in the theorem, that every primal problem is feasible and each dual minimization problem attains a finite solution \(\widehat\gamma_s\). The theorem also assumes that the extension of \(\ell\) by \(+\infty\) to negative arguments is proper, closed, and convex, and that \(\ell^{\mathrm c}\) is finite and differentiable at \(B_s(\overline M_{s-1,i},X_i)^\top\widehat\gamma_s\) for each \(D_i=d_s\), with \(B_1(X_i)\) used at \(s=1\). The proof has three steps. First, convex conjugacy and the balance constraints give the dual lower bound at the first stage and at every later stage, with the preceding weights held fixed. Second, the optimality conditions at a finite dual solution show that the weights obtained from \((\ell^{\mathrm c})'\) are nonnegative and satisfy every balance constraint. Finally, those conditions make the lower bound an equality, so the recovered weights solve the primal problem; strict convexity makes the target-group weight vector unique.

At the first stage, Algorithm 4.1 minimizes \AppendixDisplay{26}{\begin{aligned}
\min_{\{w_{1,i}\geq0:D_i=d_1\}}
&\quad \frac1N\sum_{i:D_i=d_1}\ell(w_{1,i})\\
\text{subject to}
&\quad
\left|
\frac1N\sum_{i:D_i=d_1}w_{1,i}b_{1,k}(X_i)
-\frac1N\sum_{i=1}^{N}b_{1,k}(X_i)
\right|
\leq\varepsilon_{1,k},\\
&\quad k=1,\ldots,K_1.
\end{aligned}
} At stage \(s=2,\ldots,J+1\), the weights \(\{\widehat w_{s-1,i}:D_i=d_{s-1}\}\) are fixed, and the primal problem minimizes \AppendixDisplay{27}{\begin{aligned}
\min_{\{w_{s,i}\geq0:D_i=d_s\}}
&\quad \frac1N\sum_{i:D_i=d_s}\ell(w_{s,i})\\
\text{subject to}
&\quad
\left|
\frac1N\sum_{i:D_i=d_s}w_{s,i}b_{s,k}(\overline M_{s-1,i},X_i)
-\frac1N\sum_{i:D_i=d_{s-1}}\widehat w_{s-1,i}b_{s,k}(\overline M_{s-1,i},X_i)
\right|
\leq\varepsilon_{s,k},\\
&\quad k=1,\ldots,K_s.
\end{aligned}
} Here \(B_1(X_i)=(b_{1,1}(X_i),\ldots,b_{1,K_1}(X_i))^\top\) and \(B_s(\overline M_{s-1,i},X_i)=(b_{s,1}(\overline M_{s-1,i},X_i),\allowbreak\ldots,\allowbreak b_{s,K_s}(\overline M_{s-1,i},X_i))^\top\). The constant components satisfy \(b_{s,1}=1\) and \(\varepsilon_{s,1}=0\). Consequently, the first-stage constraint fixes \(\sum_{i:D_i=d_1}w_{1,i}=N\), and each later constant constraint carries this normalization from the preceding stage. Only weights in the target treatment group are primal variables.

\subsubsection{First-stage duality by convex conjugacy}\label{app-A-5-1}

On the nonnegative domain, \(\ell^{\mathrm c}(t)=\sup_{w\geq0}\{tw-\ell(w)\}\). For \(w\geq0\) and any \(t\) at which the conjugate is finite, its definition gives \(tw-\ell(w)\leq\ell^{\mathrm c}(t)\), or equivalently \(\ell(w)+\ell^{\mathrm c}(t)\geq tw\). Because the extended loss is closed and convex, equality holds at \(w=(\ell^{\mathrm c})'(t)\) whenever the conjugate is finite and differentiable at \(t\). Fix any feasible first-stage weights and any \(\gamma_1\in\mathbb R^{K_1}\) for which the conjugate terms are finite. Applying \(\ell(w_{1,i})\geq w_{1,i}B_1(X_i)^\top\gamma_1-\ell^{\mathrm c}(B_1(X_i)^\top\gamma_1)\) and summing over the target group gives \AppendixDisplay{28}{\begin{aligned}
\frac1N\sum_{i:D_i=d_1}\ell(w_{1,i})
&\geq
\frac1N\sum_{i:D_i=d_1}w_{1,i}B_1(X_i)^\top\gamma_1
-\frac1N\sum_{i:D_i=d_1}\ell^{\mathrm c}(B_1(X_i)^\top\gamma_1)\\
&=
\gamma_1^\top\left\{\frac1N\sum_{i:D_i=d_1}w_{1,i}B_1(X_i)\right\}
-\frac1N\sum_{i:D_i=d_1}\ell^{\mathrm c}(B_1(X_i)^\top\gamma_1).
\end{aligned}
} For each component, feasibility gives \(-\varepsilon_{1,k}\leq N^{-1}\sum_{i:D_i=d_1}w_{1,i}b_{1,k}(X_i)-N^{-1}\sum_{i=1}^{N}b_{1,k}(X_i)\leq\varepsilon_{1,k}\) and hence \AppendixDisplay{29}{\gamma_{1,k}
\left\{
\frac1N\sum_{i:D_i=d_1}w_{1,i}b_{1,k}(X_i)
-\frac1N\sum_{i=1}^{N}b_{1,k}(X_i)
\right\}
\geq-\varepsilon_{1,k}|\gamma_{1,k}|.
} The \(k=1\) term is zero: \(N^{-1}\sum_{i:D_i=d_1}w_{1,i}b_{1,1}(X_i)-N^{-1}\sum_{i=1}^{N}b_{1,1}(X_i)=N^{-1}\sum_{i:D_i=d_1}w_{1,i}-1=0\). Summing the component inequalities therefore yields \AppendixDisplay{30}{\gamma_1^\top\left\{\frac1N\sum_{i:D_i=d_1}w_{1,i}B_1(X_i)\right\}
\geq
\gamma_1^\top\left\{\frac1N\sum_{i=1}^{N}B_1(X_i)\right\}
-\sum_{k=2}^{K_1}\varepsilon_{1,k}|\gamma_{1,k}|.
} Substitution gives a lower bound on the objective of every feasible first-stage weight vector: \AppendixDisplay{31}{\frac1N\sum_{i:D_i=d_1}\ell(w_{1,i})
\geq
\gamma_1^\top\left\{\frac1N\sum_{i=1}^{N}B_1(X_i)\right\}
-\frac1N\sum_{i:D_i=d_1}\ell^{\mathrm c}(B_1(X_i)^\top\gamma_1)
-\sum_{k=2}^{K_1}\varepsilon_{1,k}|\gamma_{1,k}|.
} To see that maximizing this bound is the Lagrange dual, write each absolute-value balance constraint as two inequalities with nonnegative multipliers \(\lambda_{1,k}^+\) and \(\lambda_{1,k}^-\). With \(\gamma_{1,k}=\lambda_{1,k}^--\lambda_{1,k}^+\), minimization of the Lagrangian over each \(w_{1,i}\geq0\) gives the conjugate term. For fixed \(\gamma_{1,k}\), the smallest possible multiplier sum is \(\inf_{\lambda_{1,k}^+,\lambda_{1,k}^-\geq0,\,\lambda_{1,k}^--\lambda_{1,k}^+=\gamma_{1,k}}(\lambda_{1,k}^++\lambda_{1,k}^-)=|\gamma_{1,k}|\). Thus the dual maximizes the right-hand side of the displayed bound: \AppendixDisplay{32}{\sup_{\gamma_1\in\mathbb R^{K_1}}\left[\gamma_1^\top\frac1N\sum_{i=1}^{N}B_1(X_i)-\frac1N\sum_{i:D_i=d_1}\ell^{\mathrm c}(B_1(X_i)^\top\gamma_1)-\sum_{k=2}^{K_1}\varepsilon_{1,k}|\gamma_{1,k}|\right].
} Reversing its sign gives minimization of \(N^{-1}\sum_{i:D_i=d_1}\ell^{\mathrm c}(B_1(X_i)^\top\gamma_1)-N^{-1}\sum_{i=1}^{N}B_1(X_i)^\top\gamma_1+\sum_{k=2}^{K_1}\varepsilon_{1,k}|\gamma_{1,k}|\), the first-stage dual objective stated in Theorem 5.1.

\subsubsection{Subsequent-stage duality}\label{app-A-5-2}

For \(s\geq2\), fix the preceding solution and apply the same conjugate inequality to any feasible target-group weights. It gives \AppendixDisplay{33}{\frac1N\sum_{i:D_i=d_s}\ell(w_{s,i})
\geq
\gamma_s^\top\frac1N\sum_{i:D_i=d_s}w_{s,i}B_s(\overline M_{s-1,i},X_i)
-\frac1N\sum_{i:D_i=d_s}\ell^{\mathrm c}(B_s(\overline M_{s-1,i},X_i)^\top\gamma_s).
} The balance target is now the preceding stage's weighted moment. Applying each component's tolerance bound, including the exact constant component, yields \AppendixDisplay{34}{\begin{aligned}
\frac1N\sum_{i:D_i=d_s}\ell(w_{s,i})
&\geq
\gamma_s^\top\frac1N\sum_{i:D_i=d_{s-1}}\widehat w_{s-1,i}B_s(\overline M_{s-1,i},X_i)
\\
&\qquad-\frac1N\sum_{i:D_i=d_s}\ell^{\mathrm c}(B_s(\overline M_{s-1,i},X_i)^\top\gamma_s)
-\sum_{k=2}^{K_s}\varepsilon_{s,k}|\gamma_{s,k}|.
\end{aligned}
} The same two-multiplier argument shows that maximization of this bound is the stage-\(s\) Lagrange dual. Changing its sign gives \AppendixDisplay{35}{\begin{aligned}
\widehat\gamma_s\in\operatorname*{arg\,min}_{\gamma_s\in\mathbb R^{K_s}}&\left[\frac1N\sum_{i:D_i=d_s}\ell^{\mathrm c}(B_s(\overline M_{s-1,i},X_i)^\top\gamma_s)-\gamma_s^\top\frac1N\sum_{i:D_i=d_{s-1}}\widehat w_{s-1,i}B_s(\overline M_{s-1,i},X_i)\right.\\
&\qquad\left.+\sum_{k=2}^{K_s}\varepsilon_{s,k}|\gamma_{s,k}|\right],
\end{aligned}
} which is the subsequent-stage minimization in the theorem.

\subsubsection{Feasibility of the recovered weights}\label{app-A-5-3}

Let \(\widehat\gamma_s\) solve its stage-\(s\) dual problem. At every observation with \(D_i=d_s\), define \(\widehat w_{s,i}=(\ell^{\mathrm c})'(B_s(\overline M_{s-1,i},X_i)^\top\widehat\gamma_s)\), using \(B_1(X_i)\) when \(s=1\). Because \(\ell^{\mathrm c}(t)=\sup_{w\geq0}\{tw-\ell(w)\}\) is the supremum of functions \(t\mapsto tw-\ell(w)\) with \(w\geq0\), it is nondecreasing; hence its derivative, wherever assumed to exist, is nonnegative. The differentiable part of the convex first-stage dual objective has component derivative equal to the weighted basis moment minus the full-sample basis moment. Since \(\partial|u|=\{1\}\) for \(u>0\), \([-1,1]\) for \(u=0\), and \(\{-1\}\) for \(u<0\), optimality of \(\widehat\gamma_1\) gives \(0\in N^{-1}\sum_{i:D_i=d_1}(\ell^{\mathrm c})'(B_1(X_i)^\top\widehat\gamma_1)b_{1,k}(X_i)-N^{-1}\sum_{i=1}^{N}b_{1,k}(X_i)+\varepsilon_{1,k}\partial|\widehat\gamma_{1,k}|\) for every \(k=1,\ldots,K_1\). Substituting the recovered weights gives \AppendixDisplay{36}{0\in\frac1N\sum_{i:D_i=d_1}\widehat w_{1,i}b_{1,k}(X_i)-\frac1N\sum_{i=1}^{N}b_{1,k}(X_i)+\varepsilon_{1,k}\partial|\widehat\gamma_{1,k}|.
} Because every element of \(\partial|\widehat\gamma_{1,k}|\) lies in \([-1,1]\), this condition implies, for \(k\geq2\), \AppendixDisplay{37}{\left|\frac1N\sum_{i:D_i=d_1}\widehat w_{1,i}b_{1,k}(X_i)-\frac1N\sum_{i=1}^{N}b_{1,k}(X_i)\right|\leq\varepsilon_{1,k}.
} For \(k=1\), the penalty is zero and \(b_{1,1}=1\), so \(N^{-1}\sum_{i:D_i=d_1}\widehat w_{1,i}=1\) and \(\sum_{i:D_i=d_1}\widehat w_{1,i}=N\).

At stage \(s\geq2\), the preceding weights remain fixed when the dual objective is differentiated. Its componentwise optimality condition is \AppendixDisplay{38}{\begin{aligned}
0&\in
\frac1N\sum_{i:D_i=d_s}\widehat w_{s,i}b_{s,k}(\overline M_{s-1,i},X_i)
-\frac1N\sum_{i:D_i=d_{s-1}}\widehat w_{s-1,i}b_{s,k}(\overline M_{s-1,i},X_i)
\\
&\qquad+\varepsilon_{s,k}\partial|\widehat\gamma_{s,k}|,
\quad k=1,\ldots,K_s.
\end{aligned}
} Thus all later-stage balance inequalities hold for \(k\geq2\). At \(k=1\), the exact constant constraint gives \(\sum_{i:D_i=d_s}\widehat w_{s,i}=\sum_{i:D_i=d_{s-1}}\widehat w_{s-1,i}=N\). Induction proves feasibility and normalization at every stage without assigning weights to observations outside the target group.

\subsubsection{Equality of primal and dual values}\label{app-A-5-4}

The conjugate inequality \(\ell(w)+\ell^{\mathrm c}(t)\geq tw\) becomes an equality at \(w=(\ell^{\mathrm c})'(t)\) under the stated closedness and differentiability conditions: \(\ell((\ell^{\mathrm c})'(t))+\ell^{\mathrm c}(t)=t(\ell^{\mathrm c})'(t)\). For the recovered weights \(\widehat w_{s,i}=(\ell^{\mathrm c})'(B_s(\overline M_{s-1,i},X_i)^\top\widehat\gamma_s)\), this gives \(\ell(\widehat w_{s,i})+\ell^{\mathrm c}(B_s(\overline M_{s-1,i},X_i)^\top\widehat\gamma_s)=\widehat w_{s,i}B_s(\overline M_{s-1,i},X_i)^\top\widehat\gamma_s\) at every \(D_i=d_s\). Write \(\operatorname{sgn}(t)=\mathbf1\{t>0\}-\mathbf1\{t<0\}\). For \(\widehat\gamma_{1,k}\neq0\), the first-stage optimality condition gives \AppendixDisplay{39}{\frac1N\sum_{i:D_i=d_1}\widehat w_{1,i}b_{1,k}(X_i)-\frac1N\sum_{i=1}^{N}b_{1,k}(X_i)=-\varepsilon_{1,k}\operatorname{sgn}(\widehat\gamma_{1,k}).
} Multiplying by \(\widehat\gamma_{1,k}\) gives \AppendixDisplay{40}{\widehat\gamma_{1,k}\left\{\frac1N\sum_{i:D_i=d_1}\widehat w_{1,i}b_{1,k}(X_i)-\frac1N\sum_{i=1}^{N}b_{1,k}(X_i)\right\}=-\varepsilon_{1,k}|\widehat\gamma_{1,k}|.
} When \(\widehat\gamma_{1,k}=0\), the same identity has both sides zero. The constant component contributes \(\widehat\gamma_{1,1}\{N^{-1}\sum_{i:D_i=d_1}\widehat w_{1,i}-1\}=0\). Summing over the components gives \AppendixDisplay{41}{\widehat\gamma_1^\top\left\{\frac1N\sum_{i:D_i=d_1}\widehat w_{1,i}B_1(X_i)-\frac1N\sum_{i=1}^{N}B_1(X_i)\right\}=-\sum_{k=2}^{K_1}\varepsilon_{1,k}|\widehat\gamma_{1,k}|.
} This residual identity isolates the penalty term. Rearranging it to isolate the current weighted moment for the Fenchel sum gives \AppendixDisplay{42}{\widehat\gamma_1^\top\frac1N\sum_{i:D_i=d_1}\widehat w_{1,i}B_1(X_i)=\widehat\gamma_1^\top\frac1N\sum_{i=1}^{N}B_1(X_i)-\sum_{k=2}^{K_1}\varepsilon_{1,k}|\widehat\gamma_{1,k}|.
} Summing the pointwise Fenchel equalities over the first-stage target group and substituting this identity gives \AppendixDisplay{43}{\frac1N\sum_{i:D_i=d_1}\ell(\widehat w_{1,i})
=
\widehat\gamma_1^\top\frac1N\sum_{i=1}^{N}B_1(X_i)
-\frac1N\sum_{i:D_i=d_1}\ell^{\mathrm c}(B_1(X_i)^\top\widehat\gamma_1)
-\sum_{k=2}^{K_1}\varepsilon_{1,k}|\widehat\gamma_{1,k}|.
} This is the first-stage dual lower bound evaluated at \(\widehat\gamma_1\).

At a later stage, \(uz=|u|\) for every \(z\in\partial|u|\), including \(u=0\), so the same calculation uses the preceding stage's weighted target. Multiplying the stage-\(s\) optimality conditions by \(\widehat\gamma_{s,k}\) and summing gives \AppendixDisplay{44}{\widehat\gamma_s^\top
\left\{
\frac1N\sum_{i:D_i=d_s}\widehat w_{s,i}B_s(\overline M_{s-1,i},X_i)
-\frac1N\sum_{i:D_i=d_{s-1}}\widehat w_{s-1,i}B_s(\overline M_{s-1,i},X_i)
\right\}
=-\sum_{k=2}^{K_s}\varepsilon_{s,k}|\widehat\gamma_{s,k}|.
} Substituting this identity into the sum of the stage-\(s\) Fenchel equalities yields \AppendixDisplay{45}{\begin{aligned}
\frac1N\sum_{i:D_i=d_s}\ell(\widehat w_{s,i})
&=
\widehat\gamma_s^\top\frac1N\sum_{i:D_i=d_{s-1}}\widehat w_{s-1,i}B_s(\overline M_{s-1,i},X_i)
\\
&\qquad-\frac1N\sum_{i:D_i=d_s}\ell^{\mathrm c}(B_s(\overline M_{s-1,i},X_i)^\top\widehat\gamma_s)
-\sum_{k=2}^{K_s}\varepsilon_{s,k}|\widehat\gamma_{s,k}|.
\end{aligned}
} This is the later-stage dual lower bound at \(\widehat\gamma_s\). Since every feasible primal weight vector has objective value at least its stage's dual bound, equality proves that the recovered weights minimize every primal problem. It also establishes equality of the primal and dual optimal values.

Finally, each feasible set is convex because it is defined by nonnegativity and linear balance inequalities. The objective is strictly convex in the target-group weight vector because \(\ell\) is strictly convex; specifically, \(N^{-1}\sum_{i:D_i=d_s}\ell(w_{s,i})\) is strictly convex. Two distinct feasible minimizers would have a feasible nontrivial convex combination with strictly smaller objective, a contradiction. Hence the recovered weights are the unique primal weights for the target group at every stage. In the following all-stage formula, \(B_s(\overline M_{s-1,i},X_i)\) means \(B_1(X_i)\) when \(s=1\): \AppendixDisplay{46}{\widehat w_{s,i}
=(\ell^{\mathrm c})'\!\left(B_s(\overline M_{s-1,i},X_i)^\top\widehat\gamma_s\right),
\qquad D_i=d_s,\quad s=1,\ldots,J+1.
} This completes the proof. \AppendixQED

\subsection{Proof of Theorem 5.2}\label{app-A-6}

We prove existence and uniqueness of the dual and primal solutions together with the four rates in the theorem. For each stage \(s\), the coefficient error and both root-mean-square weight errors have order \(O_p(\sum_{r=1}^{s}\{\sqrt{K_r/N}+K_r^{-r_{w,r}}\})\), while the uniform weight error has order \(O_p(\sqrt{K_s}\sum_{r=1}^{s}\{\sqrt{K_r/N}+K_r^{-r_{w,r}}\})\); all four tend to zero. The argument proceeds stage by stage. Population balance centers the smooth dual gradient at \(\gamma_s^*\), while sample Gram bounds control local curvature and the error carried from the preceding stage. These bounds give the coefficient rate; the population and same-sample root-mean-square rates follow before the induction moves to the next stage. After the induction, the basis norm gives the uniform rate, and the dual optimality conditions recover the primal weights.

\subsubsection{Sample Gram matrix bounds}\label{app-A-6-1}

We first need uniform eigenvalue bounds for the target-group and full-sample Gram matrices. For a fixed stage, independence and the deterministic basis make the centered matrices \(\mathbf1\{D_i=d_s\}B_sB_s^\top-\mathbb E[\mathbf1\{D=d_s\}B_sB_s^\top]\) independent and mean zero. The cross terms vanish in the squared Frobenius norm. Since \(\|B_sB_s^\top\|_F^2=\|B_s\|_2^4\), Assumption 5.1 gives \AppendixDisplay{47}{\begin{aligned}
&\mathbb E\left\|
\frac1N\sum_{i=1}^{N}\mathbf1\{D_i=d_s\}B_sB_s^\top
-\mathbb E[\mathbf1\{D=d_s\}B_sB_s^\top]
\right\|_F^2\\
&\quad\leq \frac1N\mathbb E\|B_s\|_2^4
\leq\frac{\sup\|B_s\|_2^2}{N}\mathbb E\|B_s\|_2^2
\leq\frac{CK_s^2}{N}.
\end{aligned}
} Indeed, \(\mathbb E\|B_s\|_2^2=\operatorname{tr}(\mathbb E[B_sB_s^\top])\leq CK_s\). The same calculation with the indicator removed bounds the full-sample Gram error by \(CK_s^2/N\). The growth condition implies \(K_s^2/N\to0\); hence \(\|A\|_{\mathrm{op}}\leq\|A\|_F\) and Markov's inequality make both Gram errors \(o_p(1)\) in operator norm. Conditioning on \((\overline M_{s-1},X)\) and using positivity gives \AppendixDisplay{48}{\mathbb E[\mathbf1\{D=d_s\}B_sB_s^\top]
=\mathbb E[\pi_{s-1}(d_s;\overline M_{s-1},X)B_sB_s^\top]
\succeq c\mathbb E[B_sB_s^\top]\succeq cI.
} The same population matrices are bounded above by \(CI\). Thus, with probability tending to one, \(cI\preceq N^{-1}\sum_i\mathbf1\{D_i=d_s\}B_sB_s^\top\preceq CI\) and \(cI\preceq N^{-1}\sum_iB_sB_s^\top\preceq CI\). Because \(J\) is fixed, these bounds hold jointly over all stages with probability tending to one. All subsequent sample Gram inequalities are taken on this joint event.

\subsubsection{Gradients at the approximation coefficients}\label{app-A-6-2}

Consider the stagewise convex objectives \AppendixDisplay{49}{Q_1(\gamma_1)=\frac1N\sum_i\{\mathbf1\{D_i=d_1\}\ell^{\mathrm c}(B_1^\top\gamma_1)-B_1^\top\gamma_1\}
+\sum_{k=2}^{K_1}\varepsilon_{1,k}|\gamma_{1,k}|,
} and, for \(s\geq2\), \AppendixDisplay{50}{Q_s(\gamma_s)=\frac1N\sum_i\{\mathbf1\{D_i=d_s\}\ell^{\mathrm c}(B_s^\top\gamma_s)
-\mathbf1\{D_i=d_{s-1}\}\widehat w_{s-1,i}B_s^\top\gamma_s\}
+\sum_{k=2}^{K_s}\varepsilon_{s,k}|\gamma_{s,k}|.
} At stage \(s\geq2\) the preceding weights are already constructed and held fixed. We differentiate only the smooth part; the absolute-value penalty is handled below.

At the first stage, adding and subtracting the population weight in that gradient gives \AppendixDisplay{51}{\begin{aligned}
&\frac1N\sum_i\{\mathbf1\{D_i=d_1\}(\ell^{\mathrm c})'(B_1^\top\gamma_1^*)-1\}B_1\\
&\quad=\underbrace{\frac1N\sum_i\{\mathbf1\{D_i=d_1\}w_1^*-1\}B_1}_{\text{(i)}}\\
&\qquad+\underbrace{\frac1N\sum_i\mathbf1\{D_i=d_1\}\{(\ell^{\mathrm c})'(B_1^\top\gamma_1^*)-w_1^*\}B_1}_{\text{(ii)}}.
\end{aligned}
} Here (i) is the sampling term and (ii) is the approximation term. Term (i) has mean zero by population balance. Positivity bounds \(w_1^*\), and the population Gram bound gives \(\mathbb E\|B_1\|_2^2\leq CK_1\). Independence therefore eliminates the cross-observation terms and yields \AppendixDisplay{52}{\mathbb E\left\|\frac1N\sum_i\{\mathbf1\{D_i=d_1\}w_1^*-1\}B_1\right\|_2^2
=\frac1N\mathbb E[\{\mathbf1\{D=d_1\}w_1^*-1\}^2\|B_1\|_2^2]
\leq\frac{CK_1}{N}.
} For Term (ii), Cauchy--Schwarz applied sample by sample and the full-sample Gram upper bound give \AppendixDisplay{53}{\begin{aligned}
&\left\|\frac1N\sum_i\mathbf1\{D_i=d_1\}
\{(\ell^{\mathrm c})'(B_1^\top\gamma_1^*)-w_1^*\}B_1\right\|_2\\
&\quad\leq\left\{\frac1N\sum_i|(\ell^{\mathrm c})'(B_1^\top\gamma_1^*)-w_1^*|^2\right\}^{1/2}
\left\|\frac1N\sum_iB_1B_1^\top\right\|_{\mathrm{op}}^{1/2}
\leq CK_1^{-r_{w,1}}.
\end{aligned}
} Here the first inequality follows by taking the supremum of the inner product with every unit vector; the indicator can be dropped from both nonnegative quadratic sums. Markov's inequality applied to the preceding second-moment bound shows that the norm of the first-stage smooth gradient at \(\gamma_1^*\) is \(O_p(\sqrt{K_1/N}+K_1^{-r_{w,1}})\).

For \(s\geq2\), suppose the preceding dual solution and its extended weight have been constructed. Adding and subtracting the two population-weight terms separates the smooth stage-\(s\) gradient at \(\gamma_s^*\) into the errors that must be controlled: \AppendixDisplay{54}{\begin{aligned}
&\frac1N\sum_i\{\mathbf1\{D_i=d_s\}(\ell^{\mathrm c})'(B_s^\top\gamma_s^*)
-\mathbf1\{D_i=d_{s-1}\}\widehat w_{s-1,i}\}B_s\\
&\quad=\underbrace{\frac1N\sum_i\mathbf1\{D_i=d_s\}
\{(\ell^{\mathrm c})'(B_s^\top\gamma_s^*)-w_s^*\}B_s}_{\text{(i)}}\\
&\qquad+\underbrace{\frac1N\sum_i\{\mathbf1\{D_i=d_s\}w_s^*
-\mathbf1\{D_i=d_{s-1}\}w_{s-1}^*\}B_s}_{\text{(ii)}}\\
&\qquad+\underbrace{\frac1N\sum_i\mathbf1\{D_i=d_{s-1}\}
(w_{s-1}^*-\widehat w_{s-1,i})B_s}_{\text{(iii)}}.
\end{aligned}
} Here (i) is the approximation term, (ii) is the sampling term, and (iii) is the preceding weight error. Term (i) has norm at most \(CK_s^{-r_{w,s}}\) by the same empirical Gram and uniform-approximation argument used at stage one. Term (ii) is centered by Proposition 3.2. Since the true weights are bounded and observations are i.i.d., its squared norm has expectation at most \(CK_s/N\); hence its norm is \(O_p(\sqrt{K_s/N})\). When \(d_s=d_{s-1}\), the population-weight recursion makes Term (ii) identically zero, although the general bound is enough here. Term (iii) contains weights estimated from this same sample, so independence across its summands cannot be used. Instead, for every realized sample the empirical Gram inequality gives \AppendixDisplay{55}{\begin{aligned}
&\left\|\frac1N\sum_i\mathbf1\{D_i=d_{s-1}\}
(w_{s-1}^*-\widehat w_{s-1,i})B_s\right\|_2\\
&\quad\leq\left\{\frac1N\sum_i|w_{s-1}^*-\widehat w_{s-1,i}|^2\right\}^{1/2}
\left\|\frac1N\sum_iB_sB_s^\top\right\|_{\mathrm{op}}^{1/2}
\leq C\left\{\frac1N\sum_i|w_{s-1}^*-\widehat w_{s-1,i}|^2\right\}^{1/2}.
\end{aligned}
} This is why the induction uses the same-sample root-mean-square error of the preceding weight. The bound multiplies this error by a constant, not by \(\sqrt{K_s}\). Combining the three sums, the norm of the smooth stage-\(s\) gradient at \(\gamma_s^*\) is at most \AppendixDisplay{56}{O_p(\sqrt{K_s/N}+K_s^{-r_{w,s}})
+C\left\{N^{-1}\sum_i|\widehat w_{s-1,i}-w_{s-1}^*|^2\right\}^{1/2}.
}

\subsubsection{Local objective growth}\label{app-A-6-3}

The coefficient error follows from a local quadratic lower bound for each \(Q_s\). Let \(\Delta\in\mathbb R^{K_s}\) be such that the segment from \(\gamma_s^*\) to \(\gamma_s^*+\Delta\) lies in \(\Gamma_s\). Assumption 5.3 bounds the second derivative of \(\ell^{\mathrm c}\) on this segment. Its first derivative is therefore Lipschitz there, and the integral Taylor formula applies. For \(s\geq2\), the resulting objective increment has three terms: \AppendixDisplay{57}{\begin{aligned}
Q_s(\gamma_s^*+\Delta)-Q_s(\gamma_s^*)
&=\underbrace{\Delta^\top\frac1N\sum_i
\{\mathbf1\{D_i=d_s\}(\ell^{\mathrm c})'(B_s^\top\gamma_s^*)
-\mathbf1\{D_i=d_{s-1}\}\widehat w_{s-1,i}\}B_s}_{\text{(I)}}\\
&\quad+\underbrace{\int_0^1(1-u)\Delta^\top
\left\{\frac1N\sum_i\mathbf1\{D_i=d_s\}
(\ell^{\mathrm c})''(B_s^\top(\gamma_s^*+u\Delta))B_sB_s^\top\right\}
\Delta\,du}_{\text{(II)}}\\
&\quad+\underbrace{\sum_{k=2}^{K_s}\varepsilon_{s,k}
\{|\gamma_{s,k}^*+\Delta_k|-|\gamma_{s,k}^*|\}}_{\text{(III)}}.
\end{aligned}
} Here (I) is the gradient evaluated above, (II) is the curvature contributed by the current stage's conjugate terms, and (III) is the change in the absolute-value penalty. The preceding fitted weights are held fixed, so their contribution to \(Q_s\) is linear and has no second derivative. At stage one the same identity holds with (I) equal to \(\Delta^\top N^{-1}\sum_i\{\mathbf1\{D_i=d_1\}(\ell^{\mathrm c})'(B_1^\top\gamma_1^*)-1\}B_1\), (II) evaluated with \(D_i=d_1\) and \(B_1\), and (III) using \(\varepsilon_1\). On the joint Gram event, Assumption 5.3 and the target-group Gram lower bound yield, for every \(u\in[0,1]\), \AppendixDisplay{58}{\begin{aligned}
&\Delta^\top\left\{\frac1N\sum_i\mathbf1\{D_i=d_s\}
(\ell^{\mathrm c})''(B_s^\top(\gamma_s^*+u\Delta))B_sB_s^\top\right\}\Delta\\
&\quad\geq c\Delta^\top\left\{\frac1N\sum_i\mathbf1\{D_i=d_s\}B_sB_s^\top\right\}\Delta
\geq c\|\Delta\|_2^2.
\end{aligned}
} Integration against \(1-u\) makes (II) at least \(c\|\Delta\|_2^2/2\); we absorb the factor \(1/2\) into \(c\) below. For (III), \(|a+b|-|a|\geq-|b|\) and Cauchy--Schwarz give the lower bound \(-\|\varepsilon_s\|_2\|\Delta\|_2\). Cauchy--Schwarz bounds (I) from below by minus \(\|\Delta\|_2\) times the corresponding smooth-gradient norm. Consequently, for \(\Delta\) in this local ball, \AppendixDisplay{59}{Q_s(\gamma_s^*+\Delta)-Q_s(\gamma_s^*)
\geq c\|\Delta\|_2^2-R_{s,N}\|\Delta\|_2,
} where \(R_{1,N}=\|N^{-1}\sum_i\{\mathbf1\{D_i=d_1\}(\ell^{\mathrm c})'(B_1^\top\gamma_1^*)-1\}B_1\|_2+\|\varepsilon_1\|_2\), and for \(s\geq2\), \(R_{s,N}=\|N^{-1}\sum_i\{\mathbf1\{D_i=d_s\}(\ell^{\mathrm c})'(B_s^\top\gamma_s^*)-\mathbf1\{D_i=d_{s-1}\}\widehat w_{s-1,i}\}B_s\|_2+\|\varepsilon_s\|_2\). These are the smooth-gradient norms plus the corresponding tolerance norms, not derivatives of the nonsmooth full objectives.

\subsubsection{Stagewise transfer of weight errors}\label{app-A-6-4}

Before applying the objective bound, we record how a solution in the small local ball constructed below yields the two root-mean-square bounds needed for induction. For arguments on the common support, the mean-value theorem and the upper bound on \((\ell^{\mathrm c})''\) give \AppendixDisplay{60}{\begin{aligned}
|\widehat w_s-w_s^*|
&\leq |(\ell^{\mathrm c})'(B_s^\top\widehat\gamma_s)
-(\ell^{\mathrm c})'(B_s^\top\gamma_s^*)|
+|(\ell^{\mathrm c})'(B_s^\top\gamma_s^*)-w_s^*|\\
&\leq C|B_s^\top(\widehat\gamma_s-\gamma_s^*)|+CK_s^{-r_{w,s}}.
\end{aligned}
} In the population norm, the fitted coefficient is held fixed while \((\overline M_{s-1},X)\) is integrated over its population distribution. Minkowski's inequality and the population Gram upper bound imply \AppendixDisplay{61}{\begin{aligned}
\left\{\int|\widehat w_s-w_s^*|^2\,dP\right\}^{1/2}
&\leq C\{(\widehat\gamma_s-\gamma_s^*)^\top
\mathbb E[B_sB_s^\top](\widehat\gamma_s-\gamma_s^*)\}^{1/2}
+CK_s^{-r_{w,s}}\\
&\leq C\|\widehat\gamma_s-\gamma_s^*\|_2+CK_s^{-r_{w,s}}.
\end{aligned}
} For the \(N\) observations used to fit the weights, the corresponding bound instead uses the full-sample Gram matrix. It is a sample-by-sample inequality and requires no independence between fitted weights and observations: \AppendixDisplay{62}{\begin{aligned}
\left\{\frac1N\sum_i|\widehat w_{s,i}-w_s^*|^2\right\}^{1/2}
&\leq C\left\{(\widehat\gamma_s-\gamma_s^*)^\top
\left(\frac1N\sum_iB_sB_s^\top\right)
(\widehat\gamma_s-\gamma_s^*)\right\}^{1/2}
+CK_s^{-r_{w,s}}\\
&\leq C\|\widehat\gamma_s-\gamma_s^*\|_2+CK_s^{-r_{w,s}}.
\end{aligned}
}

We solve the stage-one problem first, then each stage-\(s\) problem using the weights already constructed at stage \(s-1\). Choose a fixed \(\rho>0\) sufficiently small that \(\|\Delta\|_2\leq\rho/\sqrt{K_s}\) implies \(\sup|B_s^\top\Delta|\leq c\) for every stage. Thus this closed ball and all segments joining its center to its points lie in \(\Gamma_s\). At stage one the gradient calculation and Assumption 5.5 give \AppendixDisplay{63}{R_{1,N}=O_p(\sqrt{K_1/N}+K_1^{-r_{w,1}})+o_p(N^{-1/2}),
\qquad \sqrt{K_1}R_{1,N}=o_p(1),
} where the last relation is Assumption 5.4. On an event whose probability tends to one, \(\sqrt{K_1}R_{1,N}<c\rho/2\). The local quadratic bound therefore yields, simultaneously for every point on the sphere \(\|\Delta\|_2=\rho/\sqrt{K_1}\), \AppendixDisplay{64}{Q_1(\gamma_1^*+\Delta)-Q_1(\gamma_1^*)
\geq c\rho^2/K_1-R_{1,N}\rho/\sqrt{K_1}
\geq c\rho^2/(2K_1)>0.
} The dual objective is continuous on this compact ball, so it attains a minimum there. The strict boundary inequality places every such minimum in the interior. Since \(Q_1\) is convex on its entire domain, an interior minimum of the ball is also global. A point with a smaller value outside the ball would yield a nearby convex combination inside the ball with a smaller value. The Hessian of the smooth part is bounded below by \(cI\) on the ball, as shown in the curvature calculation, and the penalty is convex. Thus \(Q_1\) is strictly convex on the ball. If there were another global minimum outside it, a short segment from the interior minimum toward that point would produce a distinct global minimum inside the ball, contradicting local strict convexity. Hence \(\widehat\gamma_1\) exists, is finite, and is the unique global dual minimum. Because \(Q_1(\widehat\gamma_1)\leq Q_1(\gamma_1^*)\), the quadratic lower bound with \(\Delta=\widehat\gamma_1-\gamma_1^*\) gives \(c\|\Delta\|_2^2\leq R_{1,N}\|\Delta\|_2\). If \(\Delta=0\) the desired bound is immediate; otherwise division by \(\|\Delta\|_2\) gives \AppendixDisplay{65}{\|\widehat\gamma_1-\gamma_1^*\|_2
\leq C R_{1,N}
=O_p(\sqrt{K_1/N}+K_1^{-r_{w,1}}).
} The population and same-sample inequalities above give that same order for both first-stage root-mean-square weight errors. In particular, the first-stage weight is now defined at all observations, and its same-sample error is available for the second-stage gradient.

For the induction step, suppose stages through \(s-1\) have been constructed and their same-sample errors have the claimed rates. The three-term gradient decomposition then gives \AppendixDisplay{66}{R_{s,N}
=O_p\!\left(\sum_{r=1}^{s}\{\sqrt{K_r/N}+K_r^{-r_{w,r}}\}\right),
} where the preceding stage's same-sample root-mean-square bound controls the third term of the gradient decomposition and \(\|\varepsilon_s\|_2=o_p(N^{-1/2})\) is absorbed by the stage-\(s\) sampling rate. The local quadratic bound still gives \AppendixDisplay{67}{Q_s(\gamma_s^*+\Delta)-Q_s(\gamma_s^*)
\geq c\|\Delta\|_2^2-R_{s,N}\|\Delta\|_2.
} Because \(\sqrt{K_s}\sum_{r=1}^{s}\{\sqrt{K_r/N}+K_r^{-r_{w,r}}\}\to0\) and \(\sqrt{K_s/N}\to0\), we have \(\sqrt{K_s}R_{s,N}=o_p(1)\). On the corresponding high-probability event the same quadratic calculation makes \(Q_s(\gamma_s^*+\Delta)>Q_s(\gamma_s^*)\) throughout the sphere \(\|\Delta\|_2=\rho/\sqrt{K_s}\). Continuity on the ball, global convexity, and strict convexity within the ball give a finite, unique global dual solution \(\widehat\gamma_s\), exactly as at stage one. Substituting its coefficient difference in the quadratic lower bound gives \(\|\widehat\gamma_s-\gamma_s^*\|_2\leq C R_{s,N}\). The population and same-sample inequalities then yield \AppendixDisplay{68}{\|\widehat\gamma_s-\gamma_s^*\|_2
=O_p\!\left(\sum_{r=1}^{s}\{\sqrt{K_r/N}+K_r^{-r_{w,r}}\}\right)=o_p(1),
} and, separately, \AppendixDisplay{69}{\left\{\int|\widehat w_s-w_s^*|^2\,dP\right\}^{1/2}
=O_p\!\left(\sum_{r=1}^{s}\{\sqrt{K_r/N}+K_r^{-r_{w,r}}\}\right)=o_p(1),
} \AppendixDisplay{70}{\left\{\frac1N\sum_i|\widehat w_{s,i}-w_s^*|^2\right\}^{1/2}
=O_p\!\left(\sum_{r=1}^{s}\{\sqrt{K_r/N}+K_r^{-r_{w,r}}\}\right)=o_p(1).
} The tolerance term is absorbed because \(\|\varepsilon_s\|_2=o_p(N^{-1/2})\) and \(K_s\to\infty\). This completes the induction, in which existence and the coefficient rate at each stage precede the same-sample error bound passed to the next stage. Since \(J\) is fixed, the intersection of the stagewise Gram and gradient events still has probability tending to one.

The uniform bound is taken only after the induction. Applying \(\sup\|B_s\|_2\leq C\sqrt{K_s}\) to the pointwise weight inequality gives \AppendixDisplay{71}{\|\widehat w_s-w_s^*\|_\infty
\leq C\sqrt{K_s}\|\widehat\gamma_s-\gamma_s^*\|_2+CK_s^{-r_{w,s}}
=O_p\!\left(\sqrt{K_s}\sum_{r=1}^{s}\{\sqrt{K_r/N}+K_r^{-r_{w,r}}\}\right)=o_p(1).
} Using the same-sample root-mean-square bound rather than this uniform bound in the induction is what keeps errors from preceding stages additive.

\subsubsection{Recovery of the primal weights}\label{app-A-6-5}

It remains to identify the dual solutions with the weights in Algorithm 4.1. At the constructed \(\widehat\gamma_s\), Assumption 5.3 gives \(\widehat w_{s,i}=(\ell^{\mathrm c})'(B_s^\top\widehat\gamma_s)>0\) for every \(D_i=d_s\). The subgradient condition for the first-stage dual objective is, component by component, \AppendixDisplay{72}{\frac1N\sum_{i:D_i=d_1}\widehat w_{1,i}b_{1,k}(X_i)
-\frac1N\sum_{i=1}^{N}b_{1,k}(X_i)
\in-\varepsilon_{1,k}\partial|\widehat\gamma_{1,k}|,
\qquad k=1,\ldots,K_1.
} At each later stage the corresponding condition, with preceding weights held fixed, is \AppendixDisplay{73}{\frac1N\sum_{i:D_i=d_s}\widehat w_{s,i}b_{s,k}(\overline M_{s-1,i},X_i)
-\frac1N\sum_{i:D_i=d_{s-1}}\widehat w_{s-1,i}b_{s,k}(\overline M_{s-1,i},X_i)
\in-\varepsilon_{s,k}\partial|\widehat\gamma_{s,k}|.
} Every subgradient of \(|u|\) has magnitude at most one. Hence the recovered target-group weights satisfy all the balance inequalities. For \(k=1\), \(b_{s,1}=1\) and \(\varepsilon_{s,1}=0\), so these equalities give \(\sum_{i:D_i=d_1}\widehat w_{1,i}=N\) and recursively \(\sum_{i:D_i=d_s}\widehat w_{s,i}=\sum_{i:D_i=d_{s-1}}\widehat w_{s-1,i}=N\). The recovered weights are therefore feasible at every stage.

We verify the conjugate equality directly, since the current theorem does not assume that the loss is closed at zero. Fix a target-group observation and put \(t=B_s^\top\widehat\gamma_s\), with \(B_1(X_i)\) at stage one. Assumption 5.3 makes \(\ell^{\mathrm c}\) finite and differentiable in a neighborhood of \(t\) and gives \((\ell^{\mathrm c})'(t)=\widehat w_{s,i}>0\). Because \(b_{s,1}=1\) and \(\widehat\gamma_s\) lies in the interior of the constructed local ball, a sufficiently small shift of its first coefficient remains in \(\Gamma_s\); hence \(\ell^{\mathrm c}(t+\delta)<\infty\) for some \(\delta>0\). Take any sequence of nonnegative weights whose values in the supremum defining \(\ell^{\mathrm c}(t)\) approach that supremum. Finiteness at \(t+\delta\) bounds this sequence. Every convergent subsequence has a limit \(u\) satisfying \(\ell^{\mathrm c}(t+h)\geq\ell^{\mathrm c}(t)+hu\) for sufficiently small positive and negative \(h\); differentiability therefore forces \(u=(\ell^{\mathrm c})'(t)=\widehat w_{s,i}\). A finite convex loss is continuous at every positive weight, so the supremum is attained at this positive limit. Thus, at every target-group observation, \AppendixDisplay{74}{\ell(\widehat w_{s,i})+\ell^{\mathrm c}(B_s^\top\widehat\gamma_s)
=\widehat w_{s,i}B_s^\top\widehat\gamma_s.
} Multiplying the displayed balance residuals by their corresponding \(\widehat\gamma_{s,k}\) and summing over \(k\) gives a penalty contribution of \(-\sum_{k=2}^{K_s}\varepsilon_{s,k}|\widehat\gamma_{s,k}|\), since \(u z=|u|\) for \(z\in\partial|u|\). Summing the pointwise conjugate equalities consequently makes the recovered primal objective equal to \(-Q_s(\widehat\gamma_s)\). At stage one its linear part is \(\widehat\gamma_1^\top N^{-1}\sum_iB_1\), and at later stages it is \(\widehat\gamma_s^\top N^{-1}\sum_{i:D_i=d_{s-1}}\widehat w_{s-1,i}B_s\). Conversely, for any feasible target-group vector \(\{w_{s,i}\}\), the conjugate inequality and the componentwise balance constraints give, for \(s\geq2\), \AppendixDisplay{75}{\begin{aligned}
\frac1N\sum_{i:D_i=d_s}\ell(w_{s,i})
&\geq\widehat\gamma_s^\top\frac1N\sum_{i:D_i=d_{s-1}}\widehat w_{s-1,i}B_s
-\frac1N\sum_{i:D_i=d_s}\ell^{\mathrm c}(B_s^\top\widehat\gamma_s)\\
&\quad-\sum_{k=2}^{K_s}\varepsilon_{s,k}|\widehat\gamma_{s,k}|
=-Q_s(\widehat\gamma_s).
\end{aligned}
} At stage one, the same inequality uses \(N^{-1}\sum_iB_1\) in place of the preceding weighted basis sum. The recovered vector therefore solves the stage-\(s\) primal problem. Its feasible set is convex, and the objective is strictly convex in its target-group weights, so this primal solution is unique. We have now obtained, on a joint event whose probability tends to one, finite unique dual solutions and unique target-group primal solutions at all \(J+1\) stages. The coefficient, population root-mean-square, same-sample root-mean-square, and uniform bounds established above are exactly the four rates in Theorem 5.2. \AppendixQED

\subsection{Proof of Theorem 5.3}\label{app-A-7}

The proof compares the augmented weighting estimator with the sample mean of the true centered score. An exact sample identity separates their difference into weighted true-regression residuals and empirical balance discrepancies. The stagewise information restriction and the sample root-mean-square weight rates control the residual terms. Basis approximation controls the balance discrepancies for the true regressions, while Assumption 5.9 controls those for the estimated regressions. Combining these bounds gives the rate in Theorem 5.3; Assumption 5.8 then makes the remaining products negligible at the root-\(N\) scale.

\subsubsection{Second moments of the true score and bounded approximation coefficients}\label{app-A-7-1}

Assumption 5.6 includes \(\mathbb E[Y^2]<\infty\). Conditional Jensen's inequality and Assumption 2.3 give \AppendixDisplay{76}{\begin{aligned}
\mathbb E\!\left[\left\{\mu_J^{(d_{J+1})}(\overline M_J,X)\right\}^2\right]
&\leq \mathbb E\!\left[
\mathbb E[Y^2\mid D=d_{J+1},\overline M_J,X]
\right]\\
&=\mathbb E\!\left[
\frac{\mathbf1\{D=d_{J+1}\}Y^2}
{\pi_J(d_{J+1};\overline M_J,X)}
\right]
\leq C\mathbb E[Y^2]<\infty.
\end{aligned}
} Applying the same argument to the backward-regression recursion gives, for each \(j=1,\ldots,J\), \AppendixDisplay{77}{\begin{aligned}
\mathbb E\!\left[\left\{\mu_{j-1}^{(d_j,\ldots,d_{J+1})}\right\}^2\right]
&\leq \mathbb E\!\left[
\frac{\mathbf1\{D=d_j\}
\left\{\mu_j^{(d_{j+1},\ldots,d_{J+1})}\right\}^2}
{\pi_{j-1}(d_j;\overline M_{j-1},X)}
\right]\\
&\leq C\mathbb E\!\left[
\left\{\mu_j^{(d_{j+1},\ldots,d_{J+1})}\right\}^2
\right].
\end{aligned}
} Induction backward from the terminal regression shows that all true backward regressions have finite second moments. Their adjacent differences and the terminal outcome residual are therefore square integrable. The true centered score is \AppendixDisplay{78}{\begin{aligned}
\psi_{\boldsymbol d}(O_i)
&=\mu_0^{(d_1,\ldots,d_{J+1})}(X_i)-\theta_{\boldsymbol d}\\
&\quad+\sum_{j=1}^{J}\mathbf1\{D_i=d_j\}w_j^*
\left\{\mu_j^{(d_{j+1},\ldots,d_{J+1})}
-\mu_{j-1}^{(d_j,\ldots,d_{J+1})}\right\}\\
&\quad+\mathbf1\{D_i=d_{J+1}\}w_{J+1}^*
\left\{Y_i-\mu_J^{(d_{J+1})}\right\}.
\end{aligned}
} By the definitions of the backward regressions, each weighted residual has conditional mean zero in its treatment group given the preceding observed history. Together with \(\theta_{\boldsymbol d}=\mathbb E[\mu_0^{(d_1,\ldots,d_{J+1})}(X)]\), this yields \(\mathbb E[\psi_{\boldsymbol d}]=0\). The bounded population weights and fixed number of summands, together with the preceding second-moment bounds, imply \(V_{\boldsymbol d}=\mathbb E[\psi_{\boldsymbol d}^2]<\infty\).

Next consider the deterministic approximation coefficients. Assumption 5.7 provides \(\beta_s^*\in\mathbb R^{K_s}\) for each \(s\) such that \AppendixDisplay{79}{\sup_{\overline m_{s-1},x}
\left|
\mu_{s-1}^{(d_s,\ldots,d_{J+1})}(\overline m_{s-1},x)
-B_s(\overline m_{s-1},x)^\top\beta_s^*
\right|
\leq CK_s^{-r_{\mu,s-1}}.
} The population Gram lower bound in Assumption 5.1 then implies \AppendixDisplay{80}{\begin{aligned}
c\|\beta_s^*\|_2^2
&\leq \mathbb E[(B_s^\top\beta_s^*)^2]\\
&\leq 2\mathbb E\!\left[
\left\{\mu_{s-1}^{(d_s,\ldots,d_{J+1})}\right\}^2
\right]
+2\mathbb E\!\left[
\left\{B_s^\top\beta_s^*-
\mu_{s-1}^{(d_s,\ldots,d_{J+1})}\right\}^2
\right]\\
&\leq C+CK_s^{-2r_{\mu,s-1}}.
\end{aligned}
} Hence \(\|\beta_s^*\|_2=O(1)\). This conclusion concerns the deterministic coefficients approximating the true regressions; the later control of \(\widehat\beta_s\) comes from Assumption 5.9.

\subsubsection{Population and sample balance}\label{app-A-7-2}

For the first stage, the population identity follows by conditioning on \(X\): \(\mathbb E[\mathbf1\{D=d_1\}w_1^*(X)-1\mid X]=\pi_0(d_1;X)w_1^*(X)-1=0\). For \(s\geq2\), the weight recursion gives \(\pi_{s-1}(d_s;\overline M_{s-1},X)w_s^*=\pi_{s-1}(d_{s-1};\overline M_{s-1},X)w_{s-1}^*\). Consequently, \(\mathbb E[\mathbf1\{D=d_s\}w_s^*-\mathbf1\{D=d_{s-1}\}w_{s-1}^*\mid\overline M_{s-1},X]=0\). Multiplying by an integrable fixed function of the relevant history preserves zero population expectation.

Algorithm 4.1 imposes \(|N^{-1}\sum_i\{\mathbf1\{D_i=d_1\}\widehat w_{1,i}-1\}b_{1,k}(X_i)|\leq\varepsilon_{1,k}\) at stage one. At stage \(s\geq2\), it imposes \(|N^{-1}\sum_i\{\mathbf1\{D_i=d_s\}\widehat w_{s,i}-\mathbf1\{D_i=d_{s-1}\}\widehat w_{s-1,i}\}b_{s,k}|\leq\varepsilon_{s,k}\). Therefore, for every coefficient vector \(\beta_1\), the triangle inequality gives \AppendixDisplay{81}{\begin{aligned}
&\left|
\frac1N\sum_{i=1}^{N}
\{\mathbf1\{D_i=d_1\}\widehat w_{1,i}-1\}B_1^\top\beta_1
\right|\\
&\quad=
\left|
\sum_{k=1}^{K_1}\beta_{1,k}
\frac1N\sum_{i=1}^{N}
\{\mathbf1\{D_i=d_1\}\widehat w_{1,i}-1\}b_{1,k}
\right|
\leq\sum_{k=1}^{K_1}|\beta_{1,k}|\varepsilon_{1,k}
\leq\|\beta_1\|_1\|\varepsilon_1\|_\infty.
\end{aligned}
} The same calculation at each \(s\geq2\) gives \AppendixDisplay{82}{\begin{aligned}
&\left|
\frac1N\sum_{i=1}^{N}
\{\mathbf1\{D_i=d_s\}\widehat w_{s,i}
-\mathbf1\{D_i=d_{s-1}\}\widehat w_{s-1,i}\}B_s^\top\beta_s
\right|\\
&\quad\leq
\sum_{k=1}^{K_s}|\beta_{s,k}|\varepsilon_{s,k}
\leq\|\beta_s\|_1\|\varepsilon_s\|_\infty.
\end{aligned}
} These inequalities hold for every realized sample on which the balancing problems have solutions. Thus they also hold with the same-sample estimated coefficient vector \(\widehat\beta_s\) in place of \(\beta_s\). The true regression need not lie exactly in the basis span, so its balance discrepancy requires the approximation calculation below.

\subsubsection{Exact decomposition}\label{app-A-7-3}

The goal is the two-product convergence rate in Theorem 5.3 and, under Assumption 5.8, the asymptotic linear expansion \(\sqrt N(\widehat\theta_{\boldsymbol d}^{\mathrm{AW\text{-}BW}}-\theta_{\boldsymbol d})=N^{-1/2}\sum_{i=1}^{N}\psi_{\boldsymbol d}(O_i)+o_p(1)\). The four terms below isolate the bounds needed for both conclusions. Collecting the coefficients of each fitted regression in the estimator of Section 3.1 gives \AppendixDisplay{83}{\begin{aligned}
\widehat\theta_{\boldsymbol d}^{\mathrm{AW\text{-}BW}}
&=\underbrace{\frac1N\sum_{i=1}^{N}
\mathbf1\{D_i=d_{J+1}\}\widehat w_{J+1,i}Y_i}_{\text{(i)}}\\
&\quad-\underbrace{\frac1N\sum_{i=1}^{N}
\{\mathbf1\{D_i=d_1\}\widehat w_{1,i}-1\}
\widehat\mu_0^{(d_1,\ldots,d_{J+1})}}_{\text{(ii)}}\\
&\quad-\underbrace{\frac1N\sum_{i=1}^{N}\sum_{s=2}^{J+1}
\{\mathbf1\{D_i=d_s\}\widehat w_{s,i}
-\mathbf1\{D_i=d_{s-1}\}\widehat w_{s-1,i}\}
\widehat\mu_{s-1}^{(d_s,\ldots,d_{J+1})}}_{\text{(iii)}}
\end{aligned}
} Here (i) is the weighted outcome, (ii) is the first-stage correction, and (iii) collects the later-stage corrections. Indeed, the coefficient of \(\widehat\mu_0\) is \(1-\mathbf1\{D_i=d_1\}\widehat w_{1,i}\). For \(j=1,\ldots,J\), the adjacent correction terms give coefficient \(\mathbf1\{D_i=d_j\}\widehat w_{j,i}-\mathbf1\{D_i=d_{j+1}\}\widehat w_{j+1,i}\) for \(\widehat\mu_j\). The remaining outcome term is the final weighted mean.

Subtracting the sample mean of the true centered score, adding and subtracting the population weights, and writing each true regression as \(B_s^\top\beta_s^*\) plus its approximation residual give an exact four-term identity. Each term is grouped by the bound used below: \AppendixDisplay{84}{\begin{aligned}
&\widehat\theta_{\boldsymbol d}^{\mathrm{AW\text{-}BW}}
-\theta_{\boldsymbol d}
-\frac1N\sum_{i=1}^{N}\psi_{\boldsymbol d}(O_i)\\
&=\underbrace{\begin{aligned}[t]
&\frac1N\sum_{i=1}^{N}\sum_{j=1}^{J}
\mathbf1\{D_i=d_j\}(\widehat w_{j,i}-w_j^*)
\{\mu_j^{(d_{j+1},\ldots,d_{J+1})}
-\mu_{j-1}^{(d_j,\ldots,d_{J+1})}\}\\
&\quad+\frac1N\sum_{i=1}^{N}
\mathbf1\{D_i=d_{J+1}\}(\widehat w_{J+1,i}-w_{J+1}^*)
\{Y_i-\mu_J^{(d_{J+1})}\}
\end{aligned}}_{\text{(I)}}\\
&\quad+\underbrace{\begin{aligned}[t]
&\frac1N\sum_{i=1}^{N}
\{\mathbf1\{D_i=d_1\}\widehat w_{1,i}-1\}
B_1^\top(\beta_1^*-\widehat\beta_1)\\
&\quad+\frac1N\sum_{i=1}^{N}\sum_{s=2}^{J+1}
\{\mathbf1\{D_i=d_s\}\widehat w_{s,i}
-\mathbf1\{D_i=d_{s-1}\}\widehat w_{s-1,i}\}
B_s^\top(\beta_s^*-\widehat\beta_s)
\end{aligned}}_{\text{(II)}}\\
&\quad+\underbrace{\begin{aligned}[t]
&\frac1N\sum_{i=1}^{N}
\{\mathbf1\{D_i=d_1\}w_1^*-1\}
\{\mu_0^{(d_1,\ldots,d_{J+1})}-B_1^\top\beta_1^*\}\\
&\quad+\frac1N\sum_{i=1}^{N}\sum_{s=2}^{J+1}
\{\mathbf1\{D_i=d_s\}w_s^*
-\mathbf1\{D_i=d_{s-1}\}w_{s-1}^*\}
\{\mu_{s-1}^{(d_s,\ldots,d_{J+1})}-B_s^\top\beta_s^*\}
\end{aligned}}_{\text{(III)}}\\
&\quad+\underbrace{\begin{aligned}[t]
&\frac1N\sum_{i=1}^{N}
\mathbf1\{D_i=d_1\}(\widehat w_{1,i}-w_1^*)
\{\mu_0^{(d_1,\ldots,d_{J+1})}-B_1^\top\beta_1^*\}\\
&\quad+\frac1N\sum_{i=1}^{N}\sum_{s=2}^{J+1}
\mathbf1\{D_i=d_s\}(\widehat w_{s,i}-w_s^*)
\{\mu_{s-1}^{(d_s,\ldots,d_{J+1})}-B_s^\top\beta_s^*\}\\
&\quad-\frac1N\sum_{i=1}^{N}\sum_{s=2}^{J+1}
\mathbf1\{D_i=d_{s-1}\}(\widehat w_{s-1,i}-w_{s-1}^*)
\{\mu_{s-1}^{(d_s,\ldots,d_{J+1})}-B_s^\top\beta_s^*\}
\end{aligned}}_{\text{(IV)}}
\end{aligned}
} The identity holds for every realized sample. At each stage, \(\widehat\mu_{s-1}^{(d_s,\ldots,d_{J+1})}=B_s^\top\widehat\beta_s\) by Assumption 5.9. Thus the fitted-regression part belongs to Term (II), together with the deterministic basis approximation; the remaining true-regression approximation error is divided between Terms (III) and (IV) according to whether it multiplies population weights or weight errors.

\subsubsection{(I) Conditional variance for weight errors times true residuals}\label{app-A-7-4}

Fix \(j\in\{1,\ldots,J\}\). Algorithm 4.1 makes \(\widehat w_{j,i}\) measurable with respect to \((D_\ell,X_\ell,\overline M_{j-1,\ell})_{\ell=1}^{N}\); \(w_j^*\) is measurable with respect to the same information. The backward-regression definition gives \(\mathbb E[\mu_j^{(d_{j+1},\ldots,d_{J+1})}-\mu_{j-1}^{(d_j,\ldots,d_{J+1})}\mid D_i=d_j,\overline M_{j-1,i},X_i]=0\). Because the observations are independent, adding the other observations' preceding histories to the conditioning set preserves this equality and makes residuals from different observations conditionally uncorrelated. Assumption 5.6 bounds the corresponding conditional second moment by \(C\). Consequently, \AppendixDisplay{85}{\begin{aligned}
&\operatorname{Var}\!\left(
\frac1{\sqrt N}\sum_{i=1}^{N}
\mathbf1\{D_i=d_j\}(\widehat w_{j,i}-w_j^*)
\left\{\mu_j^{(d_{j+1},\ldots,d_{J+1})}
-\mu_{j-1}^{(d_j,\ldots,d_{J+1})}\right\}
\;\middle|\;(D_\ell,X_\ell,\overline M_{j-1,\ell})_{\ell=1}^{N}
\right)\\
&\quad=
\frac1N\sum_{i=1}^{N}
\mathbf1\{D_i=d_j\}|\widehat w_{j,i}-w_j^*|^2
\mathbb E\!\left[
\left\{\mu_j^{(d_{j+1},\ldots,d_{J+1})}
-\mu_{j-1}^{(d_j,\ldots,d_{J+1})}\right\}^2
\mid D_i=d_j,\overline M_{j-1,i},X_i
\right]\\
&\quad\leq
\frac CN\sum_{i=1}^{N}
\mathbf1\{D_i=d_j\}|\widehat w_{j,i}-w_j^*|^2
\leq
\frac CN\sum_{i=1}^{N}|\widehat w_{j,i}-w_j^*|^2.
\end{aligned}
} Theorem 5.2 makes the final empirical squared weight error \(o_p(1)\). The conditional mean of the scaled sum is zero. For any \(a,t>0\), conditional Chebyshev's inequality on the event where that error is at most \(t\), followed by an event split, gives the continuous bound \AppendixDisplay{86}{\begin{aligned}
&\Pr\!\left(\left|\frac1{\sqrt N}\sum_{i=1}^{N}
\mathbf1\{D_i=d_j\}(\widehat w_{j,i}-w_j^*)
\{\mu_j^{(d_{j+1},\ldots,d_{J+1})}-\mu_{j-1}^{(d_j,\ldots,d_{J+1})}\}\right|>a\right)\\
&=\mathbb E\!\left[
\Pr\!\left(\left|\frac1{\sqrt N}\sum_{i=1}^{N}
\mathbf1\{D_i=d_j\}(\widehat w_{j,i}-w_j^*)
\{\mu_j^{(d_{j+1},\ldots,d_{J+1})}-\mu_{j-1}^{(d_j,\ldots,d_{J+1})}\}\right|>a
\;\middle|\;(D_\ell,X_\ell,\overline M_{j-1,\ell})_{\ell=1}^{N}\right)
\right]\\
&\quad\leq\frac{C}{a^2}\mathbb E\!\left[
\mathbf1\!\left\{\frac1N\sum_{i=1}^{N}|\widehat w_{j,i}-w_j^*|^2\leq t\right\}
\frac1N\sum_{i=1}^{N}|\widehat w_{j,i}-w_j^*|^2\right]
+\Pr\!\left(\frac1N\sum_{i=1}^{N}|\widehat w_{j,i}-w_j^*|^2>t\right)\\
&\quad\leq \frac{Ct}{a^2}
+\Pr\!\left(\frac1N\sum_{i=1}^{N}|\widehat w_{j,i}-w_j^*|^2>t\right).
\end{aligned}
} The last inequality uses that the indicator restricts the squared weight error to at most \(t\). For fixed \(t>0\), the final probability tends to zero by Theorem 5.2. Taking \(N\to\infty\) and then \(t\downarrow0\) proves that the scaled stage-\(j\) sum is \(o_p(1)\).

At the final stage, condition on \((D_\ell,X_\ell,\overline M_{J,\ell})_{\ell=1}^{N}\). Assumption 5.6 supplies \(\mathbb E[\{Y-\mu_J^{(d_{J+1})}\}^2\mid D=d_{J+1},\overline M_J,X]\leq C\), and the same measurability and conditional-independence argument gives \AppendixDisplay{87}{\begin{aligned}
&\operatorname{Var}\!\left(
\frac1{\sqrt N}\sum_{i=1}^{N}
\mathbf1\{D_i=d_{J+1}\}(\widehat w_{J+1,i}-w_{J+1}^*)
\{Y_i-\mu_J^{(d_{J+1})}\}
\;\middle|\;(D_\ell,X_\ell,\overline M_{J,\ell})_{\ell=1}^{N}
\right)\\
&\quad\leq
\frac CN\sum_{i=1}^{N}|\widehat w_{J+1,i}-w_{J+1}^*|^2
=o_p(1).
\end{aligned}
} Applying the same event split to the final-stage residual and summing the fixed number of stages gives \(\text{(I)}=o_p(N^{-1/2})\).

\subsubsection{(II) Sample balance for the basis terms}\label{app-A-7-5}

Assumption 5.9 writes each fitted regression as \(B_s^\top\widehat\beta_s\). The sample balance inequalities established above hold for every coefficient vector, including the same-sample \(\widehat\beta_s\) and the deterministic \(\beta_s^*\). Their triangle inequality therefore bounds the \emph{difference} in Term (II) directly: \AppendixDisplay{88}{\begin{aligned}
|\text{(II)}|
&\leq\sum_{s=1}^{J+1}
\|\beta_s^*-\widehat\beta_s\|_1\|\varepsilon_s\|_\infty\\
&\leq\sum_{s=1}^{J+1}
\left(\sqrt{K_s}\|\beta_s^*\|_2+\|\widehat\beta_s\|_1\right)
\|\varepsilon_s\|_\infty
=o_p(N^{-1/2}).
\end{aligned}
} The deterministic coefficient norms are \(O(1)\) by the Gram and moment argument above. Assumption 5.5 states \(\|\varepsilon_s\|_\infty=o_p((NK_s)^{-1/2})\), so \(\sqrt N\sqrt{K_s}\|\beta_s^*\|_2\|\varepsilon_s\|_\infty=o_p(1)\). Assumption 5.9 states \(\sqrt N\|\widehat\beta_s\|_1\|\varepsilon_s\|_\infty=o_p(1)\). Since \(J\) is fixed, these bounds give the displayed order for the sum. Exact balance makes Term (II) zero for each realized solution.

\subsubsection{(III) Population balance for deterministic approximation residuals}\label{app-A-7-6}

The approximation residual \(\mu_{s-1}^{(d_s,\ldots,d_{J+1})}-B_s^\top\beta_s^*\) is deterministic for each \(N\). Therefore, the population balance identities center its product with the true weights. At stage one, independence across observations, bounded population weights, and uniform approximation yield \AppendixDisplay{89}{\begin{aligned}
&\mathbb E\!\left[
\left\{\frac1{\sqrt N}\sum_{i=1}^{N}
(\mathbf1\{D_i=d_1\}w_1^*-1)
(\mu_0^{(d_1,\ldots,d_{J+1})}-B_1^\top\beta_1^*)
\right\}^2
\right]\\
&\quad=
\mathbb E\!\left[
(\mathbf1\{D=d_1\}w_1^*-1)^2
(\mu_0^{(d_1,\ldots,d_{J+1})}-B_1^\top\beta_1^*)^2
\right]
\leq CK_1^{-2r_{\mu,0}}\longrightarrow0.
\end{aligned}
} At stage \(s\geq2\), the same calculation yields \AppendixDisplay{90}{\begin{aligned}
&\mathbb E\!\left[
\left\{\frac1{\sqrt N}\sum_{i=1}^{N}
(\mathbf1\{D_i=d_s\}w_s^*
-\mathbf1\{D_i=d_{s-1}\}w_{s-1}^*)
(\mu_{s-1}^{(d_s,\ldots,d_{J+1})}-B_s^\top\beta_s^*)
\right\}^2
\right]\\
&\quad=
\mathbb E\!\left[
(\mathbf1\{D=d_s\}w_s^*
-\mathbf1\{D=d_{s-1}\}w_{s-1}^*)^2
(\mu_{s-1}^{(d_s,\ldots,d_{J+1})}-B_s^\top\beta_s^*)^2
\right]\\
&\quad\leq CK_s^{-2r_{\mu,s-1}}\longrightarrow0.
\end{aligned}
} The population balance identities established above center these summands. Assumption 5.7 supplies the uniform residual bound, and Assumption 5.4 gives \(K_s\to\infty\). Markov's inequality makes Term (III) \(o_p(N^{-1/2})\) because \(J\) is fixed. This argument applies only to deterministic approximation residuals; it does not treat fitted regressions as fixed.

\subsubsection{(IV) Cauchy--Schwarz for weight errors times approximation residuals}\label{app-A-7-7}

At the first stage, Cauchy--Schwarz and the uniform approximation bound give \AppendixDisplay{91}{\begin{aligned}
&\left|\frac1N\sum_{i=1}^{N}
\mathbf1\{D_i=d_1\}(\widehat w_{1,i}-w_1^*)
\{\mu_0^{(d_1,\ldots,d_{J+1})}-B_1^\top\beta_1^*\}\right|\\
&\quad\leq
\left\{\frac1N\sum_{i=1}^{N}
\mathbf1\{D_i=d_1\}|\widehat w_{1,i}-w_1^*|^2
\right\}^{1/2}
\left\{\frac1N\sum_{i=1}^{N}
\mathbf1\{D_i=d_1\}
|\mu_0^{(d_1,\ldots,d_{J+1})}-B_1^\top\beta_1^*|^2
\right\}^{1/2}\\
&\quad\leq CK_1^{-r_{\mu,0}}
\left\{\frac1N\sum_{i=1}^{N}|\widehat w_{1,i}-w_1^*|^2\right\}^{1/2}.
\end{aligned}
} At stage \(s\geq2\), separate the current and preceding weight errors by the triangle inequality and apply the same bound to each: \AppendixDisplay{92}{\begin{aligned}
&\left|\frac1N\sum_{i=1}^{N}
\left\{\mathbf1\{D_i=d_s\}(\widehat w_{s,i}-w_s^*)
-\mathbf1\{D_i=d_{s-1}\}(\widehat w_{s-1,i}-w_{s-1}^*)\right\}
\{\mu_{s-1}^{(d_s,\ldots,d_{J+1})}-B_s^\top\beta_s^*\}\right|\\
&\quad\leq CK_s^{-r_{\mu,s-1}}
\left[
\left\{\frac1N\sum_{i=1}^{N}|\widehat w_{s,i}-w_s^*|^2\right\}^{1/2}
+\left\{\frac1N\sum_{i=1}^{N}|\widehat w_{s-1,i}-w_{s-1}^*|^2\right\}^{1/2}
\right].
\end{aligned}
} These are same-sample inequalities and require no independence between fitted weights and observations. Each stage-\(s\) weight error, for \(s\leq J\), enters its own and the next stage's imbalance; the last-stage error enters only the last imbalance. Thus collecting their coefficients gives \AppendixDisplay{93}{\begin{aligned}
|\text{(IV)}|
&\leq C\sum_{s=1}^{J}
\left(K_s^{-r_{\mu,s-1}}+K_{s+1}^{-r_{\mu,s}}\right)
\left\{\frac1N\sum_{i=1}^{N}|\widehat w_{s,i}-w_s^*|^2\right\}^{1/2}\\
&\quad+CK_{J+1}^{-r_{\mu,J}}
\left\{\frac1N\sum_{i=1}^{N}|\widehat w_{J+1,i}-w_{J+1}^*|^2\right\}^{1/2}.
\end{aligned}
} For every \(s=1,\ldots,J+1\), Theorem 5.2 gives the same-sample rate \(\{N^{-1}\sum_{i=1}^{N}|\widehat w_{s,i}-w_s^*|^2\}^{1/2}=O_p(\sum_{r=1}^{s}\{\sqrt{K_r/N}+K_r^{-r_{w,r}}\})\). Substitution gives \AppendixDisplay{94}{\begin{aligned}
|\text{(IV)}|
&=O_p\!\left(
\sum_{s=1}^{J}
\left(K_s^{-r_{\mu,s-1}}+K_{s+1}^{-r_{\mu,s}}\right)
\sum_{r=1}^{s}\left\{\sqrt{\frac{K_r}{N}}+K_r^{-r_{w,r}}\right\}
\right)\\
&\quad+O_p\!\left(
K_{J+1}^{-r_{\mu,J}}
\sum_{r=1}^{J+1}\left\{\sqrt{\frac{K_r}{N}}+K_r^{-r_{w,r}}\right\}
\right).
\end{aligned}
} Assumption 5.8 imposes \(\sqrt N\sum_{s=1}^{J}(K_s^{-r_{\mu,s-1}}+K_{s+1}^{-r_{\mu,s}})\sum_{r=1}^{s}\{\sqrt{K_r/N}+K_r^{-r_{w,r}}\}\to0\) and \(\sqrt N K_{J+1}^{-r_{\mu,J}}\sum_{r=1}^{J+1}\{\sqrt{K_r/N}+K_r^{-r_{w,r}}\}\to0\). Hence Term (IV) is \(o_p(N^{-1/2})\) under that additional assumption. Before imposing it, the displayed product rates remain in the asserted convergence rate.

\subsubsection{Rate and limiting distribution}\label{app-A-7-8}

Substitute the four bounds into the exact sample identity. Terms (I)--(III) are \(o_p(N^{-1/2})\), while Term (IV) has the product rate just obtained. Hence \AppendixDisplay{95}{\begin{aligned}
&\widehat\theta_{\boldsymbol d}^{\mathrm{AW\text{-}BW}}
-\theta_{\boldsymbol d}\\
&\quad=
\frac1N\sum_{i=1}^{N}\psi_{\boldsymbol d}(O_i)
+o_p(N^{-1/2})\\
&\qquad+
O_p\!\left(
\sum_{s=1}^{J}
\left(K_s^{-r_{\mu,s-1}}+K_{s+1}^{-r_{\mu,s}}\right)
\sum_{r=1}^{s}\left\{\sqrt{\frac{K_r}{N}}+K_r^{-r_{w,r}}\right\}
\right)\\
&\qquad+
O_p\!\left(
K_{J+1}^{-r_{\mu,J}}
\sum_{r=1}^{J+1}\left\{\sqrt{\frac{K_r}{N}}+K_r^{-r_{w,r}}\right\}
\right)
\end{aligned}
} Because the true score is i.i.d., centered, and has finite variance, \(N^{-1}\sum_i\psi_{\boldsymbol d}(O_i)=O_p(N^{-1/2})\). Therefore, \AppendixDisplay{96}{\begin{aligned}
\widehat\theta_{\boldsymbol d}^{\mathrm{AW\text{-}BW}}-\theta_{\boldsymbol d}
&=O_p(N^{-1/2})\\
&\quad+
O_p\!\left(
\sum_{s=1}^{J}
\left(K_s^{-r_{\mu,s-1}}+K_{s+1}^{-r_{\mu,s}}\right)
\sum_{r=1}^{s}\left\{\sqrt{\frac{K_r}{N}}+K_r^{-r_{w,r}}\right\}
\right)\\
&\quad+
O_p\!\left(
K_{J+1}^{-r_{\mu,J}}
\sum_{r=1}^{J+1}\left\{\sqrt{\frac{K_r}{N}}+K_r^{-r_{w,r}}\right\}
\right)
=o_p(1).
\end{aligned}
} Assumption 5.4 makes each stagewise weight rate tend to zero, and Assumption 5.7 makes each regression approximation error tend to zero; fixed \(J\) makes both product sums \(o(1)\). This proves consistency and the asserted rate without Assumption 5.8.

Now impose Assumption 5.8. Both product errors become \(o_p(N^{-1/2})\), and the decomposition reduces to \AppendixDisplay{97}{\begin{aligned}
\sqrt N\left(
\widehat\theta_{\boldsymbol d}^{\mathrm{AW\text{-}BW}}
-\theta_{\boldsymbol d}
\right)
&=
\frac1{\sqrt N}\sum_{i=1}^{N}\psi_{\boldsymbol d}(O_i)+o_p(1)\\
&\xrightarrow{d}\mathcal N(0,V_{\boldsymbol d}),
\qquad V_{\boldsymbol d}=\mathbb E[\psi_{\boldsymbol d}^2].
\end{aligned}
} The i.i.d. finite-variance central limit theorem and Slutsky's lemma give the displayed normal limit. Theorem 2.1(iii) identifies \(\psi_{\boldsymbol d}\) as the efficient influence function in the stated identification model when it has finite second moment, as established above. Its variance \(V_{\boldsymbol d}\) is therefore the efficiency bound in that model. This proves the rate, asymptotic linearity, and asymptotic distribution in Theorem 5.3. \AppendixQED

\subsection{Proof of Theorem 5.4}\label{app-A-8}

We prove that the empirical mean of the squared fitted score converges to the variance of the true centered score. The proof splits their difference into the point-estimator centering error, regression errors multiplied by fitted weights, and weight errors multiplied by true residuals. Theorem 5.3 controls the first term, full-sample regression consistency and bounded fitted weights control the second, and Theorem 5.2 with the residual moment bounds controls the third. Their empirical mean-square bounds, a square-difference inequality, and the law of large numbers give the result.

\subsubsection{Exact decomposition}\label{app-A-8-1}

The bracketed term in Theorem 5.4 is the estimated centered score \AppendixDisplay{98}{\begin{aligned}
\widehat\psi_{\boldsymbol d}(O_i)
&=\widehat\mu_0^{(d_1,\ldots,d_{J+1})}(X_i)-\widehat\theta_{\boldsymbol d}^{\mathrm{AW\text{-}BW}}\\
&\quad+\sum_{j=1}^{J}\mathbf1\{D_i=d_j\}\widehat w_{j,i}
\left\{\widehat\mu_j^{(d_{j+1},\ldots,d_{J+1})}(\overline M_{j,i},X_i)-\widehat\mu_{j-1}^{(d_j,\ldots,d_{J+1})}(\overline M_{j-1,i},X_i)\right\}\\
&\quad+\mathbf1\{D_i=d_{J+1}\}\widehat w_{J+1,i}
\left\{Y_i-\widehat\mu_J^{(d_{J+1})}(\overline M_{J,i},X_i)\right\}.
\end{aligned}
} Thus \(\widehat V_{\boldsymbol d}^{\mathrm{AW\text{-}BW}}=N^{-1}\sum_{i=1}^{N}\widehat\psi_{\boldsymbol d}(O_i)^2\). For every observation \(i\), subtract the true centered score from the estimated centered score. The exact difference is \AppendixDisplay{99}{\begin{aligned}
&\widehat\psi_{\boldsymbol d}(O_i)-\psi_{\boldsymbol d}(O_i)\\
&=\underbrace{\theta_{\boldsymbol d}-\widehat\theta_{\boldsymbol d}^{\mathrm{AW\text{-}BW}}}_{\text{(I)}}\\
&\quad+\underbrace{\begin{aligned}[t]
&\{1-\mathbf1\{D_i=d_1\}\widehat w_{1,i}\}
\{\widehat\mu_0^{(d_1,\ldots,d_{J+1})}-\mu_0^{(d_1,\ldots,d_{J+1})}\}\\
&\quad+\sum_{s=2}^{J+1}
\{\mathbf1\{D_i=d_{s-1}\}\widehat w_{s-1,i}
-\mathbf1\{D_i=d_s\}\widehat w_{s,i}\}
\{\widehat\mu_{s-1}^{(d_s,\ldots,d_{J+1})}
-\mu_{s-1}^{(d_s,\ldots,d_{J+1})}\}
\end{aligned}}_{\text{(II)}}\\
&\quad+\underbrace{\begin{aligned}[t]
&\sum_{j=1}^{J}\mathbf1\{D_i=d_j\}
(\widehat w_{j,i}-w_j^*)
\{\mu_j^{(d_{j+1},\ldots,d_{J+1})}
-\mu_{j-1}^{(d_j,\ldots,d_{J+1})}\}\\
&\quad+\mathbf1\{D_i=d_{J+1}\}
(\widehat w_{J+1,i}-w_{J+1}^*)
\{Y_i-\mu_J^{(d_{J+1})}\}
\end{aligned}}_{\text{(III)}}
\end{aligned}
} To obtain this identity, add and subtract each estimated weight times the corresponding true regression difference in the two score displays. Collecting the coefficients of each \(\widehat\mu_{s-1}-\mu_{s-1}\) yields the adjacent-stage coefficients in Term (II). The remaining intermediate and terminal residuals form Term (III). This grouping makes the three bounds independent of the stage at which a term arises.

\subsubsection{(I) Centering error}\label{app-A-8-2}

Theorem 5.3 gives \(\widehat\theta_{\boldsymbol d}^{\mathrm{AW\text{-}BW}}-\theta_{\boldsymbol d}=o_p(1)\). Term (I) is constant across observations, so its empirical root-mean-square norm is \(|\widehat\theta_{\boldsymbol d}^{\mathrm{AW\text{-}BW}}-\theta_{\boldsymbol d}|=o_p(1)\).

\subsubsection{(II) Regression errors under full-sample consistency}\label{app-A-8-3}

Theorem 5.2 and Assumption 2.3 give \(\|\widehat w_s\|_\infty=O_p(1)\) for every \(s\). The triangle inequality for the empirical root-mean-square norm, followed by the bound \(\mathbf1\{D_i=d_s\}\widehat w_{s,i}\leq\|\widehat w_s\|_\infty\) in absolute value, gives \AppendixDisplay{100}{\begin{aligned}
\left\{\frac1N\sum_{i=1}^{N}|\text{(II)}|^2\right\}^{1/2}
&\leq(1+\|\widehat w_1\|_\infty)
\left\{\frac1N\sum_{i=1}^{N}
|\widehat\mu_0^{(d_1,\ldots,d_{J+1})}-\mu_0^{(d_1,\ldots,d_{J+1})}|^2\right\}^{1/2}\\
&\quad+\sum_{s=2}^{J+1}
(\|\widehat w_{s-1}\|_\infty+\|\widehat w_s\|_\infty)
\left\{\frac1N\sum_{i=1}^{N}
|\widehat\mu_{s-1}^{(d_s,\ldots,d_{J+1})}-\mu_{s-1}^{(d_s,\ldots,d_{J+1})}|^2\right\}^{1/2}
\\
&=o_p(1).
\end{aligned}
} Each regression error in this display is evaluated at the corresponding \((\overline M_{s-1,i},X_i)\), with \(X_i\) alone when \(s=1\). Assumption 5.10 makes every root-mean-square factor \(o_p(1)\) over all \(N\) observations, including those outside the fitting treatment group. Since \(J\) is fixed, the finite sum of products with \(O_p(1)\) weight bounds is \(o_p(1)\).

\subsubsection{(III) Weight errors times true residuals}\label{app-A-8-4}

The triangle inequality and the uniform weight-error bound give \AppendixDisplay{101}{\begin{aligned}
\left\{\frac1N\sum_{i=1}^{N}|\text{(III)}|^2\right\}^{1/2}
&\leq\sum_{j=1}^{J}\|\widehat w_j-w_j^*\|_\infty
\left\{\frac1N\sum_{i=1}^{N}\mathbf1\{D_i=d_j\}
|\mu_j^{(d_{j+1},\ldots,d_{J+1})}
-\mu_{j-1}^{(d_j,\ldots,d_{J+1})}|^2\right\}^{1/2}\\
&\quad+\|\widehat w_{J+1}-w_{J+1}^*\|_\infty
\left\{\frac1N\sum_{i=1}^{N}\mathbf1\{D_i=d_{J+1}\}
|Y_i-\mu_J^{(d_{J+1})}|^2\right\}^{1/2}
=o_p(1).
\end{aligned}
} For \(j=1,\ldots,J\), Assumption 5.6 states \(\mathbb E[(\mu_j-\mu_{j-1})^2\mid D=d_j,\overline M_{j-1},X]\leq C\); at the terminal stage it states \(\mathbb E[(Y-\mu_J)^2\mid D=d_{J+1},\overline M_J,X]\leq C\). For example, conditioning on \((D,\overline M_{j-1},X)\) and then taking expectations gives \AppendixDisplay{102}{\mathbb E[\mathbf1\{D=d_j\}|\mu_j-\mu_{j-1}|^2]
=\mathbb E\!\left[\mathbf1\{D=d_j\}
\mathbb E[|\mu_j-\mu_{j-1}|^2\mid D,\overline M_{j-1},X]\right]\leq C.
} At the terminal stage the same tower calculation gives \(\mathbb E[\mathbf1\{D=d_{J+1}\}|Y-\mu_J|^2]\leq C\). Thus the indicator-weighted residual squares have finite expectations, and the law of large numbers makes each empirical root-mean-square factor \(O_p(1)\). Theorem 5.2 makes every uniform weight-error factor \(o_p(1)\). The products and their finite sum are therefore \(o_p(1)\). Applying the empirical root-mean-square triangle inequality to the exact three-term identity yields \AppendixDisplay{103}{\left\{\frac1N\sum_{i=1}^{N}
|\widehat\psi_{\boldsymbol d}(O_i)-\psi_{\boldsymbol d}(O_i)|^2\right\}^{1/2}
=o_p(1).
}

\subsubsection{Consistency of the variance estimator}\label{app-A-8-5}

Theorem 5.3 and the law of large numbers give \(N^{-1}\sum_{i=1}^{N}\psi_{\boldsymbol d}(O_i)^2=O_p(1)\). Using \(a^2-b^2=(a-b)^2+2b(a-b)\) and Cauchy--Schwarz, \AppendixDisplay{104}{\begin{aligned}
&\left|\widehat V_{\boldsymbol d}^{\mathrm{AW\text{-}BW}}-\frac1N\sum_{i=1}^{N}\psi_{\boldsymbol d}(O_i)^2\right|\\
&\quad\leq\frac1N\sum_{i=1}^{N}\left\{\widehat\psi_{\boldsymbol d}(O_i)-\psi_{\boldsymbol d}(O_i)\right\}^2\\
&\qquad+2\left\{\frac1N\sum_{i=1}^{N}\psi_{\boldsymbol d}(O_i)^2\right\}^{1/2}
\left\{\frac1N\sum_{i=1}^{N}\left\{\widehat\psi_{\boldsymbol d}(O_i)-\psi_{\boldsymbol d}(O_i)\right\}^2\right\}^{1/2}=o_p(1).
\end{aligned}
} Theorem 5.3 gives \(\mathbb E[\psi_{\boldsymbol d}^2]=V_{\boldsymbol d}<\infty\). Since the observations are i.i.d., the law of large numbers gives \(N^{-1}\sum_{i=1}^{N}\psi_{\boldsymbol d}(O_i)^2\xrightarrow{p}V_{\boldsymbol d}\). Combining these two limits proves \AppendixDisplay{105}{\widehat V_{\boldsymbol d}^{\mathrm{AW\text{-}BW}}\xrightarrow{p}V_{\boldsymbol d}.
}

Under the maintained asymptotic-normality conditions, Theorem 5.3 gives the local distributional input \AppendixDisplay{106}{\sqrt N(\widehat\theta_{\boldsymbol d}^{\mathrm{AW\text{-}BW}}-\theta_{\boldsymbol d})
=\frac1{\sqrt N}\sum_{i=1}^{N}\psi_{\boldsymbol d}(O_i)+o_p(1)
\xrightarrow{d}\mathcal N(0,V_{\boldsymbol d}).
} Thus \(\widehat V_{\boldsymbol d}^{\mathrm{AW\text{-}BW}}/N\) is the corresponding large-sample variance estimate for the estimator itself. If \(V_{\boldsymbol d}>0\), the variance consistency just proved and Slutsky's lemma yield \AppendixDisplay{107}{\frac{\sqrt N(\widehat\theta_{\boldsymbol d}^{\mathrm{AW\text{-}BW}}-\theta_{\boldsymbol d})}{\sqrt{\widehat V_{\boldsymbol d}^{\mathrm{AW\text{-}BW}}}}\xrightarrow{d}\mathcal N(0,1).
}

For two fixed generalized mediation functionals satisfying the same conditions, consider the contrast estimator \(\widehat\theta_{\boldsymbol d}^{\mathrm{AW\text{-}BW}}-\widehat\theta_{\boldsymbol d'}^{\mathrm{AW\text{-}BW}}\) for \(\theta_{\boldsymbol d}-\theta_{\boldsymbol d'}\). Subtracting their two Theorem 5.3 expansions gives \AppendixDisplay{108}{\sqrt N\left\{
(\widehat\theta_{\boldsymbol d}^{\mathrm{AW\text{-}BW}}
-\widehat\theta_{\boldsymbol d'}^{\mathrm{AW\text{-}BW}})
-(\theta_{\boldsymbol d}-\theta_{\boldsymbol d'})\right\}
=\frac1{\sqrt N}\sum_{i=1}^{N}
\{\psi_{\boldsymbol d}(O_i)-\psi_{\boldsymbol d'}(O_i)\}+o_p(1).
} The two true scores are centered and square-integrable. Their difference is therefore centered and square-integrable, because \((a-b)^2\leq2a^2+2b^2\). The i.i.d. central limit theorem applied to the displayed expansion gives a normal limit with asymptotic variance \(V_{\boldsymbol d,\boldsymbol d'}=\mathbb E[\{\psi_{\boldsymbol d}(O)-\psi_{\boldsymbol d'}(O)\}^2]\). Estimate it by \(N^{-1}\sum_{i=1}^{N}\{\widehat\psi_{\boldsymbol d}(O_i)-\widehat\psi_{\boldsymbol d'}(O_i)\}^2\). The empirical root-mean-square norm of its score error is at most the sum of the two individual score-error norms, hence is \(o_p(1)\). Applying the preceding square-difference bound to the contrast score and then the law of large numbers proves convergence to \(V_{\boldsymbol d,\boldsymbol d'}\). This construction retains the covariance between the two scores computed from the same sample. \AppendixQED

\section{Asymptotic normality under empirical-process conditions}\label{app-B}

Assumption 5.9 in the main text controls the empirical balance residual for an estimated backward regression by requiring that regression to belong to the linear span of the corresponding balancing basis. This section gives alternative sufficient conditions without that linear-span restriction. The construction of the weights, the estimator, the true backward regressions, and the true score remain the same. Only the control of the empirical balance residual for regression estimation error changes. The following assumption replaces Assumption 5.9.

\begin{assumption}[Empirical-process and product-rate control for backward regressions]\label{app-assumption-B-1}

For each \(s=1,\ldots,J+1\), assume both (a) and (b).

(a) There is a nonrandom, pointwise measurable class \(\mathcal F_s\) of functions of \((\overline M_{s-1},X)\) that does not depend on \(N\). The class has a square-integrable envelope under the corresponding population distribution, contains the true regression \(\mu_{s-1}^{(d_s,\ldots,d_{J+1})}\), and satisfies \AppendixDisplay{109}{\Pr\!\left(
\widehat\mu_{s-1}^{(d_s,\ldots,d_{J+1})}
\in\mathcal F_s
\right)\longrightarrow1.
} At the first stage, \(\mathcal F_1\) consists of functions of \(X\) alone. Define the corresponding class of functions of \(O=(Y,D,\overline M_J,X)\) at the first stage by \AppendixDisplay{110}{\mathcal G_1
=
\left\{
O\mapsto
\left\{1-\mathbf1\{D=d_1\}w_1^*(X)\right\}f(X)
:
f\in\mathcal F_1
\right\}
} and, for \(s=2,\ldots,J+1\), by \AppendixDisplay{111}{\begin{aligned}
\mathcal G_s
=
\Big\{
O\mapsto
&\left[
\mathbf1\{D=d_{s-1}\}w_{s-1}^*(\overline M_{s-2},X)
-\mathbf1\{D=d_s\}w_s^*(\overline M_{s-1},X)
\right]f(\overline M_{s-1},X)
:
f\in\mathcal F_s
\Big\}.
\end{aligned}
} Assume that \(\mathcal G_s\) is \(P\)-Donsker. In addition, the estimated regression satisfies \AppendixDisplay{112}{\left\|
\widehat\mu_{s-1}^{(d_s,\ldots,d_{J+1})}
-\mu_{s-1}^{(d_s,\ldots,d_{J+1})}
\right\|_{L_2(P_{\overline M_{s-1},X})}
=o_p(1).
} In this population norm, the realized estimated regression is held fixed and integrated against the population distribution of a new argument \((\overline m_{s-1},x)\). At the first stage, \(P_{\overline M_0,X}=P_X\). At stage \(J+1\), the regression is \(\mu_J^{(d_{J+1})}\), and the same convention applies to its estimate. The class-membership event in (a) is measurable, as required by its displayed ordinary probability. If a supremum over a function class is nonmeasurable, its probability and convergence statements are interpreted in outer probability.

(b) At each stage, the estimated regression satisfies \AppendixDisplay{113}{\begin{aligned}
&\sqrt N
\left[
\sum_{r=1}^{s}
\left\{
\sqrt{\frac{K_r}{N}}+K_r^{-r_{w,r}}
\right\}
\right]
\\
&\qquad\times
\left\{
\frac1N\sum_{i=1}^{N}
\left|
\widehat\mu_{s-1}^{(d_s,\ldots,d_{J+1})}
(\overline M_{s-1,i},X_i)
-\mu_{s-1}^{(d_s,\ldots,d_{J+1})}
(\overline M_{s-1,i},X_i)
\right|^2
\right\}^{1/2}
=o_p(1).
\end{aligned}
} The regression error in this condition is evaluated at every observation in the same full sample, including observations outside the treatment group used to learn the regression.

\end{assumption}

Conditions (a) and (b) control different terms. Condition (a) controls the sample average of regression estimation error multiplied by the true weights. Population balance centers this term. Stochastic equicontinuity over a fixed Donsker class and population \(L_2\) consistency then make it \(o_p(N^{-1/2})\). Condition (b) controls the product error that arises when the estimated weights replace the true weights. Theorem 5.2's weight-error rate includes errors propagated from preceding stages, so the product condition contains \(\sum_{r=1}^{s}\{\sqrt{K_r/N}+K_r^{-r_{w,r}}\}\).

\begin{proposition}[Asymptotic linearity under empirical-process control]\label{app-proposition-B-1}

Maintain the identification formula of Theorem 2.1, Assumption 2.3, the learning-information restrictions of Algorithm 4.1, the conditions giving the weight-convergence rates in Theorem 5.2, and the second-moment conditions in Assumption 5.6. Replace Assumption 5.9 by Assumption B.1. If the alternative assumption holds at every stage, then \AppendixDisplay{114}{\begin{aligned}
\sqrt N\left(
\widehat\theta_{\boldsymbol d}^{\mathrm{AW\text{-}BW}}
-\theta_{\boldsymbol d}
\right)
&=
\frac1{\sqrt N}\sum_{i=1}^{N}\psi_{\boldsymbol d}(O_i)
+o_p(1)\\
&\xrightarrow{d}\mathcal N(0,V_{\boldsymbol d}),
\qquad V_{\boldsymbol d}=\mathbb E[\psi_{\boldsymbol d}^2].
\end{aligned}
}

\end{proposition}

\subsection{Proof of Proposition B.1}\label{app-B-1}

\subsubsection{Exact decomposition}\label{app-B-1-1}

We begin with the uncollected augmented estimator, \AppendixDisplay{115}{\begin{aligned}
\widehat\theta_{\boldsymbol d}^{\mathrm{AW\text{-}BW}}
&=\frac1N\sum_{i=1}^{N}\Bigg[\widehat\mu_0
+\sum_{j=1}^{J}\mathbf1\{D_i=d_j\}\widehat w_{j,i}
(\widehat\mu_j-\widehat\mu_{j-1})\\
&\qquad+\mathbf1\{D_i=d_{J+1}\}\widehat w_{J+1,i}
(Y_i-\widehat\mu_J)\Bigg],
\end{aligned}
} and the sample mean of the true score plus \(\theta_{\boldsymbol d}\), \AppendixDisplay{116}{\begin{aligned}
\theta_{\boldsymbol d}+\frac1N\sum_{i=1}^{N}\psi_{\boldsymbol d}(O_i)
&=\frac1N\sum_{i=1}^{N}\Bigg[\mu_0
+\sum_{j=1}^{J}\mathbf1\{D_i=d_j\}w_j^*
(\mu_j-\mu_{j-1})\\
&\qquad+\mathbf1\{D_i=d_{J+1}\}w_{J+1}^*
(Y_i-\mu_J)\Bigg].
\end{aligned}
} Theorem 5.3 bounds its four-term identity using the balancing-basis representation in Assumption 5.9. This section instead subtracts the two displays above and groups the resulting terms by their bounding method. For every realized sample, the exact identity is \AppendixDisplay{117}{\begin{aligned}
&\widehat\theta_{\boldsymbol d}^{\mathrm{AW\text{-}BW}}
-\theta_{\boldsymbol d}
-\frac1N\sum_{i=1}^{N}\psi_{\boldsymbol d}(O_i)\\
&=\underbrace{\begin{aligned}[t]
&\frac1N\sum_{i=1}^{N}\sum_{j=1}^{J}
\mathbf1\{D_i=d_j\}(\widehat w_{j,i}-w_j^*)
\{\mu_j^{(d_{j+1},\ldots,d_{J+1})}
-\mu_{j-1}^{(d_j,\ldots,d_{J+1})}\}\\
&\quad+\frac1N\sum_{i=1}^{N}
\mathbf1\{D_i=d_{J+1}\}(\widehat w_{J+1,i}-w_{J+1}^*)
\{Y_i-\mu_J^{(d_{J+1})}\}
\end{aligned}}_{\text{(I)}}\\
&\quad+\underbrace{\begin{aligned}[t]
&\frac1N\sum_{i=1}^{N}
\{1-\mathbf1\{D_i=d_1\}w_1^*\}
\{\widehat\mu_0^{(d_1,\ldots,d_{J+1})}
-\mu_0^{(d_1,\ldots,d_{J+1})}\}\\
&\quad+\frac1N\sum_{i=1}^{N}\sum_{s=2}^{J+1}
\{\mathbf1\{D_i=d_{s-1}\}w_{s-1}^*
-\mathbf1\{D_i=d_s\}w_s^*\}
\{\widehat\mu_{s-1}^{(d_s,\ldots,d_{J+1})}
-\mu_{s-1}^{(d_s,\ldots,d_{J+1})}\}
\end{aligned}}_{\text{(II)}}\\
&\quad+\underbrace{\begin{aligned}[t]
&\frac1N\sum_{i=1}^{N}\sum_{s=2}^{J+1}
\mathbf1\{D_i=d_{s-1}\}(\widehat w_{s-1,i}-w_{s-1}^*)
\{\widehat\mu_{s-1}^{(d_s,\ldots,d_{J+1})}
-\mu_{s-1}^{(d_s,\ldots,d_{J+1})}\}\\
&\quad-\frac1N\sum_{i=1}^{N}
\mathbf1\{D_i=d_1\}(\widehat w_{1,i}-w_1^*)
\{\widehat\mu_0^{(d_1,\ldots,d_{J+1})}
-\mu_0^{(d_1,\ldots,d_{J+1})}\}\\
&\quad-\frac1N\sum_{i=1}^{N}\sum_{s=2}^{J+1}
\mathbf1\{D_i=d_s\}(\widehat w_{s,i}-w_s^*)
\{\widehat\mu_{s-1}^{(d_s,\ldots,d_{J+1})}
-\mu_{s-1}^{(d_s,\ldots,d_{J+1})}\}
\end{aligned}}_{\text{(III)}}
\end{aligned}
} To verify the signs, the first-stage regression error in the estimator has coefficient \(1-\mathbf1\{D_i=d_1\}\widehat w_{1,i}\), while the error in regression \(s-1\) has coefficient \(\mathbf1\{D_i=d_{s-1}\}\widehat w_{s-1,i}-\mathbf1\{D_i=d_s\}\widehat w_{s,i}\) for \(s\geq2\). Splitting each coefficient into its population-weight and weight-error parts yields Terms (II) and (III); the remaining true residuals yield Term (I). Each term now has one bounding method.

\subsubsection{(I) Conditional variance for weight errors times true residuals}\label{app-B-1-2}

Term (I) is identical to Term (I) in the proof of Theorem 5.3. The conditional-variance argument there gives \(\text{(I)}=o_p(N^{-1/2})\).

\subsubsection{(II) Empirical-process control for true-weight regression errors}\label{app-B-1-3}

We now show that Term (II) is \(o_p(N^{-1/2})\). Population balance centers its summands for each fixed regression function. Condition (a) requires, for every \(s=1,\ldots,J+1\), a nonrandom, pointwise measurable class \(\mathcal F_s\) of functions of \((\overline M_{s-1},X)\) that does not depend on \(N\), has a square-integrable envelope under \(P_{\overline M_{s-1},X}\), and contains the true regression \(\mu_{s-1}^{(d_s,\ldots,d_{J+1})}\). Here \(\mathcal F_1\) consists of functions of \(X\) alone. The estimated regression satisfies \AppendixDisplay{118}{\begin{aligned}
&\Pr\!\left(\widehat\mu_{s-1}^{(d_s,\ldots,d_{J+1})}\in\mathcal F_s\right)\longrightarrow1,\\
&\left\|\widehat\mu_{s-1}^{(d_s,\ldots,d_{J+1})}
-\mu_{s-1}^{(d_s,\ldots,d_{J+1})}\right\|_{L_2(P_{\overline M_{s-1},X})}=o_p(1).
\end{aligned}
} The realized estimated regression is held fixed in this population norm. Define the fixed true-weighted classes by \AppendixDisplay{119}{\mathcal G_1=\left\{O\mapsto
\{1-\mathbf1\{D=d_1\}w_1^*(X)\}f(X):f\in\mathcal F_1\right\}
} and, for \(s=2,\ldots,J+1\), \AppendixDisplay{120}{\begin{aligned}
\mathcal G_s=\Big\{O\mapsto
&\left[\mathbf1\{D=d_{s-1}\}w_{s-1}^*(\overline M_{s-2},X)
-\mathbf1\{D=d_s\}w_s^*(\overline M_{s-1},X)\right]\\
&\times f(\overline M_{s-1},X):f\in\mathcal F_s\Big\}.
\end{aligned}
} Condition (a) assumes that each \(\mathcal G_s\) is \(P\)-Donsker. The population \(L_2\) convergence above and the fixed-class stochastic equicontinuity established by this condition control the sample-selected regression.

The first population weight satisfies \(\mathbb E[\mathbf1\{D=d_1\}w_1^*(X)\mid X]=\pi_0(d_1;X)w_1^*(X)=1\). Multiplying this conditional equality by a fixed \(f(X)\) and taking expectations gives, at the first stage, for every \(f\in\mathcal F_1\), \AppendixDisplay{121}{\mathbb E\!\left[
\{1-\mathbf1\{D=d_1\}w_1^*(X)\}f(X)
\right]=0.
} For \(s\geq2\), both population weights are measurable with respect to \((\overline M_{s-1},X)\). Their recursion gives the conditional calculation \AppendixDisplay{122}{\begin{aligned}
\mathbb E[\mathbf1\{D=d_s\}w_s^*\mid\overline M_{s-1},X]
&=\pi_{s-1}(d_s;\overline M_{s-1},X)w_s^*\\
&=\pi_{s-1}(d_{s-1};\overline M_{s-1},X)w_{s-1}^*\\
&=\mathbb E[\mathbf1\{D=d_{s-1}\}w_{s-1}^*\mid\overline M_{s-1},X].
\end{aligned}
} Multiplying by a fixed \(f(\overline M_{s-1},X)\) and taking expectations yields \AppendixDisplay{123}{\mathbb E\!\left[
\{\mathbf1\{D=d_{s-1}\}w_{s-1}^*
-\mathbf1\{D=d_s\}w_s^*\}
f(\overline M_{s-1},X)
\right]=0,
\qquad f\in\mathcal F_s.
} The conditional calculations above establish population balance. Boundedness of the true weights and the square-integrable envelope ensure that the products are integrable. Thus every function in \(\mathcal G_s\) has population mean zero. On a sample for which the estimated regression belongs to \(\mathcal F_s\), the equality also holds after its realized function is fixed and substituted into the population integral. It holds that function fixed and integrates the argument of a new observation under the original distribution \(P\). In this sense, the terms in (II) are centered empirical processes.

Because the true weights are bounded, at the first stage, \AppendixDisplay{124}{\begin{aligned}
&\mathbb E_P\!\left[
\{1-\mathbf1\{D=d_1\}w_1^*(X)\}^2
\left\{
\widehat\mu_0^{(d_1,\ldots,d_{J+1})}(X)
-\mu_0^{(d_1,\ldots,d_{J+1})}(X)
\right\}^2
\right]\\
&\quad\leq
C\left\|
\widehat\mu_0^{(d_1,\ldots,d_{J+1})}
-\mu_0^{(d_1,\ldots,d_{J+1})}
\right\|_{L_2(P_X)}^2
=o_p(1).
\end{aligned}
} Likewise, for \(s\geq2\), \AppendixDisplay{125}{\begin{aligned}
&\mathbb E_P\!\left[
\{\mathbf1\{D=d_{s-1}\}w_{s-1}^*
-\mathbf1\{D=d_s\}w_s^*\}^2
\left\{
\widehat\mu_{s-1}^{(d_s,\ldots,d_{J+1})}
-\mu_{s-1}^{(d_s,\ldots,d_{J+1})}
\right\}^2
\right]\\
&\quad\leq
C\left\|
\widehat\mu_{s-1}^{(d_s,\ldots,d_{J+1})}
-\mu_{s-1}^{(d_s,\ldots,d_{J+1})}
\right\|_{L_2(P_{\overline M_{s-1},X})}^2
=o_p(1).
\end{aligned}
} Here \(\mathbb E_P\) is a population integral with the realized estimated regression held fixed. The last convergence follows from the population \(L_2\) consistency in Condition (a).

\paragraph{Stochastic equicontinuity over the fixed Donsker class}\label{app-B-1-3-p1}

For a measurable function \(g\), write \(\mathbb G_Ng=N^{-1/2}\sum_{i=1}^{N}\{g(O_i)-\mathbb E_P[g(O)]\}\) for the centered empirical process. For an event \(A\) that need not be measurable, \(P^*(A)\) is the infimum of \(P(B)\) over measurable events \(B\supseteq A\). Thus \(Z_N=o_{P^*}(1)\) means \(P^*(|Z_N|>\epsilon)\to0\) for every \(\epsilon>0\). In the potentially nonmeasurable case, the weak limit below means convergence of outer expectations \(\mathbb E^*h(Z_N)\to\mathbb E h(Z)\) for every bounded Lipschitz \(h\), with \(\mathbb E^*\) the infimum of expectations of measurable majorants. \Citet[Theorem~3.13.4]{vanDerVaart-Wellner-WeakConvergence-2023} applies to a sample-selected member of a fixed \(P\)-Donsker class when class membership has probability tending to one and the population \(L_2(P)\) distance to its fixed true member vanishes. It gives the selected empirical-process difference as \(o_{P^*}(1)\). The fixed-class stochastic-equicontinuity modulus behind that result is, for every \(a>0\), \AppendixDisplay{126}{\lim_{\delta\downarrow0}\limsup_{N\to\infty}
P^*\!\left(
\sup_{\substack{g,h\in\mathcal G_s\\
\|g-h\|_{L_2(P)}\leq\delta}}
|\mathbb G_N(g-h)|>a
\right)
=0.
}

We show the event split that applies this result to a regression chosen from the sample. Fix \(s\geq2\). The true regression belongs to \(\mathcal F_s\). If the estimated regression also belongs to that class and the \(L_2(P)\) distance between the two regressions after true-weight multiplication is at most \(\delta\), the actual empirical-process term lies in the supremum above. By the preceding population \(L_2\) bound and population centering, \AppendixDisplay{127}{\begin{aligned}
&P^*\!\left(
\left|
\frac1{\sqrt N}\sum_{i=1}^{N}
\{\mathbf1\{D_i=d_{s-1}\}w_{s-1}^*
-\mathbf1\{D_i=d_s\}w_s^*\}
\left\{
\widehat\mu_{s-1}^{(d_s,\ldots,d_{J+1})}
-\mu_{s-1}^{(d_s,\ldots,d_{J+1})}
\right\}
\right|>a
\right)\\
&\quad\leq
P^*\!\left(
\widehat\mu_{s-1}^{(d_s,\ldots,d_{J+1})}
\notin\mathcal F_s
\right)
+P^*\!\left(
C\left\|
\widehat\mu_{s-1}^{(d_s,\ldots,d_{J+1})}
-\mu_{s-1}^{(d_s,\ldots,d_{J+1})}
\right\|_{L_2(P_{\overline M_{s-1},X})}
>\delta
\right)\\
&\qquad+
P^*\!\left(
\sup_{\substack{g,h\in\mathcal G_s\\
\|g-h\|_{L_2(P)}\leq\delta}}
|\mathbb G_N(g-h)|>a
\right).
\end{aligned}
} For fixed \(\delta>0\), the first probability equals the ordinary probability of nonmembership and tends to zero by high-probability class membership; the second tends to zero by population \(L_2\) consistency. Taking the upper limit as \(N\to\infty\) for the third probability and then letting \(\delta\downarrow0\) makes it vanish by stochastic equicontinuity. Hence \AppendixDisplay{128}{\frac1{\sqrt N}\sum_{i=1}^{N}
\{\mathbf1\{D_i=d_{s-1}\}w_{s-1}^*
-\mathbf1\{D_i=d_s\}w_s^*\}
\left\{
\widehat\mu_{s-1}^{(d_s,\ldots,d_{J+1})}
-\mu_{s-1}^{(d_s,\ldots,d_{J+1})}
\right\}
=o_{P^*}(1).
}

At the first stage, the coefficient is \(1-\mathbf1\{D_i=d_1\}w_1^*\), the population distribution is \(P_X\), and the true-weighted class is \(\mathcal G_1\). Population balance centers the summand. Substituting these first-stage quantities into the same fixed-class event split gives, for every \(a,\delta>0\), \AppendixDisplay{129}{\begin{aligned}
&P^*\!\left(
\left|\mathbb G_N\!\left[
\{1-\mathbf1\{D=d_1\}w_1^*\}
\{\widehat\mu_0^{(d_1,\ldots,d_{J+1})}
-\mu_0^{(d_1,\ldots,d_{J+1})}\}
\right]\right|>a\right)\\
&\quad\leq P^*\!\left(
\widehat\mu_0^{(d_1,\ldots,d_{J+1})}\notin\mathcal F_1\right)
+P^*\!\left(C\left\|
\widehat\mu_0^{(d_1,\ldots,d_{J+1})}
-\mu_0^{(d_1,\ldots,d_{J+1})}
\right\|_{L_2(P_X)}>\delta\right)\\
&\qquad+P^*\!\left(
\sup_{\substack{g,h\in\mathcal G_1\\
\|g-h\|_{L_2(P)}\leq\delta}}
|\mathbb G_N(g-h)|>a\right).
\end{aligned}
} Condition (a) makes the first two probabilities vanish for fixed \(\delta\), and fixed-class stochastic equicontinuity makes the limiting third probability vanish as \(\delta\downarrow0\). Hence \AppendixDisplay{130}{\frac1{\sqrt N}\sum_{i=1}^{N}
\{1-\mathbf1\{D_i=d_1\}w_1^*\}
\left\{
\widehat\mu_0^{(d_1,\ldots,d_{J+1})}
-\mu_0^{(d_1,\ldots,d_{J+1})}
\right\}
=o_{P^*}(1).
} When the displayed sample sums are measurable, these statements hold as ordinary \(o_p(1)\) statements. Otherwise, all subsequent remainders are interpreted in outer probability, and the distributional conclusion in the outer weak-convergence sense defined above.

\subsubsection{(III) Cauchy--Schwarz for weight and regression estimation errors}\label{app-B-1-4}

Here \(K_r\) is the dimension of the stage-\(r\) balancing basis and \(r_{w,r}>1\) is the weight-approximation exponent from Assumption 5.2, for \(r=1,\ldots,J+1\). The full-sample product condition used here is, for each \(s=1,\ldots,J+1\), \AppendixDisplay{131}{\begin{aligned}
&\sqrt N\left[\sum_{r=1}^{s}\left\{\sqrt{K_r/N}+K_r^{-r_{w,r}}\right\}\right]
\\
&\qquad\times
\left\{\frac1N\sum_{i=1}^{N}
\left|\widehat\mu_{s-1}^{(d_s,\ldots,d_{J+1})}(\overline M_{s-1,i},X_i)
-\mu_{s-1}^{(d_s,\ldots,d_{J+1})}(\overline M_{s-1,i},X_i)\right|^2\right\}^{1/2}
=o_p(1).
\end{aligned}
} At the first stage, samplewise Cauchy--Schwarz gives \AppendixDisplay{132}{\begin{aligned}
&\left|
\frac1{\sqrt N}\sum_{i=1}^{N}
\mathbf1\{D_i=d_1\}(\widehat w_{1,i}-w_1^*)
\left\{
\widehat\mu_0^{(d_1,\ldots,d_{J+1})}
-\mu_0^{(d_1,\ldots,d_{J+1})}
\right\}
\right|\\
&\quad\leq
\sqrt N
\left\{\frac1N\sum_{i=1}^{N}|\widehat w_{1,i}-w_1^*|^2\right\}^{1/2}
\left\{\frac1N\sum_{i=1}^{N}
\left|
\widehat\mu_0^{(d_1,\ldots,d_{J+1})}
-\mu_0^{(d_1,\ldots,d_{J+1})}
\right|^2
\right\}^{1/2}.
\end{aligned}
} By Theorem 5.2, the full-sample root-mean-square weight error divided by \(\sqrt{K_1/N}+K_1^{-r_{w,1}}\) is \(O_p(1)\). Condition (b) makes the product of the full-sample root-mean-square regression error and \(\sqrt N\{\sqrt{K_1/N}+K_1^{-r_{w,1}}\}\) equal to \(o_p(1)\). The bound is therefore \(O_p(1)o_p(1)=o_p(1)\).

For \(s\geq2\), separating current and preceding weight errors gives \AppendixDisplay{133}{\begin{aligned}
&\left|
\frac1{\sqrt N}\sum_{i=1}^{N}
\left\{
\mathbf1\{D_i=d_{s-1}\}(\widehat w_{s-1,i}-w_{s-1}^*)
-\mathbf1\{D_i=d_s\}(\widehat w_{s,i}-w_s^*)
\right\}
\left\{
\widehat\mu_{s-1}^{(d_s,\ldots,d_{J+1})}
-\mu_{s-1}^{(d_s,\ldots,d_{J+1})}
\right\}
\right|\\
&\quad\leq
\sqrt N
\left[
\left\{\frac1N\sum_{i=1}^{N}|\widehat w_{s-1,i}-w_{s-1}^*|^2\right\}^{1/2}
+
\left\{\frac1N\sum_{i=1}^{N}|\widehat w_{s,i}-w_s^*|^2\right\}^{1/2}
\right]\\
&\qquad\times
\left\{\frac1N\sum_{i=1}^{N}
\left|
\widehat\mu_{s-1}^{(d_s,\ldots,d_{J+1})}
-\mu_{s-1}^{(d_s,\ldots,d_{J+1})}
\right|^2
\right\}^{1/2}.
\end{aligned}
} Theorem 5.2 and the fact that the nonnegative sum through \(s-1\) is no greater than the sum through \(s\) imply \AppendixDisplay{134}{\frac{
\left\{\frac1N\sum_{i=1}^{N}|\widehat w_{s-1,i}-w_{s-1}^*|^2\right\}^{1/2}
+
\left\{\frac1N\sum_{i=1}^{N}|\widehat w_{s,i}-w_s^*|^2\right\}^{1/2}
}{
\displaystyle\sum_{r=1}^{s}
\left\{\sqrt{\frac{K_r}{N}}+K_r^{-r_{w,r}}\right\}
}
=O_p(1).
} The preceding bound is the product of this \(O_p(1)\) quantity and the left-hand side of Condition (b), and is thus \(o_p(1)\).

\subsubsection{Limiting distribution}\label{app-B-1-5}

The stochastic-equicontinuity argument gives Term (II) as \(o_p(N^{-1/2})\), and the same-sample product bounds give Term (III) the same rate. Term (I) has that rate by the conditional-variance argument. Substituting these three bounds into this section's exact identity gives \AppendixDisplay{135}{\widehat\theta_{\boldsymbol d}^{\mathrm{AW\text{-}BW}}
-\theta_{\boldsymbol d}
-\frac1N\sum_{i=1}^{N}\psi_{\boldsymbol d}(O_i)
=o_p(N^{-1/2}).
} The true score has mean zero and finite variance \(V_{\boldsymbol d}=\mathbb E[\psi_{\boldsymbol d}(O)^2]\), so the central limit theorem and Slutsky's lemma give \AppendixDisplay{136}{\sqrt N\left(
\widehat\theta_{\boldsymbol d}^{\mathrm{AW\text{-}BW}}
-\theta_{\boldsymbol d}
\right)
\xrightarrow{d}\mathcal N(0,V_{\boldsymbol d}).
} Theorem 2.1(iii) identifies the centered score \(\psi_{\boldsymbol d}=\phi_{\boldsymbol d}-\theta_{\boldsymbol d}\) as the efficient influence function. Hence, in an observed-data model where its variance is the efficiency bound, the variance \(V_{\boldsymbol d}=\mathbb E[\psi_{\boldsymbol d}^2]\) attained by this expansion equals that bound.

\section{Implementation details}\label{app-C}

This appendix describes how the weights of Sections 6 and 7 are computed. The first subsection gives the numerical solution of the stage problems of Algorithm 4.1 and the criterion under which exact balance is regarded as not attained at a stage, the second gives the selection of the tolerances that are then introduced at that stage, and the third gives the CBPS estimation method.

\subsection{Numerical solution of the balancing problems}\label{app-C-1}

This subsection describes when BW introduces tolerances at a stage in Sections 6 and 7. BW introduces the tolerances of Section 4.2 at stage \(s\), with the weights of the earlier stages held fixed, only when no weights attain exact balance at that stage, that is, when no nonnegative weights on the \(D=d_s\) group that sum to \(N\) make the weighted mean of \(B_s(\overline M_{s-1,i},X_i)\) equal to its target in Algorithm 4.1, with \(B_1(X_i)\) in place of \(B_s(\overline M_{s-1,i},X_i)\) at \(s=1\). Such weights exist exactly when the target lies in the convex hull of \(\{B_s(\overline M_{s-1,i},X_i):D_i=d_s\}\).

BW finds these stages numerically. It minimizes the dual objective of Theorem 5.1 with the entropy dispersion \(\ell(w)=w\log w-w\) and \(\varepsilon_s=0\) by BFGS followed by Newton steps, and it regards exact balance as attained when the resulting weights are finite and sum to \(N\) and the KKT, constraint, and duality-gap residuals are at most \(2\times10^{-7}\), the latter two relative to the size of the targets and of the objective. In Sections 6 and 7, a linear program confirms that the target lies outside this convex hull at every stage where these checks fail.

Algorithm C.1 in the next subsection solves each candidate problem with positive tolerances in the same way. The dual of Theorem 5.1 then contains the penalty \(\sum_{k=2}^{K_s}\varepsilon_{s,k}|\gamma_{s,k}|\) and is minimized by L-BFGS-B followed by Newton steps, and the same checks apply with the KKT conditions of the penalized dual and with each constraint residual measured beyond its tolerance. A candidate that fails the checks is excluded from the selection. If every positive candidate fails at a stage, BW has no weights at that stage and returns no estimate, which does not occur in Sections 6 and 7.

\subsection{Selection of balance tolerances}\label{app-C-2}

This subsection gives the selection of the tolerances that Section 4.2 introduces when exact balance cannot be attained. For comparability across basis functions, let \(\overline b_{1,k}=N^{-1}\sum_{i=1}^{N}b_{1,k}(X_i)\) and \(\overline b_{s,k}=N^{-1}\sum_{i=1}^{N}b_{s,k}(\overline M_{s-1,i},X_i)\) for \(s=2,\ldots,J+1\). Define the corresponding empirical scales by \(\widehat\sigma_{1,k}^2=N^{-1}\sum_{i=1}^{N}\{b_{1,k}(X_i)-\overline b_{1,k}\}^2\) and \(\widehat\sigma_{s,k}^2=N^{-1}\sum_{i=1}^{N}\{b_{s,k}(\overline M_{s-1,i},X_i)-\overline b_{s,k}\}^2\) for \(s=2,\ldots,J+1\) and \(k=2,\ldots,K_s\). The constant component, \(k=1\), is balanced exactly and does not enter the standardized criterion below. A convenient low-dimensional parameterization uses one standardized tolerance \(\lambda_s\geq0\) at each stage, with \(\varepsilon_{s,1}(\lambda_s)=0\) and \(\varepsilon_{s,k}(\lambda_s)=\lambda_s\widehat\sigma_{s,k}\) for \(k=2,\ldots,K_s\). The scalar \(\lambda_s\) has the interpretation of a maximum standardized moment discrepancy permitted at stage \(s\). Let \(\Lambda_s\) be a finite grid of candidate values containing zero. Because an optimal grid is difficult to determine in general, a fixed finite grid is a practical choice. Sections 6 and 7 use the fixed grid \(\Lambda_s=\{0,0.005,0.01,0.025,0.05,0.10\}\).

\begin{manuscriptalgorithm}[10]{Algorithm C.1 Stagewise bootstrap selection of balance tolerances}

\begin{enumerate}
\item
  \textbf{Inputs.} Specify the candidate grids \(\Lambda_1,\ldots,\Lambda_{J+1}\), the number of bootstrap replications \(R\), and the stage-specific bases and loss used in Algorithm 4.1. Generate bootstrap count vectors \((C_1^{(r)},\ldots,C_N^{(r)})\), \(r=1,\ldots,R\), independently from the multinomial distribution with sample size \(N\) and cell probabilities \((1/N,\ldots,1/N)\). Here, \(C_i^{(r)}\) denotes the number of times observation \(i\) is selected in bootstrap sample \(r\). Using these counts is equivalent to drawing \(N\) observations with replacement from the original sample. Use the same bootstrap count vectors for all candidates at a given stage.
\item
  \textbf{Stage 1.} For each \(\lambda\in\Lambda_1\):

  \begin{enumerate}
  \item
    Set \(\varepsilon_{1,1}=0\) and \(\varepsilon_{1,k}=\lambda\widehat\sigma_{1,k}\) for \(k\geq2\).
  \item
    Solve the first-stage problem in Algorithm 4.1 on the original sample and denote the candidate weights by \(\widehat w_{1,i}(\lambda)\).
  \item
    For each bootstrap replication \(r\), hold the candidate weights fixed. For each nonconstant component \(k=2,\ldots,K_1\), compute the bootstrap imbalance, defined as the self-normalized weighted mean of \(b_{1,k}\) in replication \(r\) minus the original-sample target for that component. \AppendixDisplay{137}{\begin{aligned}
    \widehat\Delta_{1,k}^{(r)}(\lambda)
    &=
    \frac{\sum_{i=1}^{N}C_i^{(r)}\mathbf1\{D_i=d_1\}\widehat w_{1,i}(\lambda)b_{1,k}(X_i)}
    {\sum_{i=1}^{N}C_i^{(r)}\mathbf1\{D_i=d_1\}\widehat w_{1,i}(\lambda)}
    -\frac1N\sum_{i=1}^{N}b_{1,k}(X_i).
    \end{aligned}
    }
  \item
    Compute the average standardized absolute bootstrap imbalance \AppendixDisplay{138}{\widehat{\mathcal C}_1(\lambda)
    =
    \frac{1}{R(K_1-1)}
    \sum_{r=1}^{R}
    \sum_{k=2}^{K_1}
    \frac{
    \bigl|\widehat\Delta_{1,k}^{(r)}(\lambda)\bigr|
    }{
    \widehat\sigma_{1,k}
    }.
    }
  \end{enumerate}

  Select \(\widehat\lambda_1\in\operatorname*{arg\,min}_{\lambda\in\Lambda_1}\widehat{\mathcal C}_1(\lambda)\). Retain \(\widehat w_{1,i}=\widehat w_{1,i}(\widehat\lambda_1)\).
\item
  \textbf{Stage \(s\).} For each \(s=2,\ldots,J+1\), hold all previously selected weights fixed. For each \(\lambda\in\Lambda_s\):

  \begin{enumerate}
  \item
    Set \(\varepsilon_{s,1}=0\) and \(\varepsilon_{s,k}=\lambda\widehat\sigma_{s,k}\) for \(k\geq2\).
  \item
    Solve the stage-\(s\) problem in Algorithm 4.1 on the original sample using the fixed preceding weights. Denote the candidate weights by \(\widehat w_{s,i}(\lambda)\).
  \item
    Holding both the candidate weights and the preceding target weights fixed, compute \AppendixDisplay{139}{\begin{aligned}
    \widehat\Delta_{s,k}^{(r)}(\lambda)
    &=
    \frac{\sum_{i=1}^{N}C_i^{(r)}\mathbf1\{D_i=d_s\}\widehat w_{s,i}(\lambda)b_{s,k}(\overline M_{s-1,i},X_i)}
    {\sum_{i=1}^{N}C_i^{(r)}\mathbf1\{D_i=d_s\}\widehat w_{s,i}(\lambda)}\\
    &\quad-\frac1N\sum_{i:D_i=d_{s-1}}
    \widehat w_{s-1,i}
    b_{s,k}(\overline M_{s-1,i},X_i),
    \end{aligned}
    } for \(r=1,\ldots,R\) and \(k=2,\ldots,K_s\).
  \item
    Compute \AppendixDisplay{140}{\widehat{\mathcal C}_s(\lambda)
    =
    \frac{1}{R(K_s-1)}
    \sum_{r=1}^{R}
    \sum_{k=2}^{K_s}
    \frac{
    \bigl|\widehat\Delta_{s,k}^{(r)}(\lambda)\bigr|
    }{
    \widehat\sigma_{s,k}
    },
    }
  \end{enumerate}

  Select \(\widehat\lambda_s\in\operatorname*{arg\,min}_{\lambda\in\Lambda_s}\widehat{\mathcal C}_s(\lambda)\). Retain \(\widehat w_{s,i}=\widehat w_{s,i}(\widehat\lambda_s)\) and proceed to stage \(s+1\).
\item
  \textbf{Output.} Return \((\widehat\lambda_s,\widehat\varepsilon_s,\widehat w_s)\) for all \(s=1,\ldots,J+1\).
\end{enumerate}

\end{manuscriptalgorithm}

\par\medskip

The weights are estimated once on the original sample for each candidate and are not refitted within individual bootstrap replications. If the fixed candidate weights were evaluated with the original empirical frequencies, the weighted moments would satisfy the constraints imposed on that same sample by construction, with zero discrepancy under exact balance and discrepancies bounded by the tolerances under approximate balance. Such an evaluation could not assess how stable the balance is beyond the sample used to construct the weights. The bootstrap is therefore used to perturb the empirical frequency with which the observed histories enter the weighted moment.

Once \(\widehat\lambda_s\) and \(\widehat w_s\) have been selected, they are fixed before tuning the next stage. This preserves the information restriction and mirrors the recursive construction of the balancing targets. Such stagewise selection is greedy and need not minimize a single global criterion over \((\lambda_1,\ldots,\lambda_{J+1})\). BW applies the algorithm only at a stage where exact balance is regarded as not attained by the criterion of the preceding subsection, with \(\Lambda_s\) replaced by its positive values and the weights of the earlier stages held fixed, and sets \(\varepsilon_s=0\) at every other stage.

\subsection{CBPS estimation method}\label{app-C-3}

We adapt the covariate balancing propensity score (CBPS) approach of \citet{Imai-CovariateBalancingScore-2014v} to the sequential weighting representation in Section 3.1 and compute each stage with the R package CBPS \citep{R-CBPS}. The procedure uses the same stage-specific bases as Algorithm 4.1.

Fix a treatment-input vector \(\boldsymbol d=(d_{J+1},d_J,\ldots,d_1)\). For \(s=1,\ldots,J+1\), write \(B_{s,i}=B_s(\overline M_{s-1,i},X_i)\in\mathbb R^{K_s}\), where \(B_{1,i}=B_1(X_i)\) and each basis includes a constant. Specify the logistic models \AppendixDisplay{141}{\pi_{s-1}(1;\overline M_{s-1,i},X_i;\alpha_{s-1})
=
\frac{\exp(B_{s,i}^{\top}\alpha_{s-1})}
{1+\exp(B_{s,i}^{\top}\alpha_{s-1})},
\qquad s=1,\ldots,J+1,
} with \(\pi_{s-1}(0;\cdot;\alpha_{s-1})=1-\pi_{s-1}(1;\cdot;\alpha_{s-1})\). In this appendix, the final argument of each treatment probability denotes its model coefficient. Dependence of the fitted coefficients and weights on the fixed vector \(\boldsymbol d\) is suppressed.

At each stage requiring estimation, let \(\overline g_s(\alpha_{s-1})\) denote the stacked sample moments defined below. The first block contains the logistic likelihood score, and the second contains the balance discrepancies. We estimate the coefficient by minimizing \AppendixDisplay{142}{\widehat\alpha_{s-1}
\in
\operatorname*{arg\,min}_{\alpha_{s-1}\in\mathbb R^{K_s}}
\overline g_s(\alpha_{s-1})^{\top}
\widehat A_s
\overline g_s(\alpha_{s-1}).
} The weighting matrix \(\widehat A_s\) is that of the two-step option of the package. It is the generalized inverse of a matrix that the package builds from the covariances that the logistic model implies for the score and balance terms of each observation, averaged with the normalized sample weights, at a rescaled maximum likelihood estimate of the unweighted logistic regression, and it is held fixed during the minimization. Because the system is overidentified, the fitted sample moments need not all equal zero.

\begin{manuscriptalgorithm}[10]{Algorithm C.2 Sequential covariate balancing propensity scores}

\begin{enumerate}
\item
  \textbf{First stage.} Define \AppendixDisplay{143}{\overline g_1(\alpha_0)
  =
  \frac1N\sum_{i=1}^{N}
  \begin{pmatrix}
  \{D_i-\pi_0(1;X_i;\alpha_0)\}B_{1,i}
  \\[2mm]
  \left\{
  \dfrac{D_i}
  {\pi_0(1;X_i;\alpha_0)}
  -
  \dfrac{1-D_i}
  {\pi_0(0;X_i;\alpha_0)}
  \right\}B_{1,i}
  \end{pmatrix}.
  } Its balance block compares the inverse-probability-weighted treated and control groups and does not depend on \(d_1\), so one estimate \(\widehat\alpha_0\) gives the first-stage weights of both groups. Estimate \(\widehat\alpha_0\) by GMM and set \AppendixDisplay{144}{\widetilde w_{1,i}
  =
  \frac{1}{\pi_0(d_1;X_i;\widehat\alpha_0)},
  \qquad i=1,\ldots,N.
  }
\item
  \textbf{Subsequent stages.} For each \(s=2,\ldots,J+1\), hold the preceding unnormalized weights \(\{\widetilde w_{s-1,i}\}_{i=1}^{N}\) fixed.

  If \(d_s=d_{s-1}\), set \(\widetilde w_{s,i}=\widetilde w_{s-1,i}\) for all \(i\) and proceed to the next stage. The probability ratio is one, so no additional propensity-score estimation is required.

  If \(d_s\ne d_{s-1}\), define the candidate weights \AppendixDisplay{145}{\widetilde w_{s,i}(\alpha_{s-1})
  =
  \widetilde w_{s-1,i}
  \frac{
  \pi_{s-1}(d_{s-1};\overline M_{s-1,i},X_i;\alpha_{s-1})
  }{
  \pi_{s-1}(d_s;\overline M_{s-1,i},X_i;\alpha_{s-1})
  }
  } and the sample moment vector \AppendixDisplay{146}{\overline g_s(\alpha_{s-1})
  =
  \sum_{i=1}^{N}
  \begin{pmatrix}
  \dfrac{\widetilde w_{s-1,i}}{\sum_{\ell=1}^{N}\widetilde w_{s-1,\ell}}
  \{\mathbf1\{D_i=d_{s-1}\}-\pi_{s-1}(d_{s-1};\overline M_{s-1,i},X_i;\alpha_{s-1})\}B_{s,i}
  \\[2mm]
  \dfrac{
  \mathbf1\{D_i=d_{s-1}\}\widetilde w_{s-1,i}
  -
  \mathbf1\{D_i=d_s\}\widetilde w_{s,i}(\alpha_{s-1})
  }{
  \sum_{\ell:D_\ell=d_{s-1}}\widetilde w_{s-1,\ell}
  }B_{s,i}
  \end{pmatrix}.
  } The first block is the logistic score for membership in the \(D=d_{s-1}\) group, weighted by the preceding weights, and the second is the difference between the preceding weighted group and the \(D=d_s\) group under the candidate weights, divided by the total preceding weight of the \(D=d_{s-1}\) group. Estimate \(\widehat\alpha_{s-1}\) by GMM and set \(\widetilde w_{s,i}=\widetilde w_{s,i}(\widehat\alpha_{s-1})\) for all \(i=1,\ldots,N\).

  Throughout the recursion, the preceding weight function is evaluated at every observation, including observations outside the preceding target group. These weights are neither normalized nor set to zero outside that group before subsequent stages are estimated.
\item
  \textbf{Output.} After completing the recursion, normalize the weights within each target group and set them to zero outside that group: \AppendixDisplay{147}{\widehat w_{s,i}^{\mathrm{CBPS}}
  =
  \frac{
  N\mathbf1\{D_i=d_s\}\widetilde w_{s,i}
  }{
  \sum_{\ell:D_\ell=d_s}\widetilde w_{s,\ell}
  },
  \qquad
  s=1,\ldots,J+1.
  } Use these weights in the augmented estimator of Section 3.1. This final normalization ensures that the weights sum to \(N\) at each stage; including a constant among the GMM moments does not by itself ensure exact normalization.
\end{enumerate}

\end{manuscriptalgorithm}

\par\medskip

\textbf{Numerical implementation.} Sections 6 and 7 compute every stage with the function \texttt{CBPS} of the package, version 0.24, using a binary factor treatment and the default settings \texttt{method\ =\ "over"}, \texttt{twostep\ =\ TRUE}, and \texttt{iterations\ =\ 1000}. The first stage uses the option \texttt{ATT\ =\ 0}, which gives the moments in step 1. A later stage with \(d_s\neq d_{s-1}\) uses the option that takes the \(D=d_{s-1}\) group as the treated group, \texttt{ATT\ =\ 1} if \(d_{s-1}=1\) and \texttt{ATT\ =\ 2} if \(d_{s-1}=0\), and passes \(\widetilde w_{s-1,i}\) as \texttt{sample.weights}, which gives the moments in step 2. The package replaces \(B_s\) by a linear transformation that has orthonormal columns in the sample, which leaves the GMM criterion unchanged when the matrix inverted for \(\widehat A_s\) is nonsingular, and keeps the fitted probabilities within \([10^{-6},1-10^{-6}]\). It runs BFGS from the rescaled maximum likelihood estimate at which it computes \(\widehat A_s\) and from the result of minimizing a criterion based on the balance block alone, and it returns the solution with the smaller criterion value.

\section{Formulas of the balance and weight diagnostics}\label{app-D}

This appendix gives the formulas of the diagnostics in Section 7.3. For a functional \(\theta_{\boldsymbol d}\), let \(\widehat w_{s,i}\) be the weight of observation \(i\) at stage \(s\), where a stage whose treatment input equals that of the preceding stage keeps the preceding weights, and let \(n_d=\sum_{i=1}^N\mathbf1\{D_i=d\}\) be the size of the \(D=d\) group. Write \(b_{s,k,i}=b_{s,k}(\overline M_{s-1,i},X_i)\). For weights \(w\), the weighted and unweighted means of \(b_{s,k}\) in the \(D=d\) group are \(\bar b_{s,k}^{(d;w)}=(\sum_{i:D_i=d}w_ib_{s,k,i})/(\sum_{i:D_i=d}w_i)\) and \(\bar b_{s,k}^{(d)}=n_d^{-1}\sum_{i:D_i=d}b_{s,k,i}\), and its unweighted sample standard deviation is \(\widehat\sigma_{s,k}^{(d)}=\{(n_d-1)^{-1}\sum_{i:D_i=d}(b_{s,k,i}-\bar b_{s,k}^{(d)})^2\}^{1/2}\). For \(k=2,\ldots,K_s\), the TASMD is \(|\bar b_{1,k}^{(d_1;\widehat w_1)}-N^{-1}\sum_{i=1}^Nb_{1,k,i}|/\widehat\sigma_{1,k}^{(d_1)}\) at the first stage and \(|\bar b_{s,k}^{(d_s;\widehat w_s)}-\bar b_{s,k}^{(d_{s-1};\widehat w_{s-1})}|/\widehat\sigma_{s,k}^{(d_s)}\) at a later stage. The share of observation \(i\) in the total weight of the weighted group is \(q_{s,i}=\widehat w_{s,i}/(\sum_{\ell:D_\ell=d_s}\widehat w_{s,\ell})\) for \(D_i=d_s\), so that \(\sum_{i:D_i=d_s}q_{s,i}=1\). Let \(q_{s,(1)}\geq\cdots\geq q_{s,(n_{d_s})}\) be these shares sorted in descending order. The two concentration measures are \(\operatorname{MaxShare}_s=\max_{i:D_i=d_s}q_{s,i}\) and \(\operatorname{Top10Share}_s=\sum_{j=1}^{\lceil0.1n_{d_s}\rceil}q_{s,(j)}\).

\section{Additional simulations}\label{app-E}

\setcounter{table}{0}\renewcommand{\thetable}{E.\arabic{table}}

This appendix repeats the simulation of Section 6 in two other settings. Appendix E.1 estimates the backward regressions by the cross-validated elastic net of Section 7.2, and Appendix E.2 changes the strength of overlap in the treatment assignment.

\subsection{Elastic-net regressions}\label{app-E-1}

This subsection repeats the simulation of Section 6 with the backward regressions estimated by the cross-validated elastic net of Section 7.2. We first describe the regressions and then report the results.

The data-generating process, the bases \(B_1\), \(B_2\), and \(B_3\), the contrast \(\theta_{(1,1,1)}-\theta_{(1,0,1)}\), the sample sizes, and the 3,000 repetitions at each sample size are those of Section 6, and the PS, CBPS, and BW weights in each repetition are those of Table 6.1. Only the backward regressions change. Each least-squares fit in the backward chain of Section 6.2 is replaced by the regression of Section 7.2. In the training group, this regression compares unpenalized least squares with the elastic net on the nonconstant basis terms for ten equally spaced mixing parameters from 0 to 1, each with a ten-value penalty path, by out-of-fold mean squared error on a common 3-fold split, and fits the selected model to the whole training group. The outcome is thus regressed on \(B_3\) in the \(D=d_3\) group, its predictions on \(B_2\) in the \(D=d_2\) group, and those predictions on \(B_1\) in the \(D=d_1\) group. As in Section 6.2, every stage has its own regression even when consecutive treatment inputs coincide, and the two functionals share the outcome regression on \(B_3\) in the \(D=1\) group. Cross-validation selects least squares in 74\% to 88\% of the first-stage regressions and a penalized fit in more than 90\% of the second- and third-stage regressions, most often the lasso.

The three weighting methods combine their weights with these predictions in the augmented weighting estimator \(\widehat\theta_{\boldsymbol d}^{\mathrm{AW}}\) of Section 3.1, and their standard errors are the plug-in estimator of Theorem 5.4 applied to the contrast, as in Section 6.2. Regression-only estimation reports \(N^{-1}\sum_{i=1}^{N}\{\widehat\mu_0^{(1,1,1)}(Z_i)-\widehat\mu_0^{(1,0,1)}(Z_i)\}\) with the elastic-net predictions. Its sandwich standard error in Section 6.2 comes from the stacked OLS estimating equations, so for the elastic net we report its bias, variance, and RMSE only.

\subsubsection{Results}\label{app-E-1-1}

Table E.1 reports the same statistics as Table 6.1. Every method returns an estimate in every repetition.

{
\small
\setlength{\LTcapwidth}{\linewidth}
\begin{longtable}[]{@{}
  >{\raggedleft\arraybackslash}p{(\linewidth - 12\tabcolsep) * \real{0.0800}}
  >{\raggedright\arraybackslash}p{(\linewidth - 12\tabcolsep) * \real{0.3000}}
  >{\raggedleft\arraybackslash}p{(\linewidth - 12\tabcolsep) * \real{0.1200}}
  >{\raggedleft\arraybackslash}p{(\linewidth - 12\tabcolsep) * \real{0.1200}}
  >{\raggedleft\arraybackslash}p{(\linewidth - 12\tabcolsep) * \real{0.1200}}
  >{\raggedleft\arraybackslash}p{(\linewidth - 12\tabcolsep) * \real{0.1300}}
  >{\raggedleft\arraybackslash}p{(\linewidth - 12\tabcolsep) * \real{0.1300}}@{}}
\toprule\noalign{}
\begin{minipage}[b]{\linewidth}\raggedleft
\(N\)
\end{minipage} & \begin{minipage}[b]{\linewidth}\raggedright
Method
\end{minipage} & \begin{minipage}[b]{\linewidth}\raggedleft
Bias
\end{minipage} & \begin{minipage}[b]{\linewidth}\raggedleft
Variance
\end{minipage} & \begin{minipage}[b]{\linewidth}\raggedleft
RMSE
\end{minipage} & \begin{minipage}[b]{\linewidth}\raggedleft
Mean SE
\end{minipage} & \begin{minipage}[b]{\linewidth}\raggedleft
Cov.
\end{minipage} \\
\midrule\noalign{}
\endfirsthead
\toprule\noalign{}
\begin{minipage}[b]{\linewidth}\raggedleft
\(N\)
\end{minipage} & \begin{minipage}[b]{\linewidth}\raggedright
Method
\end{minipage} & \begin{minipage}[b]{\linewidth}\raggedleft
Bias
\end{minipage} & \begin{minipage}[b]{\linewidth}\raggedleft
Variance
\end{minipage} & \begin{minipage}[b]{\linewidth}\raggedleft
RMSE
\end{minipage} & \begin{minipage}[b]{\linewidth}\raggedleft
Mean SE
\end{minipage} & \begin{minipage}[b]{\linewidth}\raggedleft
Cov.
\end{minipage} \\
\midrule\noalign{}
\endhead
1,000 & PS & 0.193 & 0.089 & 0.355 & 0.263 & 81.6\% \\*
1,000 & CBPS & 0.184 & 0.038 & 0.269 & 0.230 & 86.5\% \\*
1,000 & BW & 0.155 & 0.036 & 0.245 & 0.250 & 91.6\% \\*
1,000 & Regression-only (elastic net) & 0.332 & 0.042 & 0.391 & --- & --- \\
\bottomrule\noalign{}
\noalign{\vskip\abovecaptionskip}
\caption{Estimates of \(\theta_{(1,1,1)}-\theta_{(1,0,1)}\) with the elastic-net backward regressions of Section 7.2. The true value is \(-1.133\). The samples, the weights, and the definitions of the columns are those of Table 6.1, and BW introduces the tolerances of Section 4.2 in the same 256 repetitions at \(N=265\). Regression-only (elastic net) has no standard error, and a dash marks its Mean SE and Cov.}\label{table-e1}\tabularnewline
\endlastfoot
265 & PS & 0.302 & 0.255 & 0.588 & 0.414 & 79.6\% \\*
265 & CBPS & 0.326 & 0.158 & 0.514 & 0.380 & 81.5\% \\*
265 & BW & 0.325 & 0.201 & 0.554 & 0.447 & 84.5\% \\*
265 & Regression-only (elastic net) & 0.291 & 0.171 & 0.506 & --- & --- \\
500 & PS & 0.238 & 0.151 & 0.455 & 0.335 & 81.6\% \\*
500 & CBPS & 0.246 & 0.071 & 0.363 & 0.294 & 85.4\% \\*
500 & BW & 0.214 & 0.079 & 0.353 & 0.332 & 90.0\% \\*
500 & Regression-only (elastic net) & 0.301 & 0.085 & 0.419 & --- & --- \\
\end{longtable}
}

At every sample size, BW has the highest coverage of the three weighting methods, 84.5\%, 90.0\%, and 91.6\% at \(N=265\), \(500\), and \(1{,}000\), compared with 79.6\%, 81.6\%, and 81.6\% for PS and 81.5\%, 85.4\%, and 86.5\% for CBPS. At \(N=500\) and \(1{,}000\), BW also has the smallest bias and RMSE of the three. At \(N=265\), its bias of 0.325 is close to the 0.326 of CBPS and larger than the 0.302 of PS, and its RMSE of 0.554 lies between the 0.514 of CBPS and the 0.588 of PS.

We next compare the methods at \(N=265\) separately on the repetitions with and without exact balance, as in Section 6.3. On the 2,744 repetitions with exact balance at every stage, BW has a bias of 0.313 and an RMSE of 0.533, compared with 0.295 and 0.566 for PS and 0.316 and 0.495 for CBPS, and its coverage is 85.7\%, compared with 80.4\% for PS and 82.6\% for CBPS. In the remaining 256 repetitions, in which BW introduces the tolerances of Section 4.2, BW has a bias of 0.451, compared with 0.375 for PS and 0.434 for CBPS, and its RMSE of 0.742 lies between the 0.680 of CBPS and the 0.781 of PS.

With the elastic-net regressions, BW therefore has the highest coverage of the three weighting methods at every sample size, and its bias and RMSE are comparable to those of PS and CBPS at \(N=265\) and the smallest of the three at \(N=500\) and \(1{,}000\).

\subsection{Strength of overlap}\label{app-E-2}

This subsection changes only the treatment probability of Section 6.1. The treatment is \(D=\mathbf 1\{U_D\leq\operatorname{expit}(\kappa(-Z_1+0.5Z_2-0.25Z_3-0.1Z_4))\}\) for a factor \(\kappa>0\), and Section 6 has \(\kappa=1\). We consider \(\kappa=0.5\), which strengthens overlap, and \(\kappa=1.5\), which weakens it.

The covariates, the potential mediators and outcome, the bases, the three weighting methods, the OLS backward regressions, the standard errors, the sample sizes, and the 3,000 repetitions at each sample size are those of Section 6. The treatment probability does not enter \(\theta_{(1,1,1)}-\theta_{(1,0,1)}\), so the true value remains \(-1.133\). The index \(-Z_1+0.5Z_2-0.25Z_3-0.1Z_4\) is normal with mean zero and standard deviation 1.15, so the 1st and 99th percentiles of the treatment probability are 0.21 and 0.79 for \(\kappa=0.5\), 0.06 and 0.94 for \(\kappa=1\), and 0.02 and 0.98 for \(\kappa=1.5\). In each design, half of the observations are treated in expectation.

{
\small
\setlength{\LTcapwidth}{\linewidth}
\begin{longtable}[]{@{}
  >{\raggedright\arraybackslash}p{(\linewidth - 14\tabcolsep) * \real{0.1200}}
  >{\raggedleft\arraybackslash}p{(\linewidth - 14\tabcolsep) * \real{0.0800}}
  >{\raggedright\arraybackslash}p{(\linewidth - 14\tabcolsep) * \real{0.2000}}
  >{\raggedleft\arraybackslash}p{(\linewidth - 14\tabcolsep) * \real{0.1200}}
  >{\raggedleft\arraybackslash}p{(\linewidth - 14\tabcolsep) * \real{0.1200}}
  >{\raggedleft\arraybackslash}p{(\linewidth - 14\tabcolsep) * \real{0.1200}}
  >{\raggedleft\arraybackslash}p{(\linewidth - 14\tabcolsep) * \real{0.1200}}
  >{\raggedleft\arraybackslash}p{(\linewidth - 14\tabcolsep) * \real{0.1200}}@{}}
\toprule\noalign{}
\begin{minipage}[b]{\linewidth}\raggedright
Overlap
\end{minipage} & \begin{minipage}[b]{\linewidth}\raggedleft
\(N\)
\end{minipage} & \begin{minipage}[b]{\linewidth}\raggedright
Method
\end{minipage} & \begin{minipage}[b]{\linewidth}\raggedleft
Bias
\end{minipage} & \begin{minipage}[b]{\linewidth}\raggedleft
Variance
\end{minipage} & \begin{minipage}[b]{\linewidth}\raggedleft
RMSE
\end{minipage} & \begin{minipage}[b]{\linewidth}\raggedleft
Mean SE
\end{minipage} & \begin{minipage}[b]{\linewidth}\raggedleft
Cov.
\end{minipage} \\
\midrule\noalign{}
\endfirsthead
\toprule\noalign{}
\begin{minipage}[b]{\linewidth}\raggedright
Overlap
\end{minipage} & \begin{minipage}[b]{\linewidth}\raggedleft
\(N\)
\end{minipage} & \begin{minipage}[b]{\linewidth}\raggedright
Method
\end{minipage} & \begin{minipage}[b]{\linewidth}\raggedleft
Bias
\end{minipage} & \begin{minipage}[b]{\linewidth}\raggedleft
Variance
\end{minipage} & \begin{minipage}[b]{\linewidth}\raggedleft
RMSE
\end{minipage} & \begin{minipage}[b]{\linewidth}\raggedleft
Mean SE
\end{minipage} & \begin{minipage}[b]{\linewidth}\raggedleft
Cov.
\end{minipage} \\
\midrule\noalign{}
\endhead
\(\kappa=1.5\) & 1,000 & PS & 0.244 & 0.158 & 0.467 & 0.341 & 81.3\% \\*
\(\kappa=1.5\) & 1,000 & CBPS & 0.207 & 0.054 & 0.312 & 0.278 & 88.8\% \\*
\(\kappa=1.5\) & 1,000 & BW & 0.174 & 0.050 & 0.282 & 0.312 & 94.3\% \\*
\(\kappa=1.5\) & 1,000 & Regression-only (OLS) & 0.502 & 0.061 & 0.560 & 0.233 & 42.3\% \\
\bottomrule\noalign{}
\noalign{\vskip\abovecaptionskip}
\caption{Estimates of \(\theta_{(1,1,1)}-\theta_{(1,0,1)}\) with \(\kappa=0.5\) and \(\kappa=1.5\) in the treatment probability. Only the treatment probability of Section 6.1 changes, and the true value is \(-1.133\). The definitions of the columns are those of Table 6.1. BW introduces the tolerances of Section 4.2 in 60 repetitions at \(N=265\) with \(\kappa=0.5\), and in 1,150 repetitions at \(N=265\) and 16 at \(N=500\) with \(\kappa=1.5\). In 3 repetitions with \(\kappa=1.5\) at \(N=265\), no value of \(\lambda_s\) in the grid gives BW weights at one stage, so the bias, variance, RMSE, and Mean SE of BW use the other 2,997 repetitions, and its Cov. counts these 3 repetitions as not covering the true value. The other methods return an estimate in every repetition.}\label{table-e2}\tabularnewline
\endlastfoot
\(\kappa=0.5\) & 265 & PS & 0.308 & 0.182 & 0.526 & 0.377 & 80.3\% \\*
\(\kappa=0.5\) & 265 & CBPS & 0.345 & 0.145 & 0.514 & 0.361 & 78.8\% \\*
\(\kappa=0.5\) & 265 & BW & 0.278 & 0.143 & 0.469 & 0.412 & 87.0\% \\*
\(\kappa=0.5\) & 265 & Regression-only (OLS) & 0.368 & 0.183 & 0.564 & 0.372 & 76.1\% \\
\(\kappa=0.5\) & 500 & PS & 0.217 & 0.094 & 0.375 & 0.290 & 83.1\% \\*
\(\kappa=0.5\) & 500 & CBPS & 0.242 & 0.067 & 0.355 & 0.277 & 83.6\% \\*
\(\kappa=0.5\) & 500 & BW & 0.180 & 0.064 & 0.311 & 0.299 & 90.8\% \\*
\(\kappa=0.5\) & 500 & Regression-only (OLS) & 0.303 & 0.090 & 0.426 & 0.270 & 74.8\% \\
\(\kappa=0.5\) & 1,000 & PS & 0.167 & 0.047 & 0.274 & 0.215 & 84.0\% \\*
\(\kappa=0.5\) & 1,000 & CBPS & 0.175 & 0.030 & 0.246 & 0.208 & 86.2\% \\*
\(\kappa=0.5\) & 1,000 & BW & 0.131 & 0.029 & 0.215 & 0.217 & 92.0\% \\*
\(\kappa=0.5\) & 1,000 & Regression-only (OLS) & 0.279 & 0.041 & 0.344 & 0.191 & 66.5\% \\
\(\kappa=1.5\) & 265 & PS & 0.416 & 0.416 & 0.767 & 0.484 & 75.7\% \\*
\(\kappa=1.5\) & 265 & CBPS & 0.400 & 0.280 & 0.663 & 0.469 & 81.0\% \\*
\(\kappa=1.5\) & 265 & BW & 0.359 & 0.314 & 0.666 & 0.530 & 86.2\% \\*
\(\kappa=1.5\) & 265 & Regression-only (OLS) & 0.562 & 0.287 & 0.777 & 0.459 & 72.0\% \\
\(\kappa=1.5\) & 500 & PS & 0.318 & 0.211 & 0.559 & 0.400 & 79.9\% \\*
\(\kappa=1.5\) & 500 & CBPS & 0.275 & 0.107 & 0.428 & 0.357 & 86.1\% \\*
\(\kappa=1.5\) & 500 & BW & 0.240 & 0.116 & 0.417 & 0.416 & 92.9\% \\*
\(\kappa=1.5\) & 500 & Regression-only (OLS) & 0.521 & 0.127 & 0.631 & 0.327 & 62.0\% \\
\end{longtable}
}

Table E.2 reports the same statistics as Table 6.1. At \(N=500\) and \(1{,}000\), BW has the smallest bias and RMSE and the highest coverage among the four methods for both values of \(\kappa\), as in Table 6.1. At \(N=265\), it also does so for \(\kappa=0.5\), and for \(\kappa=1.5\) it has the smallest bias and the highest coverage, while its RMSE of 0.666 is close to the 0.663 of CBPS.

At \(N=265\) with \(\kappa=1.5\), BW introduces tolerances in 1,150 of the 3,000 repetitions. On the 1,847 repetitions with exact balance at every stage, BW has a bias of 0.364 and an RMSE of 0.612, compared with 0.395 and 0.705 for PS and 0.388 and 0.585 for CBPS. On the 1,150 repetitions with tolerances, BW has a bias of 0.350 and an RMSE of 0.744, compared with 0.448 and 0.859 for PS and 0.418 and 0.773 for CBPS.

\end{document}